\documentclass[
preprint,a4paper,
secnumroman,
amssymb, amsmath,
jcp,
bibtex
]{revtex4-1}

\usepackage[english]{babel}
\babeltags{en=english}
\babeltags{eng=english}

\usepackage{CJK}

\usepackage{graphicx}
\usepackage{dcolumn}
\usepackage{siunitx}
\usepackage{bm}
\usepackage{caption}
\usepackage{subcaption}
\usepackage[usenames,dvipsnames,svgnames,table]{xcolor}

\expandafter\ifx\csname package@font\endcsname\relax\else
\expandafter\expandafter
\expandafter\usepackage
\expandafter\expandafter
\expandafter{\csname package@font\endcsname}%
\fi

\begin{document}
    
    \begin{CJK*}{GB}{}
        \title{Hydrodynamic coupling for particle-based solvent-free membrane models}
        \author{Mohsen Sadeghi}
        \thanks{Corresponding author}
        \email{mohsen.sadeghi@fu-berlin.de}
        \affiliation {Department of Mathematics and Computer Science, Freie Universit{\"a}t Berlin, Arnimallee 6, 14195 Berlin, Germany}
        \author{Frank No\'e}
        \thanks{Corresponding author}
        \email{frank.noe@fu-berlin.de}
        \affiliation {Department of Mathematics and Computer Science, Freie Universit{\"a}t Berlin, Arnimallee 6, 14195 Berlin, Germany}
        \begin{abstract}
            The great challenge with biological membrane systems is the wide range of scales involved, from nanometers and picoseconds for individual lipids, to the micrometers and beyond millisecond for cellular signalling processes. While solvent-free coarse-grained membrane models are convenient for large-scale simulations, and promising to provide insight into slow processes involving membranes, these models usually have unrealistic kinetics. One major obstacle is the lack of an equally convenient way of introducing hydrodynamic coupling without significantly increasing the computational cost of the model. To address this, we introduce a framework based on anisotropic Langevin dynamics, for which major in-plane and out-of-plane hydrodynamic effects are modeled via friction and diffusion tensors from analytical or semi-analytical solutions to Stokes hydrodynamic equations. Using this framework, we obtain accurate dispersion relations for planar membrane patches, both free-standing and in the vicinity of a wall. We also briefly discuss how non-equilibrium dynamics is affected by hydrodynamic interactions.
        \end{abstract}
        \maketitle
    \end{CJK*}

\section{Introduction}
The importance of lipid bilayers can hardly be exaggerated, considering the key roles they play in cellular transport, proliferation, motility and signal transduction \cite{Alberts2015, Kaksonen2018a, Dimou2019}. This has fueled their compelling biophysics to be extensively studied during the past four decades \cite{Brochard1975, Prost1998, Seifert1997, Shillcock2006, Marrink2009, Deserno2009, Noguchi2009, Lipowsky2018}. Numerous computational models have been proposed to simulate biomembranes at different scales \cite{Marrink2019,Friedman2018}, ranging from all-atom models \cite{Ollila2016, Poger2016, Mckiernan2016} to coarse-grained \cite{Marrink2007, Marrink2013, Deserno2014, Arnarez2015, MohamedLaradji12016} and mesoscopic variants \cite{Ayton2009, Davtyan2017, Feng2018, Sadeghi2018}. Considering length- and time-scales involved in biological processes such as exo/endocytosis \cite{Haucke2011a,Kaksonen2018a} or membrane constriction by dynamin helices \cite{Noel2019,Daumke2016,Antonny2016}, using highly granulated models often becomes a necessity  \cite{Saunders2013}. The so-called interacting particle reaction-dynamics (iPRD) models are good examples \cite{Hoffmann2019, SchoenebergEtAl_NatComm17_SNX9, Frohner2018, BiedermannEtAl_BJ15_ReaddyMM, Vijaykumar2015, Gunkel2015, SchoenebergUllrichNoe_BMC14_RDReview, SchoenebergEtAl_BJ14_PhototransductionKinetics, Ullrich2015}. Coarse-graining approaches mostly focus on reproducing equilibrium properties \cite{Saunders2013}, and generally lack realistic kinetics due to artificially smoothed energy landscape \cite{Marrink2007}, whereas membrane biology involves an abundance of non-equilibrium active processes \cite{Turlier2018, Betz2009, Qian2007} for which the kinetics is a pivotal factor.

Diverse dynamical processes associated with membranes and membrane proteins occur at rates easily spanning more than 15 orders of magnitude \cite{Gennis1989}. As we will also show in Sec. \ref{sec:intro_membrane_hydrodynamics}, large-scale membrane kinetics is mainly affected by hydrodynamic coupling between membrane and the solvent \cite{Brown2011a}. This fact renders the lack of realistic representation in the so-called solvent-free coarse-grained models much more pronounced \cite{Drouffe1991,Cooke2005a,Wang2005,Deserno2014,Arnarez2015}. To amend this, different approaches, from simple time-mapping \cite{Fritz2011}, to coarse-grained explicit solvents \cite{Ayton2006, Shkulipa2006a, Huang2012, Zgorski2016}, and more elaborately, grid-based solutions such as the lattice Boltzmann \cite{Street2006, Botan2017} and the stochastic immersed boundary methods \cite{Atzberger2007} have been proposed. Time-mapping techniques fall short when multiple timescales are present in the system, which necessarily is the case when both in-plane diffusion and out-of-plane dynamics are considered. Explicit solvent and grid-based models, while being more accurate, introduce a significant computational overhead. More importantly, these approaches necessarily force a limited and usually small length-scale to hydrodynamics due to finite simulation box sizes \cite{Venable2017, Vogele2018, Brown2011a}. 

Here, we aim to introduce a computationally efficient model that naturally integrates into particle-based models of cellular reaction kinetics and captures large-scale hydrodynamic interactions. We have recently proposed a parametric membrane model suitable for iPRD simulations, which accurately reproduces bending rigidity, area compressibility, in-plane fluidity, and budding of biomembranes \cite{Sadeghi2018}. We have already showcased the general application of the presented approach to large-scale membrane simulations based on this model \cite{Sadeghi2020}, with the emphasis on short-range hydrodynamics with less computational cost, facilitating hundreds-of-millisecond long simulations. In this paper, we extend and generalize the theory, work out special cases for different membrane geometries, and present results focused on long-range interactions.

\section{Hydrodynamic effects in membrane systems}
\label{sec:intro_membrane_hydrodynamics}
\begin{figure*}[htbp]
    \centering
    \begin{subfigure}{0.05\linewidth}
        \caption{}
        \label{fig:membrane_HI_schematics}
    \end{subfigure}
    \begin{subfigure}{0.94\linewidth}
        \includegraphics[width=\linewidth]{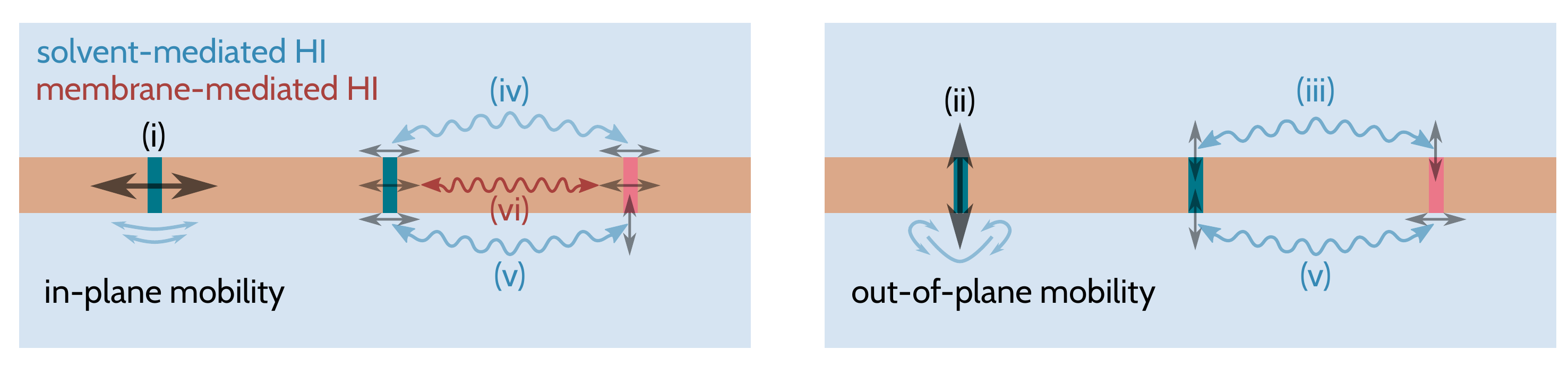}
    \end{subfigure}
    \begin{subfigure}{0.05\linewidth}
        \caption{}
        \label{fig:membrane_schematics}
    \end{subfigure}
    \begin{subfigure}{0.5\linewidth}
        \includegraphics[width=\linewidth]{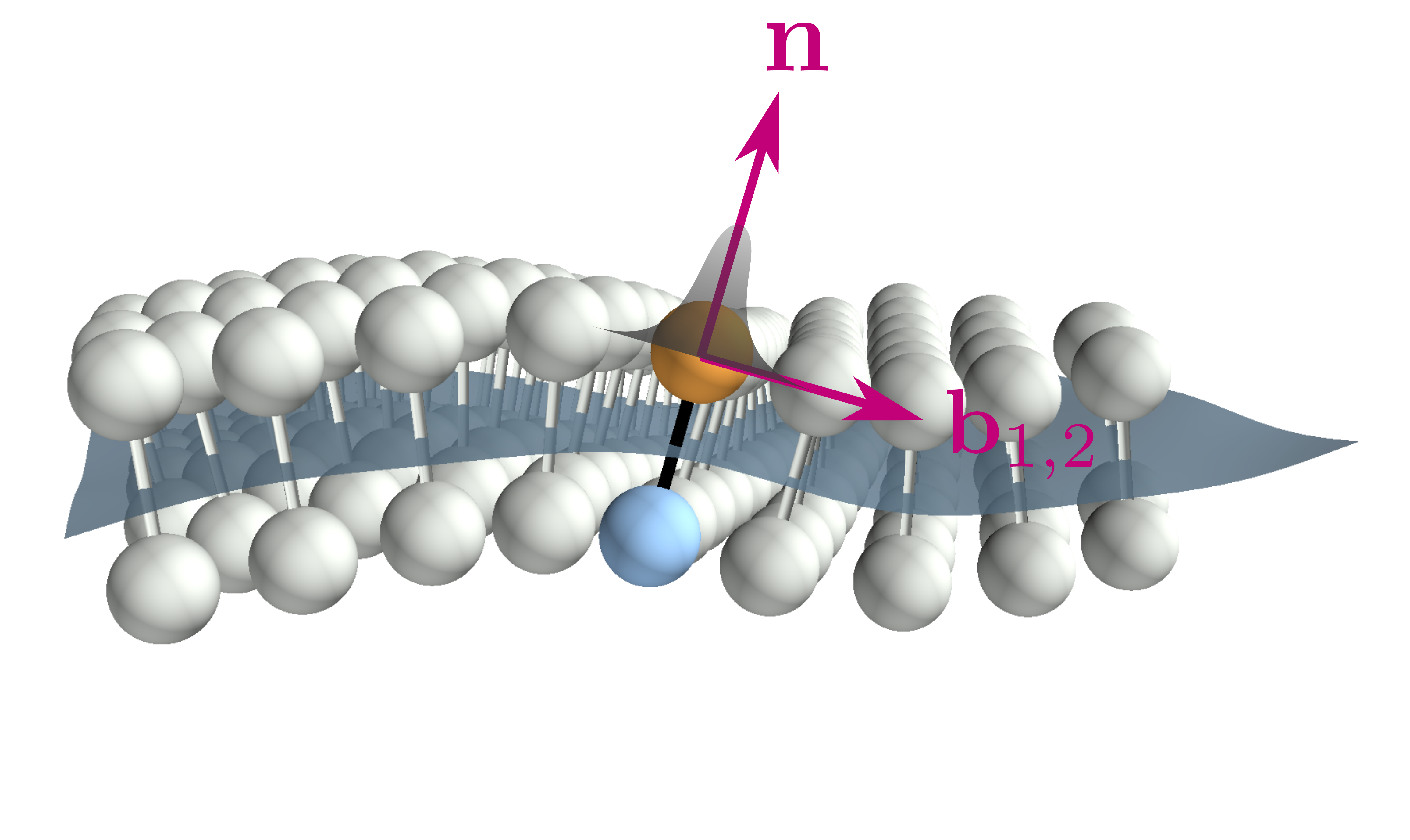}
    \end{subfigure}
    \begin{subfigure}{0.05\linewidth}
        \caption{}
        \label{fig:hydro_geometries}
    \end{subfigure}
    \begin{subfigure}{0.38\linewidth}
        \includegraphics[width=\linewidth]{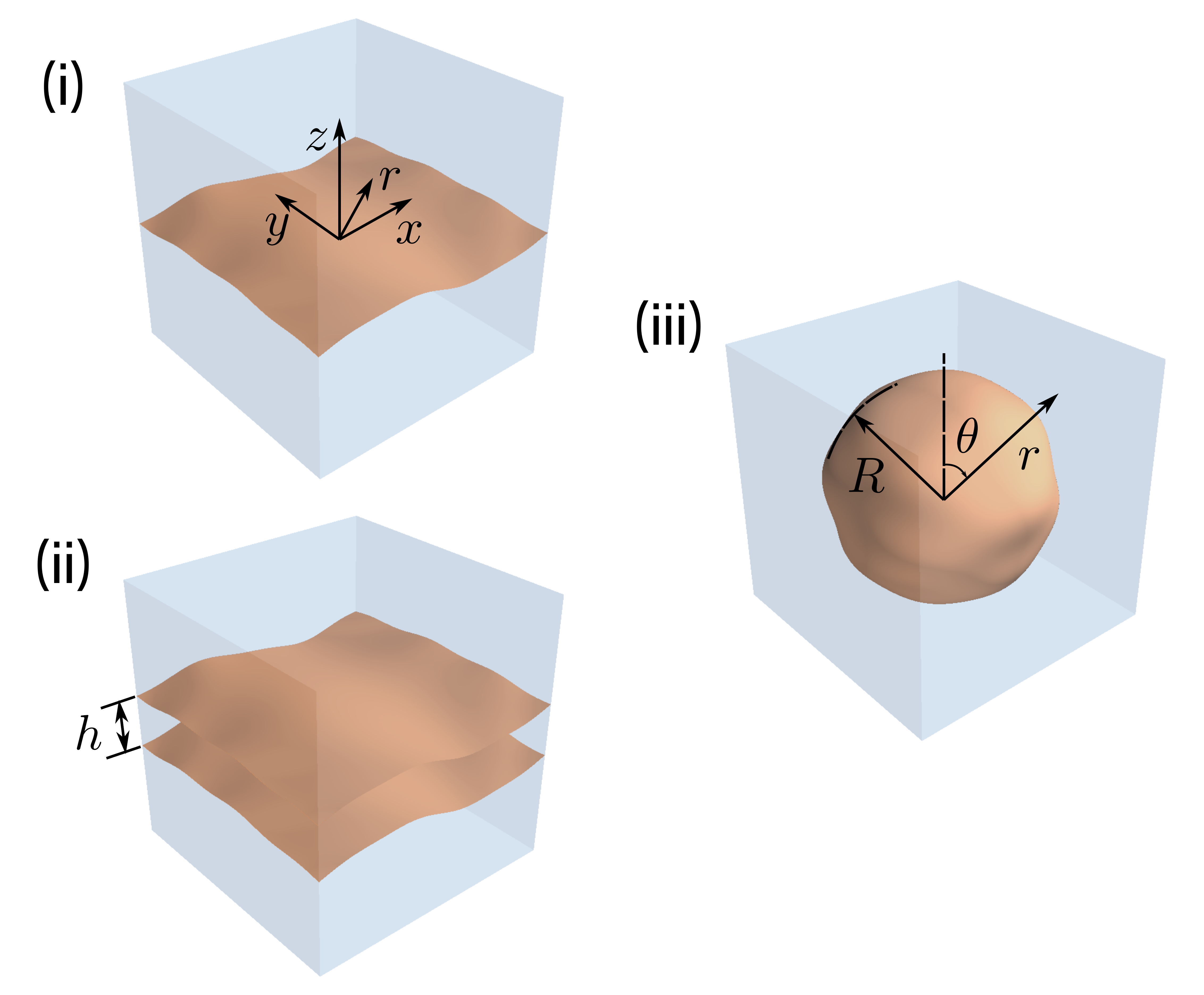}
    \end{subfigure}
    \caption{\footnotesize Schematic of the introduced framework for hydrodynamic coupling. (\textbf{a}) Components of a comprehensive description of hydrodynamic effects related to the membrane and the surrounding solvent. Distinction is made between the mobility of particles parallel to the membrane (in-plane) and perpendicular to it (out-of-plane), as well as how these mobilities are potentially coupled via hydrodynamic interactions (HI) of solvent- or membrane-mediated origin. (\textbf{b}) The particle-based membrane model composed of a close-packed lattice of representative particle-dimers. The local coordinate system describing the in-plane and out-of-plane directions, as well as a schematic of the Gaussian function used to represent velocity or stress boundary conditions per particle, are also shown for a selected particle. (\textbf{c}) Three distinct membrane geometries used in the derivation of friction and diffusion tensors: (i) single planar membrane, (ii) parallel planar membranes, (iii) spherical vesicle.}
    \label{fig:membrane_hydro_schematics}
\end{figure*}

Apart from microscopic effects such as protonation \cite{Yue2019} and hydration \cite{Schlaich2017}, presence of the solvent affects the macroscopic dynamics of the membrane via (a) incompressibility constraints (volume preservation) and (b) hydrodynamic interactions. Volumetric constraints can usually be introduced as penalizing potentials \cite{Li2018, Noguchi2005a}, but hydrodynamic interactions and their dynamical effects are much more elusive. In an admittedly arbitrary decomposition, we recognize at least six different effects innate to the hydrodynamics of membrane system (Fig. \ref{fig:membrane_HI_schematics}). For the lipids or membrane proteins, in-plane mobility is governed by the viscosity of the membrane as well as the corresponding transverse flow in the solvent (effect (i) in Fig. \ref{fig:membrane_HI_schematics}). This is the focus of the pioneering work of Saffman and Delbr\"uck \cite{Saffman1975, Saffman1976} (see Eq. (\ref{eq:saffman_delbruck})). This in-plane effect is also accompanied by long-range interactions pertaining to in-plane hydrodynamics of the membrane as a 2D fluid, as well as propagation of shear through the solvent (respectively, effects (vi) and (iv) in Fig. \ref{fig:membrane_HI_schematics}). These effects are extremely important in the dynamics of aggregation and phase separation (see \cite{Brown2011a} and references within), lipid diffusion \cite{Panzuela2018}, fluctuation of lipid domains \cite{Camley2010}, and dynamics of embedded inclusions \cite{Sorkin2020, Camley2019, Oppenheimer2010, Oppenheimer2011}. Finally, the in-plane hydrodynamics has also been shown to affect the macroscopic dynamics of the membrane \cite{Arroyo2010}.  

On the other hand, as important, but much less studied, are hydrodynamic effects due to out-of-plane motion, which will be our main focus. To clarify, by out-of-plane motion we refer to bending of the membrane as result of collective motion of constituents, and not the protrusion of individual lipids (effect (ii) in Fig. \ref{fig:membrane_HI_schematics}). Out-of-plane motions are also subject to solvent-mediated hydrodynamic interactions (effect (iii) in Fig. \ref{fig:membrane_HI_schematics}).  These interactions can be rigorously studied in the context of continuum membrane models via Green's function kernels \cite{Kramer1971a, Brochard1975, Seifert1994, Seifert1994a, Prost1998, Brown2011a}. Although successful in describing macroscopic experimental observations \cite{Pfeiffer1993, Kaizuka2006, Peukes2014}, as previously observed by other researchers \cite{Ayton2006}, these solutions are not readily applicable to particle-based simulations. Towards tackling this challenge, we propose anisotropic stochastic dynamics to be used in conjunction with the hydrodynamics obtained from Stokes equations in the form of the fluid response to localized Gaussian displacements or forces. In the presented work, we propose a model that includes effects (i), (ii), and (iii) (Fig. \ref{fig:membrane_HI_schematics}). The assumed decoupling of in-plane and out-of-plane hydrodynamic interactions is justified for small out-of-plane displacements \cite{Brown2011a}. However, within the presented framework, a fully coupled treatment is also possible.

Before continuing, and to emphasize the importance of the out-of-plane kinetics, it is instructive to examine a scaling argument. Consider a membrane with bending rigidity $\kappa$ involved in a budding process through an aperture of radius $R$. The solvent resists the outward motion of the bud with a force proportional to $\eta R^2 \dot{\theta}/\sin\theta$ where $\theta$ is half the central angle facing the bud. Deformation of the membrane into the spherical cap results in a returning force proportional to $\kappa\sin\theta/R$. Thus, the timescale governing the formation of the bud is $\eta R^3/\kappa$. The $R^3$ dependence guarantees that the out-of-plane hydrodynamic dissipation overshadows other mechanisms at a large enough scale, and becomes the sole factor determining membrane kinetics. This fact also explains the lack of attention to out-of-plane hydrodynamics for most particle-based models. In most investigations, small patches of membrane are considered and trajectories are not long enough for the membrane undulations and their kinetics to have a detectable effect \cite{Marrink2019}. But the inevitable move to large-scale simulations that elucidate slow dynamics in processes such as membranes remodeling renders this study indispensable.
\section{Stochastic dynamics of a particle-based membrane model}
\label{sec:stochastic_dynamics}
Dynamics of particles floating in a fluid environment is very well described by the Langevin equation \cite{Ermak1978},
\begin{equation}
    \label{eq:langevin}
    m_i \dot{\mathbf{v}}_i\left(t\right)=\mathbf{f}_i\left(t\right) - \sum_{j} \boldsymbol{\zeta}_{ij}\cdot \mathbf{v}_j\left(t\right) + \sum_{j} \mathbf{c}_{ij} \cdot\boldsymbol{\nu}_j\left(t\right)
\end{equation}
where $m$ and $\mathbf{v}$ respectively denote particle mass and velocity, $\mathbf{f}_i$ is the sum of forces on the $i$-th particle and $\boldsymbol{\zeta}_{ij}$'s are tensors describing pairwise friction. It is to be noted that in this form, the friction tensor simultaneously encodes dissipation and pairwise hydrodynamic interactions. Random forces are represented by the $\boldsymbol{\nu}_j\left(t\right)$ as outcomes of Gaussian processes with $\langle \boldsymbol{\nu}_i \left(t\right)\rangle=0$ and $\langle \boldsymbol{\nu}_i \left(t\right) \boldsymbol{\nu}_j \left(t^\prime \right)\rangle = 2 \delta_{ij} \delta \left(t - t^\prime \right) \mathbf{I}$, having $\boldsymbol{\zeta}_{ij} = \frac{1}{kT} \sum_{l} \mathbf{c}_{il} \, \cdot\mathbf{c}_{jl}$. If we consider such a description in the over-damped regime, the following discretized equation can be used for updating the particle positions \cite{Ermak1978},
\begin{equation}
\label{eq:overdamped_langevin}
\Delta \mathbf{r}_i=\Delta t \sum_j \boldsymbol{\nabla}_j \cdot \mathbf{D}_{ij} +\frac{\Delta t}{kT} \sum_j \mathbf{D}_{ij}\cdot \mathbf{f}_j\left(t\right) + \boldsymbol{\chi}_i \left(\Delta t\right)
\end{equation}
where $\mathbf{D}_{ij}$ is the pairwise diffusion tensor, $\boldsymbol{\nabla}_j \cdot$ denotes the divergence with respect to the position of the particle $j$, and the noise term, $\boldsymbol{\chi}_i \left(\Delta t\right)$, is described by the moments,
\begin{subequations}
\label{eq:noise_term}
\begin{align}
\langle \boldsymbol{\chi}_i \left(\Delta t\right) \rangle &= 0\\
\langle \boldsymbol{\chi}_i \left(\Delta t\right) \boldsymbol{\chi}_j \left(\Delta t\right) \rangle &= 2 \mathbf{D}_{ij} \Delta t
\end{align}
\end{subequations}
with no correlation between subsequent time steps. For spherical particles dispersed in dilute solutions, several approximations of the $\mathbf{D}_{ij}$ tensor are available. The simplest, and most widely used approach, which completely neglects hydrodynamic interactions, is the Stokes-Einstein formula, $\mathbf{D}_{ij}=\frac{kT}{6\pi \eta R}\delta_{ij}\mathbf{I}$ \cite{Stokes1851, Einstein1905}, with $k$ being the Boltzmann constant, $T$ the temperature, $\eta$ the viscosity of the solvent, and $R$ the particle radius. Hydrodynamic interactions can be additionally modeled via finding the point-force solution to Stokes equations (Eq. (\ref{eq:stokes})), i.e. Stokeslet or the Oseen tensor \cite{HiromiYamakawa}. This adds $\mathbf{D}_{ij}=\frac{kT}{8\pi \eta r_{ij}} \left(\mathbf{I}+\frac{\mathbf{r}_{ij} \mathbf{r}_{ij}}{r_{ij}^2} \right)$ with $i\neq j$ for hydrodynamic interactions between particle pairs. Further improvements to this model are also available in the form of Rotne-Prager \cite{Rotne1969} and Rotne-Prager-Yamakawa \cite{Yamakawa1970} tensors.

These hydrodynamic models are only valid in the limit of dilute solutions, with interactions calculated for a pair of particles in an infinite fluid domain, unaffected by the rest of the system. Unfortunately, it is not trivial to include these higher order contributions in the hydrodynamic model, and also, the implied assumption that the hydrodynamic interactions are pairwise additive does not generally hold. Extending models such as the Oseen tensor to the dense assembly of particles in a coarse-grained membrane model is thus not advisable. Even when hydrodynamic interactions are neglected, as will be shown, the out-of-plane mobility of membrane particles are poorly described by the Stokes-Einstein formula.

Consider a membrane model, as schematically shown in Fig. \ref{fig:membrane_schematics}, with local orthonormal bases at the outer surface of the leaflets, and the displacement of each particle decomposed as the sum of in-plane and out-of-plane contributions, respectively projected on the $\mathbf{n}$- and $\mathbf{b}_{1,2}$-vectors. We propose the following general form for the diffusion tensor,
\begin{equation}
\begin{split}
\mathbf{D}_{ii}&=D^{\parallel}_i \mathbf{I} + (D^{\perp}_{ii} - D^{\parallel}_i)\mathbf{n}_i\mathbf{n}_i\\
\mathbf{D}_{ij}&=D^{\perp}_{ij}\mathbf{n}_i\mathbf{n}_j\,\,\,\, , \,\,\,\, i \neq j
\end{split}
\label{eq:general_diffusion_tensor}
\end{equation}
with $D^{\parallel}$ and $D^{\perp}$ respectively representing the in-plane and the out-of-plane diffusion coefficients and $\mathbf{n}_i$ being the normal vector at the position of the $i$-th particle (Fig. \ref{fig:membrane_schematics}). A similar description holds for friction tensors with the $\zeta^{\parallel}$ and $\zeta^{\perp}$ components. In proposing Eq. (\ref{eq:general_diffusion_tensor}), we have made the following assumptions: (a) each particle experiences anisotropic diffusion due to the difference between in-plane and out-of-plane mobilities, (b) hydrodynamic interactions mediated by the solvent exist between pairs of particles, but are limited to forces acting along the membrane normals. Before moving forward, we can inspect the divergence terms, $\boldsymbol{\nabla}_j \cdot \mathbf{D}_{ij}$ in Eq. (\ref{eq:overdamped_langevin}). Unlike the Oseen or Rotne-Prager tensors, these terms do not vanish identically, and are instead given by,
\begin{equation}
\begin{split}
\boldsymbol{\nabla}_i \cdot \mathbf{D}_{ii}&=(D^{\perp}_{ii} - D^{\parallel}_i) \left(\boldsymbol{\nabla}_i \cdot \mathbf{n}_i\right)\mathbf{n}_i\\
\boldsymbol{\nabla}_j \cdot \mathbf{D}_{ij}&=\left(\mathbf{n}_i \cdot \boldsymbol{\nabla}_j D^{\perp}_{ij} \right)\mathbf{n}_j + D^{\perp}_{ij} \mathbf{n}_i \cdot \boldsymbol{\nabla}_j\mathbf{n}_j
\,\,\,\, , \,\,\,\, i \neq j
\end{split}
\label{eq:divergence_terms}
\end{equation}
It can readily be verified that for flat membranes, both expressions are identically zero (note that $\mathbf{n}_i \cdot \boldsymbol{\nabla}_j$ is the directional derivative along the $\mathbf{n}_i$, and we will derive $D^{\perp}_{ij}$ as a sole function of in-plane distance between $i$ and $j$ particles). It is also reasonable to expect them to be negligible for small membrane curvatures. But most importantly, in proposing Eq. (\ref{eq:general_diffusion_tensor}), we have assumed the membrane to have non-zero thickness, with diffusion tensors used for the two leaflets separately. If we compress the membrane to zero thickness, both expressions in Eq. (\ref{eq:divergence_terms}) become identically zero and the diffusion tensor becomes incompressible.  

For the in-plane mobility of particles, in order to locally include transverse solvent effects (Fig. \ref{fig:membrane_HI_schematics}-(i)), we use the well-established Saffman-Delbr\"{u}ck model of the diffusion of cylindrical inclusions in fluid sheets \cite{Saffman1975,Saffman1976},
\begin{equation}
D^{\parallel}_i= \frac{kT}{4\pi\,\mu_\mathrm{m} d_\mathrm{m}} \left[\ln\left(\frac{\mu_\mathrm{m} d_\mathrm{m}}{\eta R_i} \right) - \gamma \right]
\label{eq:saffman_delbruck}
\end{equation}
with $\gamma\approx 0.577$ being the Euler--Mascheroni constant. Now, if we consider our recently developed membrane model \cite{Sadeghi2018}, which shares some aspects of triangulated membrane models \cite{Noguchi2005, Noguchi2005a,Bahrami2017}, we can consider membrane particles as disk-like object diffusing in a medium with a predefined viscosity $\mu_\mathrm{m}$. We consider the radius $R_i$ as half the lattice parameter of the model, and $d_\mathrm{m}$ the effective thickness attributed to the membrane. 
\section {Hydrodynamics of the fluid domain in the vicinity of membranes}
\label{sec:solvent_hydrodynamics}
We derive expressions for $D^{\perp}_{ij}$ (or $\zeta^{\perp}_{ij}$) through finding the response of the fluid domain to prescribed velocity and stress boundary conditions on the membrane surface. On the scales of interest, the viscous forces dominate and the inertia-less Stokes equations hold:
\begin{subequations}
    \label{eq:stokes}
    \begin{align}
    \label{eq:stokes_momentum}
    \eta\nabla^2\mathbf{v} &=\boldsymbol{\nabla} p\\
    \label{eq:stokes_continuity}
    \boldsymbol{\nabla}\cdot\mathbf{v} &=0
    \end{align}
\end{subequations}
where $\mathbf{v}\left(\mathbf{r}\right)$ and $p\left(\mathbf{r}\right)$ denote velocity and pressure fields. Considering the solvent to be an incompressible Newtonian fluid, the stress tensor is given by:
\begin{equation}
\label{eq:stress}
\boldsymbol{\sigma} =\eta \left[ \left(\boldsymbol{\nabla} \mathbf{v}\right) + \left(\boldsymbol{\nabla} \mathbf{v}\right)^\mathrm{T}\right] - p \mathbf{I}
\end{equation}
We consider membranes with the three geometries shown in Fig. \ref{fig:hydro_geometries}, which instead of point forces or displacements, are subject to Gaussian velocity or stress boundary conditions of the following form,
\begin{equation}
v^{\perp}\left(s\right) = \frac{W}{4\pi\alpha^2} \,e^{-{s ^ 2}/{4\alpha^2}}\,\,\,\, ,\,\,\,\, 
\sigma^{\perp}\left(s\right) = -\frac{F}{4\pi\alpha^2}\,e^{-{s ^ 2}/{4\alpha^2}}
\label{eq:gaussian_bc}
\end{equation}
centered at the position of an individual particle (Fig. \ref{fig:membrane_schematics}), with the length-scale parameter $\alpha$ giving the ``size'' of particles from the perspective of the fluid domain. We generally assume the fluid domain to be infinite, and the out-of-plane fluctuations of the membrane to be negligible in comparison.
\subsection {Single planar membrane}
\label{sec:method_hydrodynamics_single_planar_membrane}
For this geometry, we are looking for solutions in the $\mathbb{R}^3_{z^+}$ half-space for prescribed velocities or stresses at $z=0$ (Fig. \ref{fig:hydro_geometries}-(i)). Consider the Fourier transform of the velocity field only in the $x$ and $y$ directions,
\begin{equation}
\label{eq:fourier_vel}
\mathbf{v} (x, y, z) = \frac{1}{\left(2\pi\right)^2}\int{d^2 q \,\,\tilde{\mathbf{v}}\left(q_1, q_2, z\right) \exp(i\mathbf{q} \cdot \mathbf{r})}
\end{equation}
where $\mathbf{q} = \left(q_1, q_2, 0\right)$. Similar Fourier transforms of the pressure and the stress field are denoted by $\tilde{p}\left(q_1, q_2, z\right)$ and $\tilde{\sigma}\left(q_1, q_2, z\right)$. Additionally, we use the orthonormal basis $\hat{\mathbf{z}}$, $\hat{\mathbf{q}}$, and $\hat{\mathbf{z}}\times\hat{\mathbf{q}}$ in the Fourier space \cite{Kramer1971a}, with the respective components of the $\tilde{\mathbf{v}}$ vector field given by $\tilde{v}_{\perp} = \left(\hat{\mathbf{z}}\times\hat{\mathbf{q}}\right) \cdot \tilde{\mathbf{v}}$, $\tilde{v}_{\parallel} = \hat{\mathbf{q}} \cdot \tilde{\mathbf{v}}$, and $\tilde{v}_{z} = \hat{\mathbf{z}} \cdot \tilde{\mathbf{v}}$. Thus transforming both sides of the continuity equation (Eq. (\ref{eq:stokes_continuity})) yields,
\begin{equation}
\label{eq:fourier_continuity}
i q \tilde{v}_{\parallel} + \frac{\partial \tilde{v}_z}{\partial z}=0
\end{equation}
Similarly, for the momentum diffusion equation (Eq. (\ref{eq:stokes_momentum})):
\begin{equation}
\label{eq:fourier_momentum1}
\eta \left (-q^2 + \frac{\partial^2}{\partial z^2} \right) \tilde{\mathbf{v}} =\left(i\mathbf{q} + \hat{\mathbf{z}}\frac{\partial}{\partial z}\right)\tilde{p}
\end{equation}
which in combination, yield the general solution,
\begin{subequations}
\label{eq:fourier_solutions}
\begin{align}
\label{eq:fourier_solution_pressure}
\tilde{p}&=A_1\left(\mathbf{q}\right) \exp (-q z) + A_2\left(\mathbf{q}\right) \exp (q z)\\
\label{eq:fourier_solution_v_perp}
\tilde{v}_{\perp} &= B_1\left(\mathbf{q}\right) \exp (-q z) + B_2\left(\mathbf{q}\right) \exp (q z)\\
\label{eq:fourier_solution_v_para}
\tilde{v}_{\parallel}=&i\,\left[\frac{A_1\left(\mathbf{q}\right)}{2\eta}\left( \frac{1}{q} - z \right) - C_1\left(\mathbf{q}\right)\right] \exp (-q z)\nonumber\\&+i\,\left[\frac{A_2\left(\mathbf{q}\right)}{2\eta}\left(\frac{1}{2q} + z \right) + C_2\left(\mathbf{q}\right)\right] \exp (q z)\\
\label{eq:fourier_solution_v_z}
\tilde{v}_z=&\left[\frac{A_1\left(\mathbf{q}\right)}{2\eta}\, z + C_1\left(\mathbf{q}\right)\right] \exp (-q z) \nonumber\\&+\left[\frac{A_2\left(\mathbf{q}\right)}{2\eta}\left(-\frac{1}{2q} + z \right) + C_2\left(\mathbf{q}\right) \right] \exp (q z)
\end{align}
\end{subequations}
Using Eq. (\ref{eq:stress}), the normal stress in the $z$ direction is
\begin{equation}
\label{eq:fourier_stress_z}
\begin{split}
\tilde{\sigma}_{zz}=&-\left[A_1\left(\mathbf{q}\right) z + 2\eta C_1\left(\mathbf{q}\right) \right] \, q\, \exp (-q z)\\
&+\left[A_2\left(\mathbf{q}\right) \left(-\frac{1}{2q} + z\right) + 2\eta C_2\left(\mathbf{q}\right) \right] \, q\, \exp (q z)
\end{split}
\end{equation}
The velocity and the pressure fields should remain bounded as $z\rightarrow \infty$, leading to $A_2\left(\mathbf{q}\right) = B_2\left(\mathbf{q}\right)= C_2\left(\mathbf{q}\right) = 0$. On the $z=0$ boundary, the solution reduces to:
\begin{subequations}
    \begin{align}
        \label{eq:velocity_bc_z}
        \tilde{v}_z\left(z=0\right) &= C_1\left(\mathbf{q}\right)\\
        \label{eq:velocity_bc_parallel}
        \tilde{v}_{\parallel}\left(z=0\right) &=i\,\left[\frac{A_1\left(\mathbf{q}\right)}{2\eta q} - C_1\left(\mathbf{q}\right)\right]\\
        \label{eq:stress_bc_zz}
        \tilde{\sigma}_{zz}\left(z=0\right) &= - 2\eta C_1\left(\mathbf{q}\right) \, q
    \end{align}
\end{subequations}
The boundary at $z=0$ is formed by the particle-based membrane. If the no-slip condition is assumed at the interface, the fluid velocity field and the surface velocity distribution, $w\left(x,y\right)$, which is dictated by the motion of membrane particles, should coincide. Assuming the membrane to also follow the continuity condition of an incompressible fluid, we have  $i\,q\,\tilde{w}_{\parallel}\left(\mathbf{q}\right)=0$. Thus, at the boundary, $\tilde{v}_{\parallel}\left(\mathbf{q}, z=0\right)=\tilde{w}_{\parallel}\left(\mathbf{q}\right)=0$ and $A_1\left(\mathbf{q}\right)=2\eta \, q\, C_1\left(\mathbf{q}\right)$. It is worthwhile to pause and consider the fact that we have actively included the membrane incompressibility condition into the solution. This condition is missing from well-known Green's function solutions for out-of-plane hydrodynamics \cite{Brown2011a, Granek1997}, which reduces them to the $zz$ components of the Oseen tensor. As we will show in Sec. \ref{sec:friction_and_diffusion}, our solution only converges to the Oseen tensor for particles far apart, which means the membrane incompressibility condition is important for particle pairs in close vicinity.

If the velocity distribution on the boundary is given by $v_z^*\left(r\right)$ with $r=\sqrt{x^2 + y^2}$ and the origin coinciding with the position of a particle, the coefficient $C_1\left(\mathbf{q}\right)$ is simply given by the Fourier transform of $v_z^*\left(r\right)$ (Eq. (\ref{eq:velocity_bc_z})), which due to its rotational symmetry is related to the Hankel transform of order zero as $\tilde{v}_z^*\left(q\right)=2\pi\mathcal{H}_0\left[v_z^*\left(r\right)\right]$, with the transform pair \cite{Piessens2000,Duffy1994}
\begin{equation}
\label{eq:hankel_transform}
\begin{split}
F\left(q\right) &= \mathcal{H}_0\left[f\left(r\right)\right] = \int_{0}^{\infty} {f\left(r\right) J_0 \left (q r\right)r\,dr}\\
f\left(r\right) &= \mathcal{H}^{-1}_0\left[F\left(q\right)\right] = \int_{0}^{\infty} {F\left(q\right) J_0 \left (q r\right)q\,dq}
\end{split}
\end{equation}
where $J_0$ is the zeroth order Bessel function of the first kind. The resulting stress distribution on the boundary is obtained using the Eq. (\ref{eq:stress_bc_zz}). Making use of the connection between Fourier and Hankel transforms, we have:
\begin{equation}
\label{eq:vel_bc_surface_stress}
\sigma_{zz}\left(r, z=0\right) = -2 \eta \mathcal{H}^{-1}_0 \left[q\, \mathcal{H}_0\left[v_z^*\left(r\right)\right]\right]
\end{equation}
Similarly, if the stress is prescribed on the boundary by the function $\sigma^{*}_{zz}\left(r\right)$, the resulting velocity on the boundary is
\begin{equation}
\label{eq:stress_bc_surface_velocity}
v_z\left(r, z=0\right) = -\frac{1}{2 \eta} \mathcal{H}^{-1}_0 \left[\frac{1}{q}\, \mathcal{H}_0\left[\sigma^{*}_{zz}\left(r\right)\right]\right]
\end{equation}
For the velocity boundary condition given by Eq. (\ref{eq:gaussian_bc}), with $v_z^*\left(r\right) = v^\perp\left(r\right)$, $W$ acts as an effective flux, such that $W=A_p v_p$ with $A_p$ being the area per particle and $v_p$, an effective particle velocity. Performing the Hankel transforms in Eq. (\ref{eq:vel_bc_surface_stress}) \cite{Piessens2000,Duffy1994}, we get (Fig. \ref{fig:single_membrane}),
\begin{equation}
\label{eq:surface_stress_gaussian_2}
\begin{split}
\sigma_{zz}\left(r, z=0\right) &= -\frac{\eta W}{4 \sqrt{\pi} \alpha^3} \,\times\, e ^ {-\xi} \left[\left(1-2 \xi\right)I_0 \left(\xi \right) + 2 \xi I_1 \left(\xi\right) \right]
\end{split}
\end{equation}
where $\xi = \frac{r^2}{8\alpha^2}$ and $I_0$ and $I_1$ are the modified Bessel functions of the first kind. Similarly, applying the Gaussian stress boundary condition of Eq. (\ref{eq:gaussian_bc}) to Eq. (\ref{eq:stress_bc_surface_velocity}) yields (Fig. \ref{fig:single_membrane}),
\begin{equation}
\label{eq:surface_velocity_gaussian_2}
\begin{split}
v_z\left(r, z=0\right) &= \frac{F}{8\sqrt{\pi}\eta \alpha} \,\times\, e ^ {-\xi} I_0 \left(\xi \right)
\end{split}
\end{equation}
where $F$ now represents the total force exerted on the membrane by the fluid. Noteworthy is that integrating the normal stress on the boundary in the former case yields zero total force.
\subsection{Parallel planar membranes}
\label{sec:method_hydrodynamics_parallel_planar_membranes}
For this case, we consider a pair of parallel planar membranes separated by a distance $h$, such that one lies on the $z=0$ and the opposing one on $z=h$ planes (Fig. \ref{fig:hydro_geometries}-(ii)). The fields given by Eqs. (\ref{eq:fourier_solution_v_para}), (\ref{eq:fourier_solution_v_z}) and (\ref{eq:fourier_stress_z}) are still valid, and only suitable boundary conditions need be applied,
\begin{equation}
\label{eq:parallel_membranes_bc}
\begin{split}
\tilde{v}_\parallel \left (\mathbf{q}, z=0\right) =& \tilde{v}_\parallel \left (\mathbf{q}, z=h\right) = 0\\
\tilde{v}_z \left (\mathbf{q}, z=0\right) = \tilde{v}^{*}_z \left (q\right)\,\, &, \,\,
\tilde{v}_z \left (\mathbf{q}, z=h\right) = 0\\
&or\\
\tilde{\sigma}_{zz} \left (\mathbf{q}, z=0\right) = \tilde{\sigma}^{*}_{zz} \left (q\right)\,\, &, \,\,
\tilde{\sigma}_{zz} \left (\mathbf{q}, z=h\right) = 0\\
\end{split}
\end{equation}
for the two scenarios, where either Gaussian velocity or stress distributions are used. Without explicitly giving the expressions for the coefficients, we reproduce the final results. For velocity boundary conditions of Eq. (\ref{eq:gaussian_bc}) applied at $z=0$,
\begin{subequations}
    \begin{align}
        \label{eq:parallel_membranes_sig_z_0}
        \tilde{\sigma}_{zz}\left(q, z=0\right) &= -2\eta\, q \left[\frac{1 + 2 \epsilon \,e^{-\epsilon} - e^{-2 \epsilon}}{1 - \left(\epsilon^2 + 2\right) e^{-\epsilon} + e^{-2 \epsilon}}\right]\tilde{v}^{*}_z\left(q\right) \\
        \label{eq:parallel_membranes_sig_z_h}
        \tilde{\sigma}_{zz}\left(q,z=h\right) &= -2\eta\,q \left[\frac{e^{\frac{-\epsilon}{2}}\left[ 2 + \epsilon + \left(\epsilon - 2\right)\,e^{-\epsilon}\right]}{1 - \left(\epsilon^2 + 2\right) e^{-\epsilon} + e^{-2 \epsilon}}\right]\tilde{v}^{*}_z\left(q\right)
    \end{align}
\end{subequations}
where $\epsilon=2\,q\,h$. It can be seen that if $q > 0$, with $h\rightarrow\infty$, the solution given in Eq. (\ref{eq:parallel_membranes_sig_z_0}) converges to the one given for a single planar membrane in Eq. (\ref{eq:stress_bc_zz}). The $q=0$ case is a special exception arising from the fact that in this pure Dirichlet problem, the pressure and consequently, stress, are undetermined up to a constant. This translates to the solutions given in Eqs. (\ref{eq:parallel_membranes_sig_z_0}) and (\ref{eq:parallel_membranes_sig_z_h}) being singular at $q=0$ in Fourier space.

For the second case, where the stress boundary conditions of Eq. (\ref{eq:gaussian_bc}) are applied at $z=0$, we have,
\begin{subequations}
    \begin{align}
        \label{eq:parallel_membranes_v_z_0}
        \tilde{v}_z\left(q, z=0\right) &= -\frac{1}{2\eta}\,\frac{1}{q} \left[\frac{1 + 2 \epsilon \,e^{-\epsilon} - e^{-2 \epsilon}}{1 - 2 e^{-\epsilon} + e^{-2 \epsilon}}\right]\tilde{\sigma}^{*}_{zz}\left(q\right) \\
        \label{eq:parallel_membranes_v_z_h}
        \tilde{v}_z\left(q,z=h\right) &= -\frac{1}{2\eta}\,\frac{1}{q} \left[\frac{e^{\frac{-\epsilon}{2}}\left(2 + \epsilon + \left(\epsilon - 2\right)\,e^{-\epsilon}\right)}{1 - 2 e^{-\epsilon} + e^{-2 \epsilon}}\right]\tilde{\sigma}^{*}_{zz}\left(q\right)
    \end{align}
\end{subequations}
Given the complexity of these expressions, it is only possible to calculate the Hankel transforms numerically. To do so, we have employed the discrete Hankel transform \cite{Johnson1987, Lemoine1994} as implemented in the GNU Scientific Library (GSL) \cite{GSLref}. To also remove the effect of the indeterminate constant pressure field in case of velocity boundary conditions, we set the resulting force on each of the two membranes to zero, in accordance with the solution for single planar membrane (Fig. \ref{fig:parallel_membranes}).

The semi-analytically obtained stress and velocity distributions on the two membranes, in response to velocity boundary conditions applied on one (Fig. \ref{fig:parallel_membranes}) are valid for hydrodynamic interaction between two parallel planar membranes as well as between a single membrane and a rigid wall in its vicinity, similar to what has been extensively investigated for continuum membrane models \cite{Brochard1975,Seifert1994,Gov2004,Kaizuka2006}. As expected, the solution for parallel membranes converges to that of the single membrane as $h\rightarrow\infty$ (Fig. \ref{fig:parallel_membranes}). Although, for values of $h$ as large as $20\alpha$, there still exists a non-negligible deviation from a single-membrane solution, especially when the stress boundary conditions are considered.
\subsection{Spherical vesicle}
\label{sec:method_hydrodynamics_spherical_vesicle}
Finally, we consider the hydrodynamics predicted by the Stokes equations around a spherical vesicle (Fig. \ref{fig:hydro_geometries}-(iii)). This is especially interesting to investigate the effect of membrane curvature on the hydrodynamics. In spherical coordinates, axisymmetric solutions to the Stokes equations are found through the divergence-free stream function, $\psi \left (r, \theta \right)$, defined such that \cite{Brenn2017},
\begin{equation}
        \label{eq:streamfunc}
        v_r\left(r, \theta \right) = -\frac{1}{r^2 \sin \left(\theta \right)}\frac{\partial \psi}{\partial \theta}\,\,\,\, , \,\,\,\,
        v_\theta\left(r, \theta \right) = \frac{1}{r \sin \left(\theta \right)}\frac{\partial \psi}{\partial r}
\end{equation}
which leads to the general solution to the Stokes equations as
\begin{equation}
\label{eq:stream_function_spherical}
\psi = \sum_{m=0}^{\infty}f_m\left(r\right)\left[A_m P_m^{\prime} \left(\cos \theta \right)\right]
\end{equation}
in which $P_m$ denotes the Legendre polynomial of order $m$ and $f_m\left(r\right)=\sum_i c_i r^k_i$ with $k_i$'s being the roots of the polynomial $k\left(k-3\right)\left[\left(k-1\right)\left(k-2\right) - 2 m \left(m + 1\right)\right]=-m \left(m + 1\right)\left(m - 2\right) \left(m + 3\right)$. Using this stream function, we obtain the velocity and stress fields as
\begin{subequations}
    \begin{align}
        \label{eq:v_r_spherical}
        v_{r} \left(r,\theta\right)&=\sum_{m=0}^{\infty}\frac{-A_m P_m \left(\cos \theta \right)}{r^2} \Big[m\left(m + 1\right)f_m\left(r\right)\Big]\\
        \label{eq:v_th_spherical}
        v_{\theta} \left(r,\theta\right)&=\sum_{m=0}^{\infty}\frac{A_m P^{\prime}_m \left(\cos \theta \right)}{r}\Big[f^{\prime}_m\left(r\right)\sin\left(\theta\right)\Big]\\
        \label{eq:stress_r_spherical}
        \begin{split}
            \sigma_{rr} \left(r,\theta\right)&=\eta \, \sum_{m=0}^{\infty}\frac{A_m P_m \left(\cos \theta \right)}{r^3}\Big[r^3 f_m^{\prime\prime\prime}\left(r\right)\\ &- 3m\left(m + 1\right) r f_m^{\prime}\left(r\right) + 6m\left(m + 1\right) f_m \left(r\right)\Big]
        \end{split}
    \end{align}
\end{subequations}
Subject to the following boundary conditions,
\begin{equation}
\label{eq:bc_spherical}
\begin{split}
v_{\theta}\left(r=R,\theta\right)&= 0\\
v_r\left(r=R,\theta\right)&=\sum_{m=0}^{\infty}v_m P_m \left(\cos \theta \right)\\
&or\\
\sigma_{rr}\left(r=R,\theta\right)&=\sum_{m=0}^{\infty}s_m P_m \left(\cos \theta \right)
\end{split}
\end{equation}
As a result of the axisymmetric flow assumption, it is not possible to apply boundary conditions reflecting divergence-free in-plane flow of membrane particles, as was used for planar membranes. Instead, we have used the stronger condition of membrane particles being frozen in-plane, with $v_{\theta}\left(r=R,\theta\right)= 0$, as well as $v_{\phi}\left(r,\theta\right)= 0$ due to the axisymmetry ($\phi$ is the azimuth angle, not shown in Fig. \ref{fig:hydro_geometries}-(iii)). We use a similar Gaussian distributed velocity or stress boundary condition at the zenith ($\theta = 0$) to satisfy rotational symmetry. We expand the Gaussian in terms of Legendre polynomials up to degree 87 to obtain values of $v_m$ or $s_m$. To conform with previous results, the distance between particles, $r_{ij}$, is considered along a great circle passing the two (Fig. \ref{fig:curved_membrane}). As expected, the resulting stress and velocity distributions approach those of a single planar membrane with increasing vesicle radius (Fig. \ref{fig:curved_membrane}). Yet, they are in surprising agreement, even for small radii. This points to the fact that the curvature in general has little effect on the hydrodynamic interactions across the membrane, when only contributions normal to the membrane surface are considered.
\section{Out-of-plane components of friction and diffusion tensors}
\label{sec:friction_and_diffusion}
To calculate the out-of-plane components of friction and diffusion tensors, $\zeta^{\perp}_{ij}$ and $D^{\perp}_{ij}$, we interpret the boundary conditions of Eq. (\ref{eq:gaussian_bc}) as test inputs, and numerically integrate the resulting fields over discrete patches on the membrane to obtain the local response. We consider the input Gaussians to be centered at the position of the particle $j$, and we take the integration domain $\Omega_i$ with area $A_p$, around particle $i$, to have,
\begin{equation}
\zeta^{\perp}_{ij} = \frac{F^{\perp}_i}{v^{\perp}_j} = \frac{-\int_{\Omega_i}{\boldsymbol{\sigma}_{n}\cdot d\mathbf{S}}}{W/A_p} \,\,\,\, , \,\,\,\,
\frac{D^{\perp}_{ij}}{kT} = \frac{v^{\perp}_i}{F^{\perp}_j} = \frac{\int_{\Omega_i}{\mathbf{v} \cdot d\mathbf{S}}}{A_p F}
\label{eq:friction_diffusion_perp_def}
\end{equation}
For a single planar membrane, analytical expression for $\zeta^{\perp}_{ii}$ and $D^{\perp}_{ii}$ components can be obtained by integrating on disks,
\begin{subequations}
    \begin{align}
    \zeta^{\perp}_{ii} &=16\pi^{3/2} \eta \alpha \times \xi_p^2 e ^ {-\xi_p} \left[I_0 \left(\xi_p \right) - I_1 \left(\xi_p\right) \right]
    \label{eq:friction_ii_single_membrane}
    \\
    D^{\perp}_{ii} &=\frac{kT}{8\sqrt{\pi}\eta \alpha} \times e ^ {-\xi_p} \left[I_0 \left(\xi_p \right) + I_1 \left(\xi_p \right)\right]
    \label{eq:diffusion_ii_single_membrane}
    \end{align}
    \label{eq:friction_diffusion_ii_single_membrane}
\end{subequations}
where $\xi_p = \frac{r_p^2}{8\alpha^2} = \frac{A_p}{8\pi\alpha^2}$. When $\alpha\rightarrow 0$, the Gaussian boundary conditions (Eq. (\ref{eq:gaussian_bc})) approach delta functions, making the solutions equivalent to Stokeslets,
\begin{equation}
    \lim_{\alpha\rightarrow 0}\zeta^{\perp}_{ii} =2 \pi \eta r_p \,\,\,\, , \,\,\,\,
    \lim_{\alpha\rightarrow 0}D^{\perp}_{ii} =\frac{kT}{2 \pi \eta r_p}
    \label{eq:friction_diffusion_ii_single_membrane_limit}
\end{equation}
Eqs. (\ref{eq:friction_diffusion_ii_single_membrane}) and (\ref{eq:friction_diffusion_ii_single_membrane_limit}) provide expressions most similar in nature to the Stokes-Einstein formula $D = kT / 6\pi \eta r_p$. Interestingly, the limiting case shows that membrane particles experience a threefold decrease in friction compared to free-floating spherical particles, which is the result of the symmetry-breaking in the fluid domain due to the presence of the membrane. Also, the solvent cannot permeate the membrane, and does not engulf the particles, but only affects them from one side. Using the approximation that for particles far apart, the expression under the integral in Eqs. (\ref{eq:friction_diffusion_perp_def}) only weakly depends on radial separation, we can also find an approximation for the $D^{\perp}_{ij}$,
\begin{equation}
    \label{eq:diffusion_ij_single_membrane}
    \lim_{\substack{r_{ij}\rightarrow \infty\\ r_{ij} \gg r_c}} D^{\perp}_{ij} \approx \frac{kT}{4 \pi \eta r_{ij}}
\end{equation} 
Which is half the $zz$ component of the Oseen tensor. In our description, particles on each leaflet experience hydrodynamic effects from a half-space of fluid. From the perspective of particles far apart, the sum of the effects corresponding to the two leaflets would thus reproduce the Oseen tensor.

Apart from the given cases, it is not in general trivial to find closed-form expressions for friction and diffusion tensors. It is however straightforward to obtain numerical results using Eqs. (\ref{eq:friction_diffusion_perp_def}). As we are interested in application of the introduced method in our particle-based membrane model \cite{Sadeghi2018, Sadeghi2020}, we have carried out the calculations for a hexagonal lattice of points with the constant lattice parameter $a$. We have used Gauss quadrature on disks with the area of $A_p$ around each particle \cite{Kim1997}, with the area per particle calculated based on the surface densities. The ratio $\alpha / a$ would serve here as an effective scaling factor. To demonstrate, we have chosen $\alpha / a = 0.1$ and $\alpha / a = 0.5$, and we have calculated $\zeta^{\perp}_{ij}$ and $D^{\perp}_{ij}$ values for (i) a single planar membrane, (ii) two systems of parallel planar membranes, respectively distanced $5.0 a$ and $10.0 a$ apart, and (iii) two spherical vesicles with the respective radii of $6.7 a$, and $13.3 a$ (Figs. \ref{fig:compiled_friction_0.1}-\ref{fig:compiled_diffusion_0.5}).
\section{Large-scale membrane kinetics from particle-based simulations}
In order to investigate the kinetics prescribed by the presented approach, we have performed simulations using our membrane model \cite{Sadeghi2018}, in which the bilayer is modeled by particle-dimers in a close-packed arrangement, with a lattice parameter of $a=$ \SI{10}{\nano\meter}. The potential functions for bonded interactions are as follows \cite{Sadeghi2018},
\begin{subequations}
    \label{eq:potential_model}
    \begin{align}
        \label{eq:potential_model_Us}
        U_\mathrm{s}\left(r_{ij}\right) &= D_e\left[1 - \exp\left(-\alpha\left(r_{ij}-r_\mathrm{eq}\right)\right)\right]^2\\
        \label{eq:potential_model_Ua}
        U_\mathrm{a}\left(\theta_{i^\prime i j}\right) &= K_a \left(\theta_{i^\prime i j}-\theta_\mathrm{eq}\right)^2\\
        \label{eq:potential_model_Ud}
        U_\mathrm{d}\left(d_{i i^\prime}\right) &= K_d \left(d_{i i^\prime}-d_\mathrm{eq}\right)^2
    \end{align}
\end{subequations}
Particles belonging to each leaflet are connected to their nearest-neighbor counterparts via Morse-type bonds (Eq. (\ref{eq:potential_model_Us})). Also, harmonic angle-bending potentials given by Eq. (\ref{eq:potential_model_Ua}) act against the out-of-plane rotations of these bonds (the primed index designates the opposing particle in a dimer). Finally, particles in a dimer are connected via harmonic bonds (Eq. (\ref{eq:potential_model_Ud})). Parameter-space optimization based on the energy density of the membrane, using the properties listed in Tab. \ref{tab:membrane_properties}, yields force field parameters (Tab. \ref{tab:pot_param}) \cite{Sadeghi2018}.

We obtain particle trajectories via updating the position of particles according to Eq. (\ref{eq:overdamped_langevin}). We have used a rather large timestep of \SI{0.5}{\nano\second} which offers a substantial improvement in the accessible simulation time when compared with the timestep of \SI{0.02}{\nano\second} used with the same model with a deterministic integrator \cite{Sadeghi2018}. We update diffusion tensors, $\mathbf{D}_{ij}$, in every 10 steps of integration, based on Eqs. (\ref{eq:general_diffusion_tensor}) and instantaneous normal vectors. We calculate normal vectors for triangles formed between in-plane bonds with contributions from neighboring triangles summed up for each particle. Diffusion tensors for pairs of particles are assembled into a global $3N\times3N$ diffusion matrix, which accounts for all degrees of freedom. In order to construct the correlated random displacements given by Eq. (\ref{eq:noise_term}), the traditional approach is to use either the Cholesky decomposition \cite{Ermak1978}, or the square root of the assembled diffusion matrix \cite{Fixman1986}. But these $\mathcal{O}\left(N^3\right)$ operations are computationally expensive, and thus, we have resorted to the approximate method developed by Geyer and Winter \cite{Geyer2009}. More elaborate approximate methods that split up hydrodynamic interactions into near- and far-field contributions are also available \cite{Banchio2003}. 

All the simulations are performed at $T=$ \SI{298}{\kelvin} and we have chosen water with the viscosity of \SI{0.890}{\milli\pascal\second} as the solvent. Two-dimensional periodic boundary conditions are applied parallel to the membrane, and in-plane degrees of freedom are coupled to the Langevin piston barostat, resulting in a tension-free membrane \cite{Feller1995}.
\begin{table}[h]
    \centering
    \caption{
        Properties of the membrane used for the parametrization of the membrane model used with simulations in Secs. \ref{sec:dispersion_free_membrane} and \ref{sec:dispersion_near_surface}. Values of the bending rigidity, $\kappa$, and Gaussian curvature modulus, $\bar{\kappa}$,  are based on data given in \cite{Marsh2006, Hu2012, Nagle2013,Dimova2014}, while for area compressibility modulus, $K_\mathrm{area}$, data from \cite{Janosi2010,Klauda2010, Raghunathan2012, Braun2013} have been considered.}
    \label{tab:membrane_properties}
    \begin{tabular}{*4{>{\centering\arraybackslash}p{0.2\columnwidth}}}
        \hline
        $d_\mathrm{m} \, (\mathrm{nm})$ & $\kappa \, (\mathrm{kT})$ & $\bar{\kappa} \, (\mathrm{kT})$ & $K_\mathrm{area} (\mathrm{N}\mathrm{m}^{-1})$\\
        \hline
        4.0 & 18.73 & -14.98 & 0.270 \\
        \hline
    \end{tabular}
\end{table}
\begin{table}[h]
    \centering
    \caption{
        Membrane force field parameters for the interactions given in Eqs. (\ref{eq:potential_model}).}
    \label{tab:pot_param}
    \begin{tabular}{*4{>{\centering\arraybackslash}p{0.2\columnwidth}}}
        \hline
        $r_\mathrm{eq}$\newline$(\mathrm{nm})$&$\theta_\mathrm{eq}$\newline$(\mathrm{rad})$&$d_\mathrm{eq}$\newline$(\mathrm{nm})$&\\
        \hline
        10.0&${\pi}/{2}$&4.0&\\
        \hline
        $D_{\mathrm{e}}$\newline$(\mathrm{kJ}/\mathrm{mol})$&$\alpha$\newline$(1/\mathrm{nm})$&$K_{\mathrm{b}}$\newline$(\mathrm{kJ}/\mathrm{mol})$& $K_{\mathrm{d}}$\newline$(\mathrm{kJ}/\mathrm{mol}\,\mathrm{nm}^2)$\\
        \hline
        9.91&0.12&20.74&6.19\\
        \hline
    \end{tabular}
\end{table}
\subsection{Dispersion relations of free-standing planar membrane patches}
\label{sec:dispersion_free_membrane}
We consider membrane patches with the side length of $L=$ \SI{0.5}{\micro\meter} to study the kinetics of their equilibrium thermal fluctuations. It is well-known that hydrodynamic interactions are in general long-range, and similar to electrostatics, methods are available for tracting their long-range contributions \cite{Ando2013}. Here, in order to reduce the complexity of the model, we have used a finite cut-off for hydrodynamic interactions, which at most coincides with half the size of the simulation box, $L/2$. Though because we have derived all the hydrodynamic effects for infinite domains, the cut-off only affects the number of contributing particles and not the accuracy of the hydrodynamics itself.

\begin{figure*}[htbp]
    \centering
    \begin{subfigure}{0.01\linewidth}
        \caption{}
        \label{fig:hydro_dispersion_cutoff}
    \end{subfigure}
    \begin{subfigure}{0.45\linewidth}
        \includegraphics[width=0.92\linewidth]{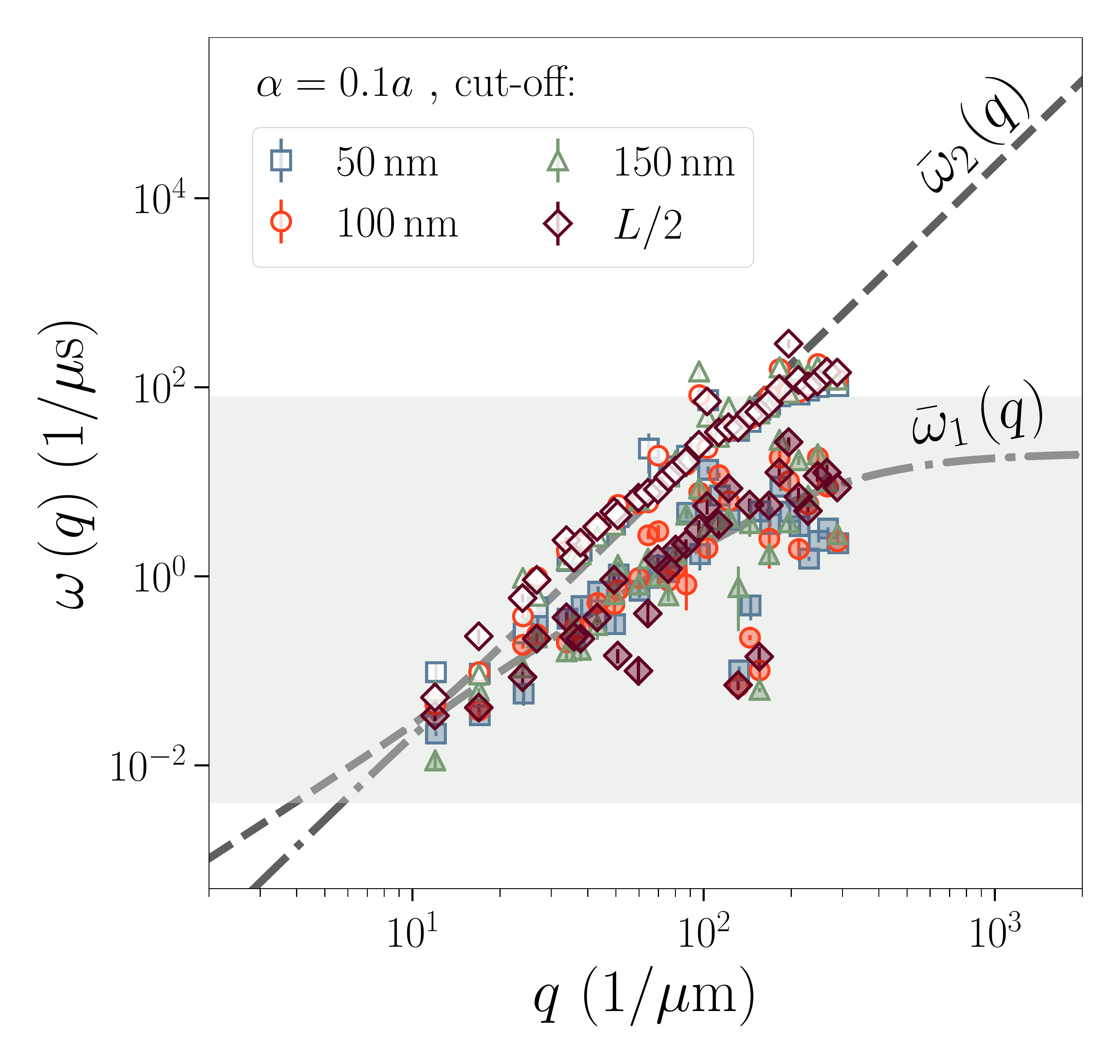}
    \end{subfigure}
    \begin{subfigure}{0.01\linewidth}
        \caption{}
        \label{fig:hydro_dispersion_alpha}
    \end{subfigure}
    \begin{subfigure}{0.45\linewidth}
        \includegraphics[width=0.92\linewidth]{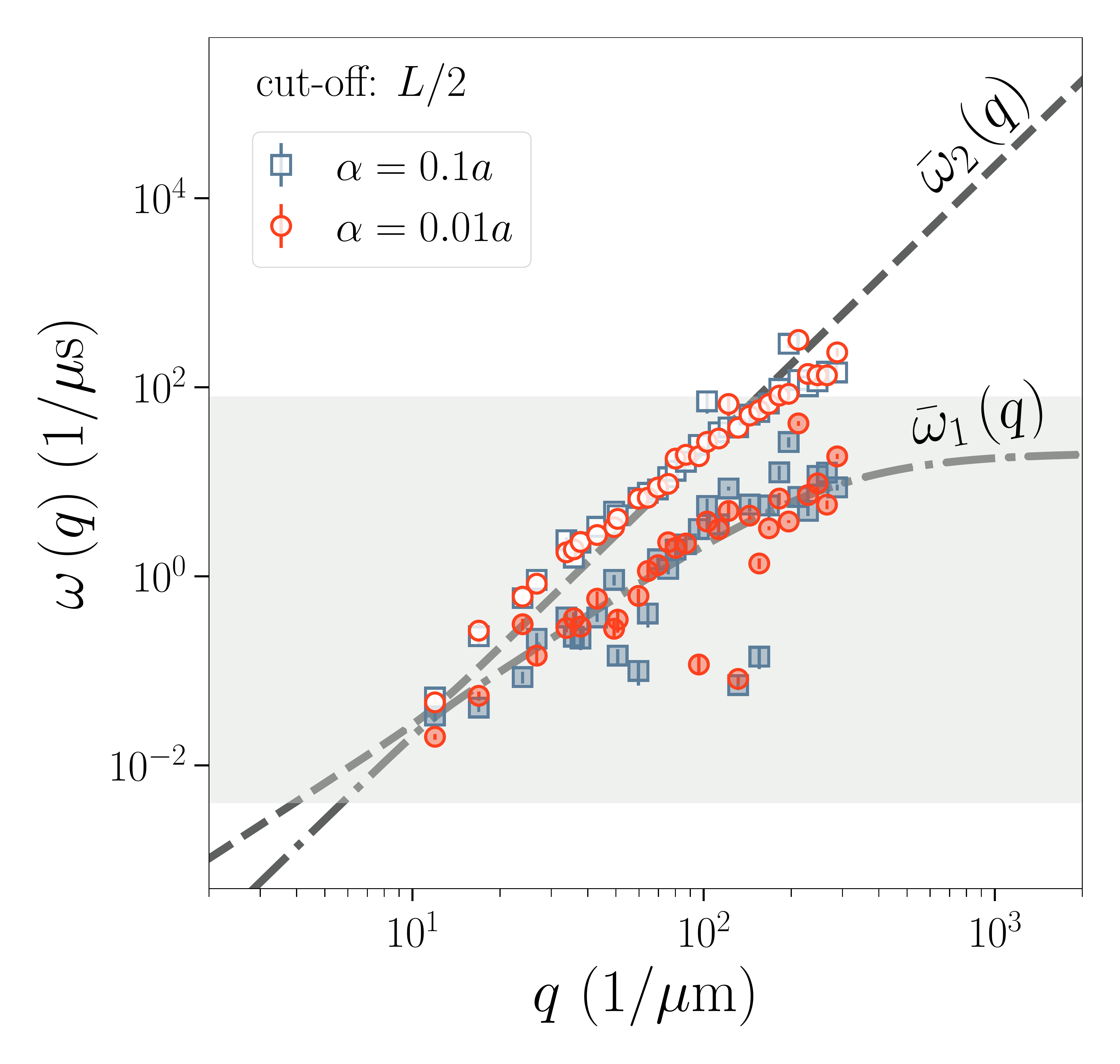}
    \end{subfigure}\\
    \begin{subfigure}{0.01\linewidth}
        \caption{}
        \label{fig:thermal_power_spectrum}
    \end{subfigure}
    \begin{subfigure}{0.45\linewidth}
    \includegraphics[width=0.92\linewidth]{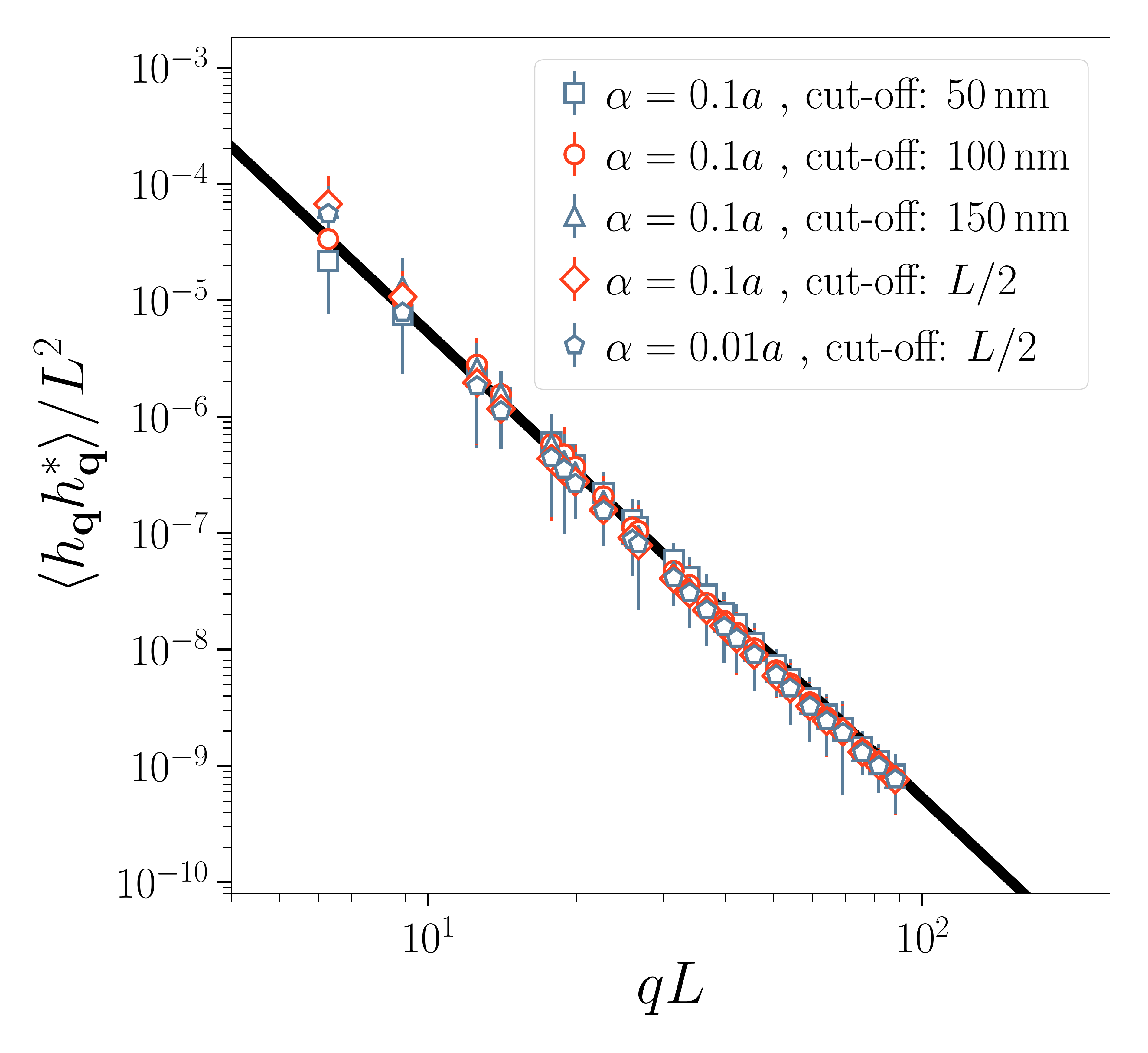}
    \end{subfigure}
    \begin{subfigure}{0.01\linewidth}
        \caption{}
        \label{fig:nonequilibrium_convergence}
    \end{subfigure}
    \begin{subfigure}{0.45\linewidth}
    \includegraphics[width=0.92\linewidth]{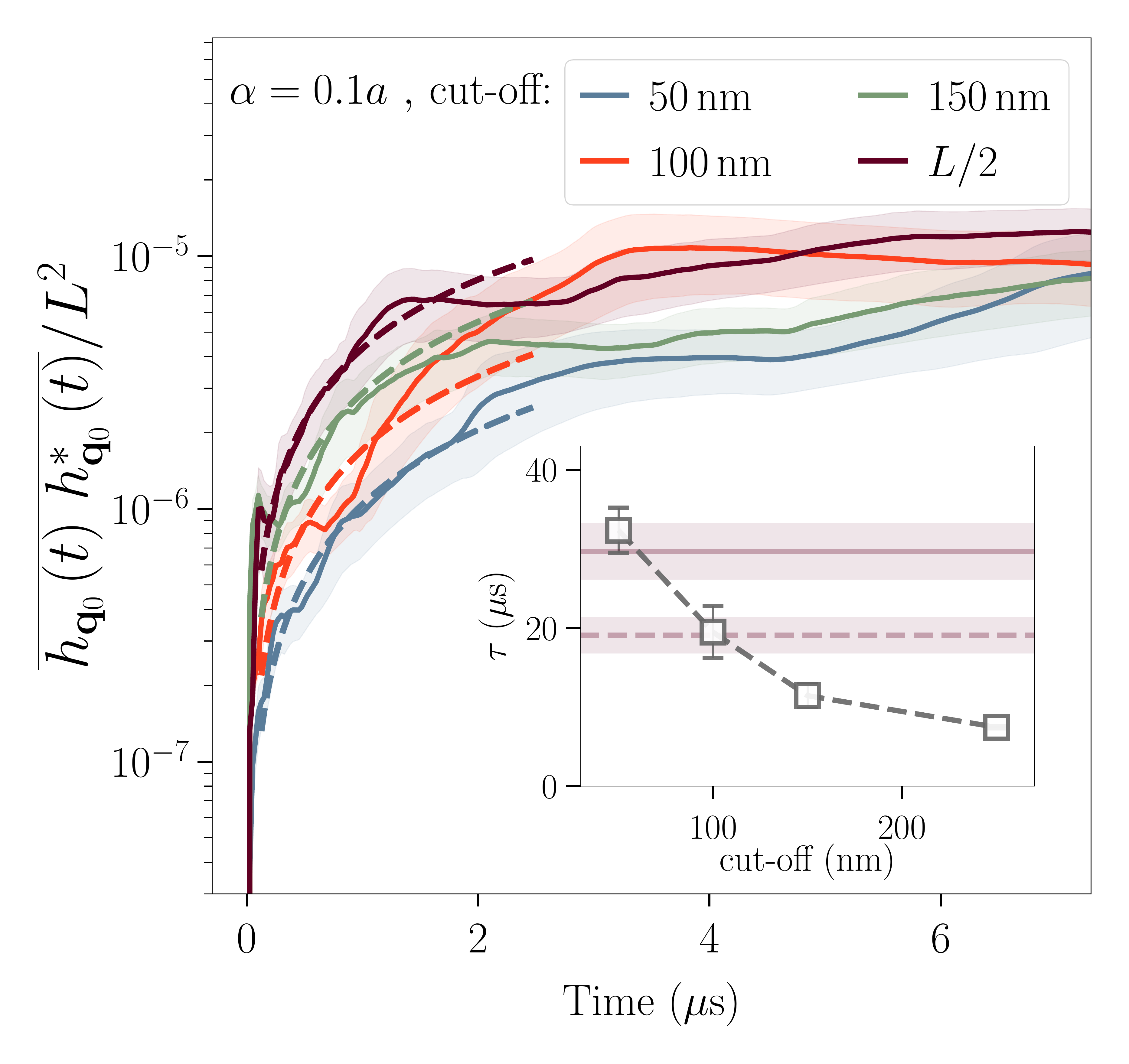}
    \end{subfigure}
    \caption{\footnotesize Kinetics of a free-standing planar membrane patch of lateral size $L=$ \SI{0.5}{\micro\meter}, suspended in water, modeled with the particle-based membrane model with the lattice parameter of $a=$ \SI{10}{\nano\meter}. (\textbf{a}) Dispersion relations when different cut-off radii are used in the treatment of hydrodynamic interactions. Empty and filled symbols respectively denote fast (hydrodynamic) and slow (slipping) relaxation modes. Predictions of the continuum-based model ($\bar{\omega}_1$ and $\bar{\omega}_2$) are included for comparison. The gray region shows the range of meaningful frequencies corresponding to the sampling rate and length of trajectories. (\textbf{b}) Similar to (a), for different choices of the scaling factor $\alpha/a$.  (\textbf{c}) Power spectrum of thermal undulations of all the membrane patches for which the dispersion relations are given in (a) and (b). The solid black line is the prediction of the continuum model (Eq. (\ref{eq:thermal_undulation})). (\textbf{d}) Irreversible relaxation of the energy of the largest undulation mode to the equilibrium value. Results are given for different choices of the cut-off radii. Dashed lines are fitted exponential functions (Eq. (\ref{eq:non_equilibrium_relaxation})). The inset plot shows the implied timescale as a function of the cut-off radius. The horizontal lines and shaded regions in the inset plot correspond to the mean and standard deviation of the two timescales of the same mode, obtained from equilibrium simulations with the largest cut-off.}
    \label{fig:dispersion}
\end{figure*}

Considering $h_\mathbf{q}\left(t\right)$ to denote the amplitude of the undulation mode with the wave vector $\mathbf{q}$, in the absence of in-plane tension, we have, \cite{Seifert1993,Seifert1994,Seifert1997}
\begin{subequations}
    \begin{align}
    \label{eq:thermal_undulation}
    \frac{1}{L^2}\langle h_\mathbf{q} h^*_\mathbf{q}\rangle &= \frac{k T}{\kappa\,\left(qL\right)^4}\\
    \label{eq:mode_relaxation}
    \langle h_\mathbf{q}\left(t\right) h^*_\mathbf{q}\left(0\right)\rangle &= A_1e^{-\omega_1\left(q\right) t} + A_2e^{-\omega_2\left(q\right) t} 
    \end{align}
\end{subequations}
where $\kappa$ is the bending modulus of the membrane. Seifert et al. have provided theoretical values for the relaxation frequencies, $\omega_{1,2} \left(q\right)$, based on a continuum elastic membrane with inter-leaflet friction and fluctuating lipid densities which experiences hydrodynamics interactions described by the Oseen tensor \cite{Seifert1993,Seifert1994,Seifert1997}. These values, denoted here as $\bar{\omega}_{1,2} \left(q\right)$, are the eigenvalues of the time evolution operator, $-\boldsymbol{\Gamma}\left(q\right) \mathbf{E}\left(q\right)$, with the following definition,
\begin{equation}
\begin{split}
    \label{eq:time_evolution_operator}
        \frac{\partial}{\partial t}
        \left(
        \begin{array}{c}
        h_\mathbf{q} \\
        \rho_\mathbf{q}
        \end{array}
        \right)
        &= -\boldsymbol{\Gamma}\left(\mathbf{q}\right) \mathbf{E}\left(\mathbf{q}\right)
        \left(
        \begin{array}{c}
        h_\mathbf{q} \\
        \rho_\mathbf{q}
        \end{array}
        \right)
        \\
        \boldsymbol{\Gamma}\left(\mathbf{q}\right) &= 
        \left(
        \begin{array}{cc}
        \frac{1}{4\eta q} & 0\\
        0 & \frac{q^2}{2\left(2b + 2\eta q + \eta_\mathrm{m} q^2/d_\mathrm{m} \right)}
        \end{array}
        \right)\\
        \mathbf{E}\left(\mathbf{q}\right) &= 
        \left(
        \begin{array}{cc}
        \tilde{\kappa}q^4 & -\frac{d_\mathrm{m}}{4} K_\mathrm{area} q^2\\
         -\frac{d_\mathrm{m}}{4} K_\mathrm{area} q^2 & K_\mathrm{area}
        \end{array}
        \right)
\end{split}
\end{equation}
with $K_\mathrm{area}$ being the area compressibility modulus of the membrane, $\tilde{\kappa} = \kappa + \frac{1}{16} d_\mathrm{m}^2 K_\mathrm{area}$ the effective bending modulus and $b$ a phenomenological inter-leaflet friction coefficient \cite{Seifert1993, Seifert1994a, Seifert1997}. In our range of inspection, the smaller eigenvalue $\bar{\omega}_1$ (slower dynamics) corresponds the so-called ``slipping mode'' resulting from in-plane density fluctuations and the friction between the two leaflets. The faster ``hydrodynamic'' mode with the larger eigenvalue $\bar{\omega}_2$ corresponds the viscous loss in the fluid. Note that these two modes also mix, and for $q$ below the closest approach of $\bar{\omega}_1$ and $\bar{\omega}_2$ curves in Fig. \ref{fig:dispersion}, they switch places \cite{Seifert1994a, Shkulipa2006a}, confirming our scaling argument in Sec. \ref{sec:intro_membrane_hydrodynamics}.

In using Eq. (\ref{eq:time_evolution_operator}), the viscosity of the solvent, as well as the thickness, viscosity, bending rigidity, and area compressibility modulus of the membrane, are all a priori values used in the parametrization of the model and calculation of the diffusion tensors. The only remaining parameter is the inter-leaflet friction, $b$. Experimental values of $b$ are in the range \SIrange[range-units = single]{e8}{e9}{\newton \second \per \meter\cubed} \cite{Evans1994, Chizmadzhev1999}, while coarse-grained membrane simulations predict much smaller value of \SI{1.4e6}{\newton \second \per\meter\cubed} \cite{Shkulipa2005a}. It is to be expected that the inter-leaflet friction coefficient be highly sensitive to the resolution with which the lipids are modeled. Here, we have found a value of $b=$ \SI{e6}{\newton \second \per \meter\cubed} to give a good match between the theoretical frequencies and simulation results (Fig. \ref{fig:dispersion}). However, in our range of inspection, the inter-leaflet friction mostly affects the slipping mode and has little effect on the hydrodynamic mode, which is the focus of our investigation.

We have used two different values of the scaling factor $\alpha$ equal to $0.01 a$ and $0.1 a$ to obtain diffusion tensors. Also, for the case of $\alpha=0.1 a$, we have performed simulations with cut-off radii of \SI{50}{\nano\meter}, \SI{100}{\nano\meter}, \SI{150}{\nano\meter} and the maximum possible cut-off of $L/2$. We have performed each simulation for a total time of \SI{0.5}{\milli\second}. For each sampled frame, the height function, $h\left(x, y, t\right)$, is obtained by mapping the vertical position of particles to a regular grid. Fast Fourier transform is used to obtain values of $h_\mathbf{q}\left(t\right)$, which are used to calculate the ensemble averages and the auto-correlation functions in Eqs. (\ref{eq:thermal_undulation}) and (\ref{eq:mode_relaxation}). The resulting dispersion relations are obtained by fitting biexponential functions to the autocorrelations (Figs. \ref{fig:hydro_dispersion_cutoff} and \ref{fig:hydro_dispersion_alpha}).

The power spectra of thermal undulations (Fig. \ref{fig:thermal_power_spectrum}), which very well follows the expected behavior given by Eq. (\ref{eq:thermal_undulation}), demonstrates that the correct equilibrium distribution is achieved in all cases, serving as a sanity check for the equilibrium sampling offered by the dynamical method developed here. 

In all the cases considered, very good approximations of the desired dispersion relations, especially, the fast mode, $\bar{\omega}_2$, are achieved (Figs. \ref{fig:hydro_dispersion_cutoff} and \ref{fig:hydro_dispersion_alpha}). Including longer ranges of interactions slightly improves the prediction, giving generally better results when $L/2$ is used (Fig. \ref{fig:hydro_dispersion_alpha}). The choice of the hydrodynamic scale parameter, $\alpha$, mostly affects the slow mode, $\bar{\omega}_1$, with smaller $\alpha$ yielding better agreement with the continuum model. The slow mode is affected by in-plane density fluctuations and would thus be sensitive to small changes in the interactions between adjacent particles.
\subsection{Non-equilibrium relaxation dynamics of planar membrane patches}
Apart from equilibrium fluctuations, the kinetics of which is described by Eq. (\ref{eq:mode_relaxation}), we can also look at the irreversible relaxation dynamics when the system evolves from a non-equilibrium distribution towards equilibrium. Specifically, we look at how membrane undulations develop from an initially flat configuration (which is in-effect a delta distribution around the minimum energy) to the equilibrium distribution given by Eq. (\ref{eq:thermal_undulation}). We observe that the cut-off radius of hydrodynamic interactions has a significant effect on the kinetics of this process. To quantify this, we have considered the time evolution of the undulation mode with the largest wavelength, and have tracked the energy of this mode as a function of time (Fig. \ref{fig:nonequilibrium_convergence}). Dashed lines in Fig. \ref{fig:nonequilibrium_convergence} are fits of the equation,
\begin{equation}
\label{eq:non_equilibrium_relaxation}
\frac{k T}{\kappa\,\left(qL\right)^4}\left(1 - \exp\left(-t/\tau\right) \right)
\end{equation}
to the initial segment of these time series. We are not aware of a rigorous expression for $\tau$ in Eq. (\ref{eq:non_equilibrium_relaxation}), and arguably no reliable statistical theory exists to describe the kinetics in this scenario \cite{Ciccotti2018}. The timescales are generally comparable with the equilibrium counterparts (in-set pane in Fig. \ref{fig:nonequilibrium_convergence}), which is reminiscent of regression relations \cite{Onsager1931a}. However, with the increase in the hydrodynamic interaction range, we also observe kinetics faster than the equilibrium fluctuations.
\subsection{Dispersion relations of a membrane near a wall}
\label{sec:dispersion_near_surface}
Finally, we consider the equilibrium fluctuations of a planar membrane patch hovering in the vicinity of a solid wall. Considering the velocity boundary conditions given in Eq. (\ref{eq:parallel_membranes_bc}), the friction tensors derived for parallel membranes are also appropriate for this case. That is, however, not the case with the directly derived diffusion tensors based on the stress boundary conditions. Thus, instead of directly finding the components of the global diffusion matrix, we obtain these components using the inverse of the friction matrix. The matrix inversion is done for the system of two opposing membranes and we take only the components corresponding to the hydrodynamic interactions across one of the membranes (see Figs. \ref{fig:parallel_membranes}, \ref{fig:compiled_diffusion_0.1} and \ref{fig:compiled_diffusion_0.5}). Also, for the membrane leaflet not facing the wall, we have used the diffusion tensors for a single planar membrane, as in Sec. \ref{sec:dispersion_free_membrane}. We have performed simulations with a square membrane patch of $\sim$\SI{150}{\nano\meter} side length, and we have considered cases where the patch is distanced \SI{10}{}, \SI{20}{}, \SI{50}{}, and \SI{100}{\nano\meter} away from the wall. Except from what is implicitly described by the hydrodynamic model, we have not included any other interactions with the wall. Also, we have refrained from including any volume preservation constraints. 

As the reference continuum model, we use the derivation of Seifert \cite{Seifert1994}, for the dynamics of a membrane bound to preserve a mean distance $\bar{z}$ with a solid wall,
\begin{equation}
\begin{split}
    \label{eq:dynamics_near_wall}
        \frac{\partial}{\partial t} h_\mathbf{q} &= -\boldsymbol{\Gamma}\left(\mathbf{q}\right) \mathbf{E}\left(\mathbf{q}\right)h_\mathbf{q}
        \\
        \boldsymbol{\Gamma}\left(\mathbf{q}\right) &= \frac{1}{2 \eta q} \, \frac{\sinh ^ 2 \left(q\bar {z}\right) - \left(q\bar {z}\right)^2}{\sinh ^ 2 \left(q\bar {z}\right) - \left(q\bar {z}\right) ^ 2 + \sinh \left(q\bar {z}\right) \cosh \left(q\bar {z}\right) + q\bar {z}}\\
        \mathbf{E}\left(\mathbf{q}\right) &= 
        \kappa q^4
\end{split}
\end{equation}
This model predicts only one ``hydrodynamic'' dispersion mode. But we anticipated that the model would potentially contain two other time scales similar to the free-standing membrane, as the two leaflets experience different solvent hydrodynamics and the in-plane density fluctuations are also present. Thus, we have obtained dispersion relations from simulation trajectories via fitting a triple exponential decay to the relaxation of undulation modes.

\begin{figure*}[htbp]
    \centering
    \includegraphics[width=0.45\linewidth]{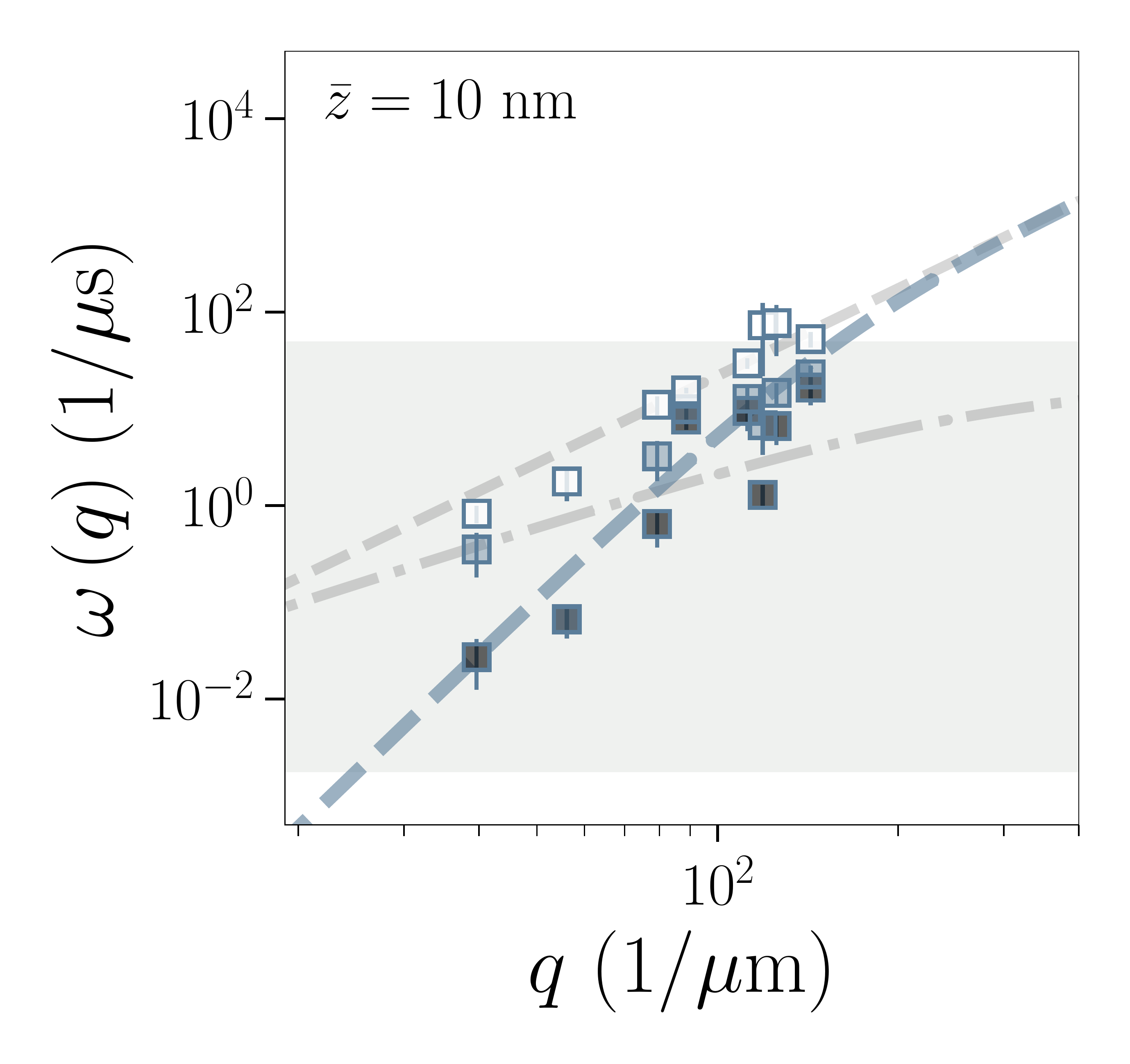}
    \includegraphics[width=0.45\linewidth]{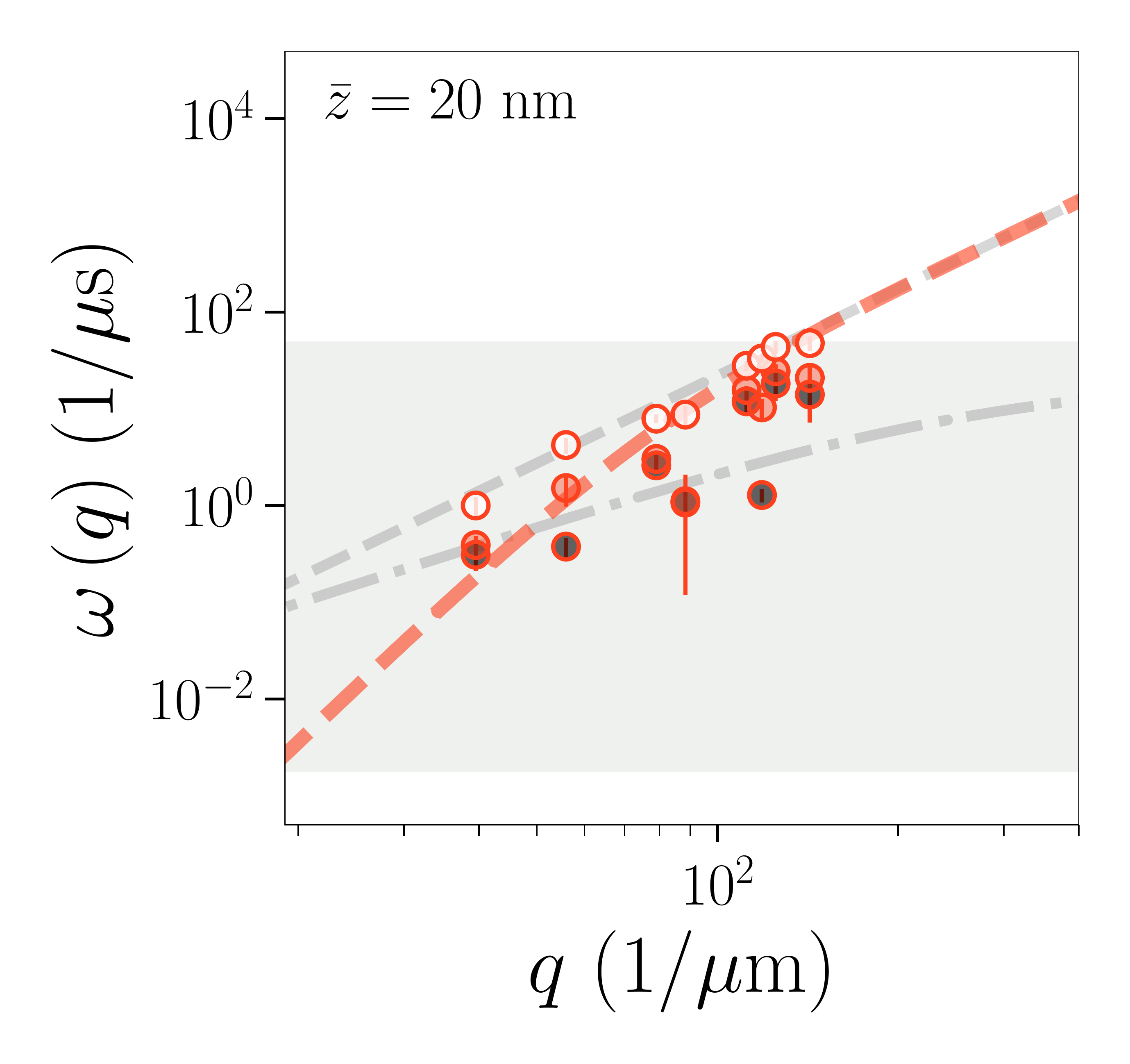}
    \includegraphics[width=0.45\linewidth]{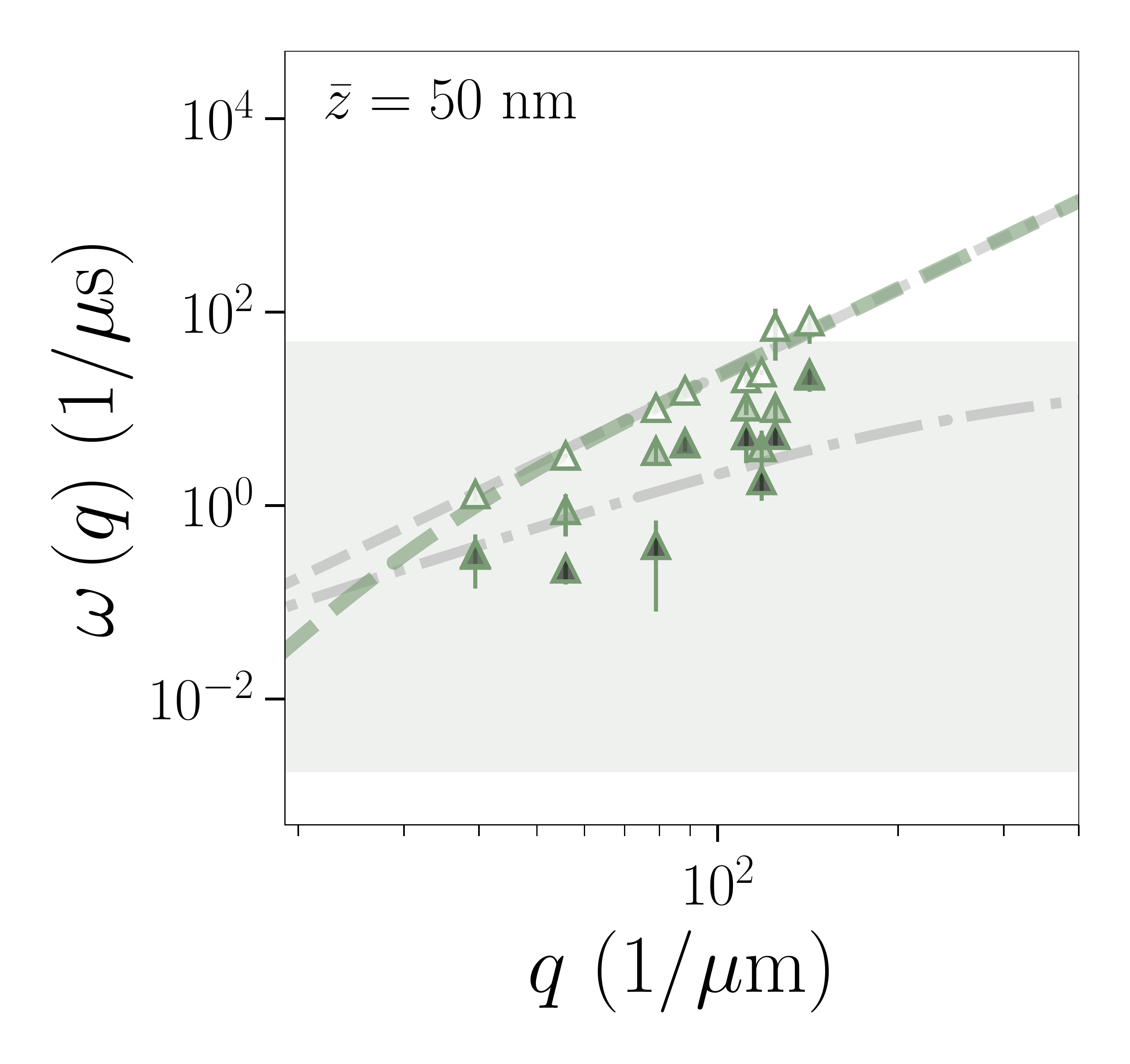}
    \includegraphics[width=0.45\linewidth]{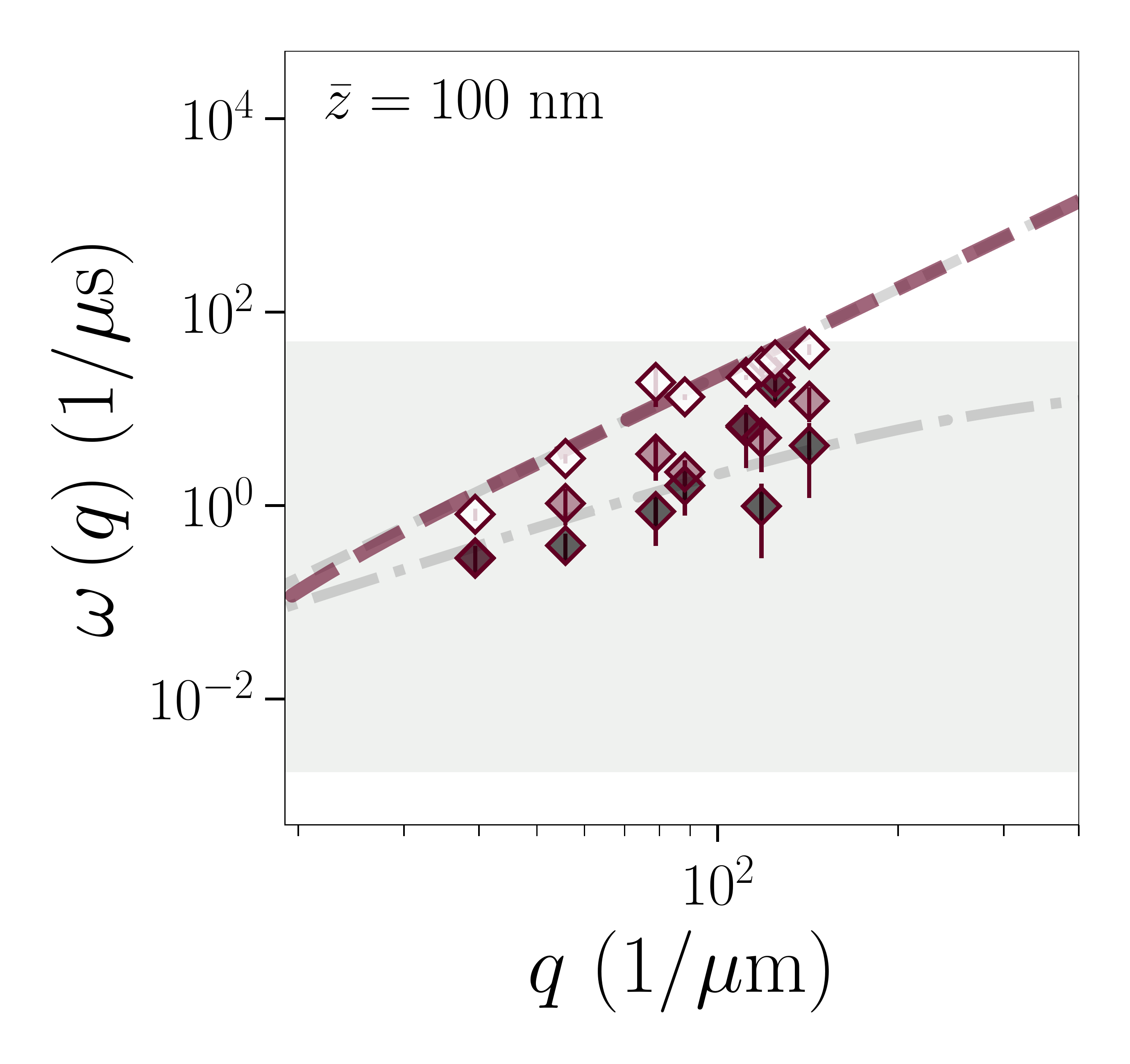}
    \caption{\footnotesize Dispersion relations for planar membrane patches in the vicinity of a wall. The three relaxation frequencies are denoted by empty, color-filled, and gray-filled symbols. Results are shown for different mean distances between the membrane and the wall. Colored dashed lines are predictions of the continuum model for a bound membrane (Eq. (\ref{eq:dynamics_near_wall})), while gray dashed and dot-dashed lines are the same as in Fig. \ref{fig:dispersion}.}
    \label{fig:dispersion_near_wall}
\end{figure*}

Dispersion relations obtained from particle-based simulations match the prediction of Eq. (\ref{eq:dynamics_near_wall}) interestingly well (Fig. \ref{fig:dispersion_near_wall}). Specifically, a clear timescale separation is apparent between membranes closer to the wall with $\bar{z}=$\SI{10}{} and \SI{20}{\nano\meter} and those farther away. In the former, a third slowest timescale clearly persists, while for the latter, the kinetics of the included undulation modes are almost indistinguishable from free-standing membranes (Fig. \ref{fig:dispersion_near_wall}).
\section{Conclusion}
We have introduced a framework for coupling coarse-grained membrane models to solvent hydrodynamics via anisotropic stochastic dynamics and a general form of friction or diffusion tensors. Using exact solutions of Stokes hydrodynamic equations in idealized geometries, we derived expressions or numerical results for the components of these tensors, such that they describe selected in-plane and out-of-plane hydrodynamic effects. This approach offers a simple, robust, and computationally efficient means of tackling multiple time-scale kinetics of membranes. Consulting rule-of-thumb categorization of available spatiotemporal scales with different membrane models proves the timestep and the trajectory lengths possible with our approach to go well beyond the usual particle-based coarse-grained models \cite{Ramakrishnan2014}. 

Using the proposed framework, we investigated dispersion relations for planar membrane patches, both in a free-standing state and in the vicinity of a wall. We showed how our first-principle approach to hydrodynamics leads to realistic large-scale kinetics in both cases, relying solely on properties such as bending modulus of the membrane and viscosity of the solvent as input, realistic timescales are predicted, removing the need for ad hoc corrections after the fact. We further investigated the effect of the range to which hydrodynamic interactions are included, and showed how this parameter affects the irreversible kinetics of membranes relaxing to equilibrium. Our finding suggests fast equilibration of large membrane systems follows long-range hydrodynamic coupling. This is an admittedly preliminary result that hints at the importance of hydrodynamic interaction on the kinetics of non-equilibrium processes.

We believe the approach described here to be applicable to a wide range of coarse-grained membrane models with little adjustments. Our approach to hydrodynamics includes a length-scale contained in the parameter $\alpha$. In the range of inspection with our membrane model, the results were consistent for values of $\alpha$ differing one order of magnitude. Further investigation of this hydrodynamic length-scale and how it adapts to model resolution is of course necessary.

The dynamical picture emerging from the application of the presented framework with computationally efficient coarse-grained models offers realistic kinetics for all the constituents of the system. It opens the door to large-scale high-performance dynamical models of biomembranes and membrane-associated proteins. We believe this to be the only viable means to reliably investigate complex, membrane-involved biological processes close to their native time-scales.

\section*{Conflicts of interest}
There are no conflicts of interest to declare for this study.

\section*{Acknowledgements}
This research has been funded by Deutsche Forschungsgemeinschaft (DFG) through grant SFB 958/Project A04 ``Spatiotemporal model of neuronal signalling and its regulation by presynaptic membrane scaffolds'', SFB 1114/Project C03 ``Multiscale modelling and simulation for spatiotemporal master equations'', and European Research Commission, ERC CoG 772230 ``ScaleCell''.

\section*{Data availability}
The data that support the findings of this study are available from the corresponding authors upon reasonable request.

%%%END OF MAIN TEXT%%%


%\balance

\renewcommand\refname{References}

%%%REFERENCES%%%
\bibliography{library.bib} %You need to replace "rsc" on this line with the name of your .bib file

\begin{thebibliography}{100}

\bibitem{Alberts2015}
{\sc B.~Alberts}, {\sc A.~Johnson}, {\sc J.~Lewis}, {\sc D.~Morgan}, {\sc
  M.~Raff}, {\sc K.~Roberts}, and {\sc P.~Walter},
\newblock {\em {Molecular Biology of the Cell}},
\newblock Garland Science, Taylor {\&} Francis Group, LLC, New York, 6 edition,
  2015.

\bibitem{Kaksonen2018a}
{\sc M.~Kaksonen} and {\sc A.~Roux},
\newblock {\em Nat. Rev. Mol. Cell Biol.} {\bf 19}, 313 (2018).

\bibitem{Dimou2019}
{\sc E.~Dimou}, {\sc K.~Cosentino}, {\sc E.~Platonova}, {\sc U.~Ros}, {\sc
  M.~Sadeghi}, {\sc P.~Kashyap}, {\sc T.~Katsinelos}, {\sc S.~Wegehingel}, {\sc
  F.~No{\'{e}}}, {\sc A.~J. Garc{\'{i}}a-S{\'{a}}ez}, {\sc H.~Ewers}, and {\sc
  W.~Nickel},
\newblock {\em J. Cell Biol.} {\bf 218}, 683 (2019).

\bibitem{Brochard1975}
{\sc F.~Brochard} and {\sc J.~Lennon},
\newblock {\em J. Phys.} {\bf 36}, 1035 (1975).

\bibitem{Prost1998}
{\sc J.~Prost}, {\sc J.-B. Manneville}, and {\sc R.~Bruinsma},
\newblock {\em Eur. Phys. J. B} {\bf 1}, 465 (1998).

\bibitem{Seifert1997}
{\sc U.~Seifert},
\newblock {\em Adv. Phys.} {\bf 46}, 13 (1997).

\bibitem{Shillcock2006}
{\sc J.~C. Shillcock} and {\sc R.~Lipowsky},
\newblock {\em J. Phys. Condens. Matter} {\bf 18}, S1191 (2006).

\bibitem{Marrink2009}
{\sc S.~J. Marrink}, {\sc A.~H. de~Vries}, and {\sc D.~P. Tieleman},
\newblock {\em Biochim. Biophys. Acta - Biomembr.} {\bf 1788}, 149 (2009).

\bibitem{Deserno2009}
{\sc M.~Deserno},
\newblock {\em Macromol. Rapid Commun.} {\bf 30}, 752 (2009).

\bibitem{Noguchi2009}
{\sc H.~Noguchi},
\newblock {\em J. Phys. Soc. Japan} {\bf 78}, 041007 (2009).

\bibitem{Lipowsky2018}
{\sc R.~Lipowsky},
\newblock {Understanding Membranes and Vesicles: A Personal Recollection of the
  Last Two Decades},
\newblock in {\em Phys. Biol. Membr.}, pp. 3--44, Springer International
  Publishing, Cham, 2018.

\bibitem{Marrink2019}
{\sc S.~J. Marrink}, {\sc V.~Corradi}, {\sc P.~C. Souza}, {\sc H.~I.
  Ing{\'{o}}lfsson}, {\sc D.~P. Tieleman}, and {\sc M.~S. Sansom},
\newblock {\em Chem. Rev.} {\bf 119}, 6184 (2019).

\bibitem{Friedman2018}
{\sc R.~Friedman}, {\sc S.~Khalid}, {\sc C.~Aponte-Santamar{\'{i}}a}, {\sc
  E.~Arutyunova}, {\sc M.~Becker}, {\sc K.~J. Boyd}, {\sc M.~Christensen}, {\sc
  J.~T. Coimbra}, {\sc S.~Concilio}, {\sc C.~Daday}, {\sc F.~J. van Eerden},
  {\sc P.~A. Fernandes}, {\sc F.~Gr{\"{a}}ter}, {\sc D.~Hakobyan}, {\sc
  A.~Heuer}, {\sc K.~Karathanou}, {\sc F.~Keller}, {\sc M.~J. Lemieux}, {\sc
  S.~J. Marrink}, {\sc E.~R. May}, {\sc A.~Mazumdar}, {\sc R.~Naftalin}, {\sc
  M.~Pickholz}, {\sc S.~Piotto}, {\sc P.~Pohl}, {\sc P.~Quinn}, {\sc M.~J.
  Ramos}, {\sc B.~Schi{\o}tt}, {\sc D.~Sengupta}, {\sc L.~Sessa}, {\sc
  S.~Vanni}, {\sc T.~Zeppelin}, {\sc V.~Zoni}, {\sc A.~N. Bondar}, and {\sc
  C.~Domene},
\newblock {Understanding Conformational Dynamics of Complex Lipid Mixtures
  Relevant to Biology}, 2018.

\bibitem{Ollila2016}
{\sc O.~H. Ollila} and {\sc G.~Pabst},
\newblock {\em Biochim. Biophys. Acta - Biomembr.} {\bf 1858}, 2512 (2016).

\bibitem{Poger2016}
{\sc D.~Poger}, {\sc B.~Caron}, and {\sc A.~E. Mark},
\newblock {\em Biochim. Biophys. Acta - Biomembr.} {\bf 1858}, 1556 (2016).

\bibitem{Mckiernan2016}
{\sc K.~A. Mckiernan}, {\sc L.-P. Wang}, and {\sc V.~S. Pande},
\newblock {\em J. Chem. Theory Comput.} {\bf 12}, 5960 (2016).

\bibitem{Marrink2007}
{\sc S.~J. Marrink}, {\sc H.~J. Risselada}, {\sc S.~Yefimov}, {\sc D.~P.
  Tieleman}, and {\sc A.~H. {De Vries}},
\newblock {\em J. Phys. Chem. B} {\bf 111}, 7812 (2007).

\bibitem{Marrink2013}
{\sc S.~J. Marrink} and {\sc D.~P. Tieleman},
\newblock {\em Chem. Soc. Rev.} {\bf 42}, 6801 (2013).

\bibitem{Deserno2014}
{\sc M.~Deserno}, {\sc K.~Kremer}, {\sc H.~Paulsen}, {\sc C.~Peter}, and {\sc
  F.~Schmid},
\newblock {\em {Computational Studies of Biomembrane Systems: Theoretical
  Considerations, Simulation Models, and Applications}}, pp. 237--283,
\newblock Springer International Publishing, Cham, 2014.

\bibitem{Arnarez2015}
{\sc C.~Arnarez}, {\sc J.~J. Uusitalo}, {\sc M.~F. Masman}, {\sc H.~I.
  Ing{\'{o}}lfsson}, {\sc D.~H. {De Jong}}, {\sc M.~N. Melo}, {\sc X.~Periole},
  {\sc A.~H. {De Vries}}, and {\sc S.~J. Marrink},
\newblock {\em J. Chem. Theory Comput.} {\bf 11}, 260 (2015).

\bibitem{MohamedLaradji12016}
{\sc M.~Laradji}, {\sc P.~B.~S. Kumar}, and {\sc E.~J. Spangler},
\newblock {\em J. Phys. D. Appl. Phys.} {\bf 49}, 293001 (2016).

\bibitem{Ayton2009}
{\sc G.~S. Ayton}, {\sc E.~Lyman}, {\sc V.~Krishna}, {\sc R.~D. Swenson}, {\sc
  C.~Mim}, {\sc V.~M. Unger}, and {\sc G.~A. Voth},
\newblock {\em Biophys. J.} {\bf 97}, 1616 (2009).

\bibitem{Davtyan2017}
{\sc A.~Davtyan}, {\sc M.~Simunovic}, and {\sc G.~A. Voth},
\newblock {\em J. Chem. Phys.} {\bf 147}, 044101 (2017).

\bibitem{Feng2018}
{\sc S.~Feng}, {\sc Y.~Hu}, and {\sc H.~Liang},
\newblock {\em J. Chem. Phys.} {\bf 148}, 164705 (2018).

\bibitem{Sadeghi2018}
{\sc M.~Sadeghi}, {\sc T.~R. Weikl}, and {\sc F.~No{\'{e}}},
\newblock {\em J. Chem. Phys.} {\bf 148}, 044901 (2018).

\bibitem{Haucke2011a}
{\sc V.~Haucke}, {\sc E.~Neher}, and {\sc S.~J. Sigrist},
\newblock {\em Nat. Rev. Neurosci.} {\bf 12}, 127 (2011).

\bibitem{Noel2019}
{\sc J.~K. Noel}, {\sc F.~No{\'{e}}}, {\sc O.~Daumke}, and {\sc A.~S.
  Mikhailov},
\newblock {\em Biophys. J.} {\bf 117}, 1870 (2019).

\bibitem{Daumke2016}
{\sc O.~Daumke} and {\sc G.~J. Praefcke},
\newblock {\em Biopolymers} {\bf 105}, 580 (2016).

\bibitem{Antonny2016}
{\sc B.~Antonny}, {\sc C.~Burd}, {\sc P.~{De Camilli}}, {\sc E.~Chen}, {\sc
  O.~Daumke}, {\sc K.~Faelber}, {\sc M.~Ford}, {\sc V.~A. Frolov}, {\sc
  A.~Frost}, {\sc J.~E. Hinshaw}, {\sc M.~M. Kozlov}, {\sc M.~Lenz}, {\sc H.~H.
  Low}, {\sc H.~Mcmahon}, {\sc C.~Merrifield}, {\sc T.~D. Pollard}, and {\sc
  P.~J. Robinson},
\newblock {\em EMBO J.} {\bf 35}, 2270 (2016).

\bibitem{Saunders2013}
{\sc M.~G. Saunders} and {\sc G.~A. Voth},
\newblock {\em Annu. Rev. Biophys.} {\bf 42}, 73 (2013).

\bibitem{Hoffmann2019}
{\sc M.~Hoffmann}, {\sc C.~Fr{\"{o}}hner}, and {\sc F.~No{\'{e}}},
\newblock {\em PLoS Comput. Biol.} {\bf 15}, e1006830 (2019).

\bibitem{SchoenebergEtAl_NatComm17_SNX9}
{\sc J.~Sch{\"{o}}neberg}, {\sc M.~Lehmann}, {\sc A.~Ullrich}, {\sc Y.~Posor},
  {\sc W.-T. Lo}, {\sc G.~Lichtner}, {\sc J.~Schmoranzer}, {\sc V.~Haucke}, and
  {\sc F.~No{\'{e}}},
\newblock {\em Nat. Commun.}  (2017).

\bibitem{Frohner2018}
{\sc C.~Fr{\"{o}}hner} and {\sc F.~No{\'{e}}},
\newblock {\em J. Phys. Chem. B} , acs.jpcb.8b06981 (2018).

\bibitem{BiedermannEtAl_BJ15_ReaddyMM}
{\sc J.~Biedermann}, {\sc A.~Ullrich}, {\sc J.~Sch{\"{o}}neberg}, and {\sc
  F.~No{\'{e}}},
\newblock {\em Biophys. J.} {\bf 108}, 457 (2015).

\bibitem{Vijaykumar2015}
{\sc A.~Vijaykumar}, {\sc P.~G. Bolhuis}, {\sc P.~Rein}, {\sc A.~Vijaykumar},
  {\sc P.~G. Bolhuis}, and {\sc P.~Rein},
\newblock {\em J. Chem. Phys.} {\bf 214102}, 0 (2015).

\bibitem{Gunkel2015}
{\sc M.~Gunkel}, {\sc J.~Sch{\"{o}}neberg}, {\sc W.~Alkhaldi}, {\sc S.~Irsen},
  {\sc F.~No{\'{e}}}, {\sc U.~B. Kaupp}, and {\sc A.~Al-Amoudi},
\newblock {\em Structure} {\bf 23}, 628 (2015).

\bibitem{SchoenebergUllrichNoe_BMC14_RDReview}
{\sc J.~Sch{\"{o}}neberg}, {\sc A.~Ullrich}, and {\sc F.~No{\'{e}}},
\newblock {\em BMC Biophys.} {\bf 7}, 11 (2014).

\bibitem{SchoenebergEtAl_BJ14_PhototransductionKinetics}
{\sc J.~Sch{\"{o}}neberg}, {\sc M.~Heck}, {\sc K.~P. Hofmann}, and {\sc
  F.~No{\'{e}}},
\newblock {\em Biophys. J.} {\bf 107}, 1042 (2014).

\bibitem{Ullrich2015}
{\sc A.~Ullrich}, {\sc M.~A. B{\"{o}}hme}, {\sc J.~Sch{\"{o}}neberg}, {\sc
  H.~Depner}, {\sc S.~J. Sigrist}, and {\sc F.~No{\'{e}}},
\newblock {\em PLoS Comput. Biol.} {\bf 11}, e1004407 (2015).

\bibitem{Turlier2018}
{\sc H.~Turlier} and {\sc T.~Betz},
\newblock {Fluctuations in Active Membranes},
\newblock in {\em Phys. Biol. Membr.}, edited by {\sc P.~Bassereau} and {\sc
  P.~Sens}, pp. 581--619, Springer, Cham, 2018.

\bibitem{Betz2009}
{\sc T.~Betz}, {\sc M.~Lenz}, {\sc J.-F. Joanny}, and {\sc C.~Sykes},
\newblock {\em Proc. Natl. Acad. Sci.} {\bf 106}, 15320 (2009).

\bibitem{Qian2007}
{\sc H.~Qian},
\newblock {\em Annu. Rev. Phys. Chem.} {\bf 58}, 113 (2007).

\bibitem{Gennis1989}
{\sc R.~B. Gennis},
\newblock {\em {Biomembranes}},
\newblock Springer Advanced Texts in Chemistry, Springer New York, New York,
  NY, 1989.

\bibitem{Brown2011a}
{\sc F.~L. Brown},
\newblock {\em Q. Rev. Biophys.} {\bf 44}, 391 (2011).

\bibitem{Drouffe1991}
{\sc J.~M. Drouffe}, {\sc A.~C. Maggs}, and {\sc S.~Leibler},
\newblock {\em Science} {\bf 254}, 1353 (1991).

\bibitem{Cooke2005a}
{\sc I.~R. Cooke}, {\sc K.~Kremer}, and {\sc M.~Deserno},
\newblock {\em Phys. Rev. E} {\bf 72}, 011506 (2005).

\bibitem{Wang2005}
{\sc Z.~J. Wang} and {\sc D.~Frenkel},
\newblock {\em J. Chem. Phys.} {\bf 122}, 234711 (2005).

\bibitem{Fritz2011}
{\sc D.~Fritz}, {\sc K.~Koschke}, {\sc V.~A. Harmandaris}, {\sc N.~F. {Van Der
  Vegt}}, and {\sc K.~Kremer},
\newblock {\em Phys. Chem. Chem. Phys.} {\bf 13}, 10412 (2011).

\bibitem{Ayton2006}
{\sc G.~S. Ayton}, {\sc J.~L. McWhirter}, and {\sc G.~A. Voth},
\newblock {\em J. Chem. Phys.} {\bf 124}, 64906 (2006).

\bibitem{Shkulipa2006a}
{\sc S.~A. Shkulipa}, {\sc W.~K. {Den Otter}}, and {\sc W.~J. Briels},
\newblock {\em Phys. Rev. Lett.} {\bf 96}, 178302 (2006).

\bibitem{Huang2012}
{\sc M.~J. Huang}, {\sc R.~Kapral}, {\sc A.~S. Mikhailov}, and {\sc H.~Y.
  Chen},
\newblock {\em J. Chem. Phys.} {\bf 137}, 055101 (2012).

\bibitem{Zgorski2016}
{\sc A.~Zgorski} and {\sc E.~Lyman},
\newblock {\em Biophys. J.} {\bf 111}, 2689 (2016).

\bibitem{Street2006}
{\sc H.~Street} and {\sc U.~Kingdom},
\newblock {\em arXiv:cond-mat} , 0607382 (2006).

\bibitem{Botan2017}
{\sc V.~Botan}, {\sc V.~D. Ustach}, {\sc K.~Leonhard}, and {\sc R.~Faller},
\newblock {\em J. Phys. Chem. B} {\bf 121}, 10394 (2017).

\bibitem{Atzberger2007}
{\sc P.~J. Atzberger}, {\sc P.~R. Kramer}, and {\sc C.~S. Peskin},
\newblock {\em J. Comput. Phys.} {\bf 224}, 1255 (2007).

\bibitem{Venable2017}
{\sc R.~M. Venable}, {\sc H.~I. Ing{\'{o}}lfsson}, {\sc M.~G. Lerner}, {\sc
  B.~S. Perrin}, {\sc B.~A. Camley}, {\sc S.~J. Marrink}, {\sc F.~L. Brown},
  and {\sc R.~W. Pastor},
\newblock {\em J. Phys. Chem. B} {\bf 121}, 3443 (2017).

\bibitem{Vogele2018}
{\sc M.~V{\"{o}}gele}, {\sc J.~K{\"{o}}finger}, and {\sc G.~Hummer},
\newblock {\em Phys. Rev. Lett.} {\bf 120}, 268104 (2018).

\bibitem{Sadeghi2020}
{\sc M.~Sadeghi} and {\sc F.~No{\'{e}}},
\newblock {\em Nat. Commun.} {\bf 11}, 2951 (2020).

\bibitem{Yue2019}
{\sc Z.~Yue}, {\sc C.~Li}, {\sc G.~A. Voth}, and {\sc J.~M. Swanson},
\newblock {\em J. Am. Chem. Soc.} {\bf 141}, 13421 (2019).

\bibitem{Schlaich2017}
{\sc A.~Schlaich}, {\sc J.~Kappler}, and {\sc R.~R. Netz},
\newblock {\em Nano Lett.} {\bf 17}, 5969 (2017).

\bibitem{Li2018}
{\sc H.~Li}, {\sc H.~Y. Chang}, {\sc J.~Yang}, {\sc L.~Lu}, {\sc Y.~H. Tang},
  and {\sc G.~Lykotrafitis},
\newblock {\em Appl. Math. Mech. (English Ed.} {\bf 39}, 3 (2018).

\bibitem{Noguchi2005a}
{\sc H.~Noguchi} and {\sc G.~Gompper},
\newblock {\em Phys. Rev. E - Stat. Nonlinear, Soft Matter Phys.} {\bf 72}
  (2005).

\bibitem{Saffman1975}
{\sc P.~G. Saffman}, {\sc M.~Delbruck}, and {\sc M.~Delbr{\"{u}}ck},
\newblock {\em Proc Natl Acad Sci USA} {\bf 72}, 3111 (1975).

\bibitem{Saffman1976}
{\sc P.~G. Saffman},
\newblock {\em J. Fluid Mech.} {\bf 73}, 593 (1976).

\bibitem{Panzuela2018}
{\sc S.~Panzuela} and {\sc R.~Delgado-Buscalioni},
\newblock {\em Phys. Rev. Lett.} {\bf 121}, 048101 (2018).

\bibitem{Camley2010}
{\sc B.~A. Camley}, {\sc C.~Esposito}, {\sc T.~Baumgart}, and {\sc F.~L.~H.
  Brown},
\newblock {\em Biophys. J.} {\bf 99}, L44 (2010).

\bibitem{Sorkin2020}
{\sc B.~Sorkin} and {\sc H.~Diamant},
\newblock {Persistent collective motion of a dispersing membrane domain}, 2020.

\bibitem{Camley2019}
{\sc B.~A. Camley} and {\sc F.~L. Brown},
\newblock {\em J. Chem. Phys.} {\bf 151}, 124104 (2019).

\bibitem{Oppenheimer2010}
{\sc N.~Oppenheimer} and {\sc H.~Diamant},
\newblock {\em Phys. Rev. E - Stat. Nonlinear, Soft Matter Phys.} {\bf 82}
  (2010).

\bibitem{Oppenheimer2011}
{\sc N.~Oppenheimer} and {\sc H.~Diamant},
\newblock {\em Phys. Rev. Lett.} {\bf 107} (2011).

\bibitem{Arroyo2010}
{\sc M.~Arroyo}, {\sc A.~DeSimone}, and {\sc L.~Heltai},
\newblock {\em arXiv}  (2010).

\bibitem{Kramer1971a}
{\sc L.~Kramer},
\newblock {\em J. Chem. Phys.} {\bf 55}, 2097 (1971).

\bibitem{Seifert1994}
{\sc U.~Seifert},
\newblock {\em Phys. Rev. E} {\bf 49}, 3124 (1994).

\bibitem{Seifert1994a}
{\sc U.~Seifert} and {\sc S.~A. Langer},
\newblock {\em Biophys. Chem.} {\bf 49}, 13 (1994).

\bibitem{Pfeiffer1993}
{\sc W.~Pfeiffer}, {\sc S.~K{\"{o}}nig}, {\sc J.~F. Legrand}, {\sc T.~Bayerl},
  {\sc D.~Richter}, and {\sc E.~Sackmann},
\newblock {\em Europhys. Lett.} {\bf 23}, 457 (1993).

\bibitem{Kaizuka2006}
{\sc Y.~Kaizuka} and {\sc J.~T. Groves},
\newblock {\em Phys. Rev. Lett.} {\bf 96} (2006).

\bibitem{Peukes2014}
{\sc J.~Peukes} and {\sc T.~Betz},
\newblock {\em Biophys. J.} {\bf 107}, 1810 (2014).

\bibitem{Ermak1978}
{\sc D.~L. Ermak} and {\sc J.~A. McCammon},
\newblock {\em J. Chem. Phys.} {\bf 69}, 1352 (1978).

\bibitem{Stokes1851}
{\sc G.~G. Stokes},
\newblock {\em Trans. Cambridge Philos. Soc.} {\bf 9}, 8 (1851).

\bibitem{Einstein1905}
{\sc A.~Einstein},
\newblock {\em Ann. d. Phys.} {\bf 322}, 549 (1905).

\bibitem{HiromiYamakawa}
{\sc {Hiromi Yamakawa}},
\newblock {\em {Modern Theory of Polymer Solutions}},
\newblock Harper {\&} Row Publishers, New York, 1971.

\bibitem{Rotne1969}
{\sc J.~Rotne} and {\sc S.~Prager},
\newblock {\em J. Chem. Phys.} {\bf 50}, 4831 (1969).

\bibitem{Yamakawa1970}
{\sc H.~Yamakawa},
\newblock {\em J. Chem. Phys.} {\bf 53}, 436 (1970).

\bibitem{Noguchi2005}
{\sc H.~Noguchi} and {\sc G.~Gompper},
\newblock {\em Proc. Natl. Acad. Sci. U. S. A.} {\bf 102}, 14159 (2005).

\bibitem{Bahrami2017}
{\sc A.~H. Bahrami} and {\sc G.~Hummer},
\newblock {\em ACS Nano} {\bf 11}, 9558 (2017).

\bibitem{Granek1997}
{\sc R.~Granek},
\newblock {\em J. Phys. II} {\bf 7}, 1761 (1997).

\bibitem{Piessens2000}
{\sc R.~Piessens},
\newblock {The Hankel Transform},
\newblock in {\em Transform. Appl. Handb.}, edited by {\sc {Ed. Alexander}} and
  {\sc {D. Poularikas}}, chapter~9, p. 30 pages, CRC Press LLC, Boca Raton, 2
  edition, 2000.

\bibitem{Duffy1994}
{\sc D.~Duffy},
\newblock {\em {Transform Methods for Solving Partial Differential Equations}},
\newblock Chapman {\&} Hall/CRC, 2 edition, 1994.

\bibitem{Johnson1987}
{\sc H.~F. Johnson},
\newblock {\em Comput. Phys. Commun.} {\bf 43}, 181 (1987).

\bibitem{Lemoine1994}
{\sc D.~Lemoine},
\newblock {\em J. Chem. Phys.} {\bf 101}, 3936 (1994).

\bibitem{GSLref}
{\sc M.~Galassi}, {\sc J.~Davies}, {\sc B.~Gough}, {\sc G.~Jungman}, {\sc
  P.~Alken}, and {\sc R.~Ulerich},
\newblock {GNU Scientific Library Reference},
\newblock Technical Report July, 2013.

\bibitem{Gov2004}
{\sc N.~Gov}, {\sc A.~G. Zilman}, and {\sc S.~Safran},
\newblock {\em Phys. Rev. E - Stat. Nonlinear, Soft Matter Phys.} {\bf 70}
  (2004).

\bibitem{Brenn2017}
{\sc G.~Brenn},
\newblock {\em {Analytical Solutions for Transport Processes}},
\newblock Mathematical Engineering, Springer Berlin Heidelberg, Berlin,
  Heidelberg, 2017.

\bibitem{Kim1997}
{\sc K.~Kim} and {\sc M.~Song},
\newblock {\em Korean J. Comput. Appl. Math.} {\bf 4}, 179 (1997).

\bibitem{Fixman1986}
{\sc M.~Fixman},
\newblock {\em Macromolecules} {\bf 19}, 1204 (1986).

\bibitem{Geyer2009}
{\sc T.~Geyer} and {\sc U.~Winter},
\newblock {\em J. Chem. Phys.} {\bf 130}, 114905 (2009).

\bibitem{Banchio2003}
{\sc A.~J. Banchio} and {\sc J.~F. Brady},
\newblock {\em J. Chem. Phys.} {\bf 118}, 10323 (2003).

\bibitem{Feller1995}
{\sc S.~E. Feller}, {\sc Y.~Zhang}, {\sc R.~W. Pastor}, and {\sc B.~R. Brooks},
\newblock {\em J. Chem. Phys.} {\bf 103}, 4613 (1995).

\bibitem{Marsh2006}
{\sc D.~Marsh},
\newblock {\em Chem. Phys. Lipids} {\bf 144}, 146 (2006).

\bibitem{Hu2012}
{\sc M.~Hu}, {\sc J.~J. Briguglio}, and {\sc M.~Deserno},
\newblock {\em Biophys. J.} {\bf 102}, 1403 (2012).

\bibitem{Nagle2013}
{\sc J.~F. Nagle},
\newblock {\em Faraday Discuss.} {\bf 161}, 11 (2013).

\bibitem{Dimova2014}
{\sc R.~Dimova},
\newblock {\em Adv. Colloid Interface Sci.} {\bf 208}, 225 (2014).

\bibitem{Janosi2010}
{\sc L.~Janosi} and {\sc A.~A. Gorfe},
\newblock {\em J. Chem. Theory Comput.} {\bf 6}, 3267 (2010).

\bibitem{Klauda2010}
{\sc J.~B. Klauda}, {\sc R.~M. Venable}, {\sc J.~A. Freites}, {\sc J.~W.
  O'Connor}, {\sc D.~J. Tobias}, {\sc C.~Mondragon-Ramirez}, {\sc I.~Vorobyov},
  {\sc A.~D. MacKerell}, and {\sc R.~W. Pastor},
\newblock {\em J. Phys. Chem. B} {\bf 114}, 7830 (2010).

\bibitem{Raghunathan2012}
{\sc M.~Raghunathan}, {\sc Y.~Zubovski}, {\sc R.~M. Venable}, {\sc R.~W.
  Pastor}, {\sc J.~F. Nagle}, and {\sc S.~Tristram-Nagle},
\newblock {\em J. Phys. Chem. B} {\bf 116}, 3918 (2012).

\bibitem{Braun2013}
{\sc A.~R. Braun}, {\sc J.~N. Sachs}, and {\sc J.~F. Nagle},
\newblock {\em J. Phys. Chem. B} {\bf 117}, 5065 (2013).

\bibitem{Ando2013}
{\sc T.~Ando}, {\sc E.~Chow}, and {\sc J.~Skolnick},
\newblock {\em J. Chem. Phys.} {\bf 139}, 121922 (2013).

\bibitem{Seifert1993}
{\sc U.~Seifert} and {\sc S.~K. Langer},
\newblock {\em Euro. Phys. Lett.} {\bf 23}, 71 (1993).

\bibitem{Evans1994}
{\sc E.~Evans} and {\sc A.~Yeung},
\newblock {\em Chem. Phys. Lipids} {\bf 73}, 39 (1994).

\bibitem{Chizmadzhev1999}
{\sc Y.~A. Chizmadzhev}, {\sc D.~A. Kumenko}, {\sc P.~I. Kuzmin}, {\sc L.~V.
  Chernomordik}, {\sc J.~Zimmerberg}, and {\sc F.~S. Cohen},
\newblock {\em Biophys. J.} {\bf 76}, 2951 (1999).

\bibitem{Shkulipa2005a}
{\sc S.~A. Shkulipa}, {\sc W.~K. {Den Otter}}, and {\sc W.~J. Briels},
\newblock {\em Biophys. J.} {\bf 89}, 823 (2005).

\bibitem{Ciccotti2018}
{\sc G.~Ciccotti}, {\sc M.~Ferrario}, and {\sc C.~Sch{\"{u}}tte},
\newblock {\em Entropy} {\bf 20}, 348 (2018).

\bibitem{Onsager1931a}
{\sc L.~Onsager},
\newblock {\em Phys. Rev.} {\bf 38}, 2265 (1931).

\bibitem{Ramakrishnan2014}
{\sc N.~Ramakrishnan}, {\sc P.~B. {Sunil Kumar}}, and {\sc R.~Radhakrishnan},
\newblock {\em Phys. Rep.} {\bf 543}, 1 (2014).

\end{thebibliography}

%\clearpage
\linespread{1}

%\clearpage
\newcommand{\barsize}{0.79\linewidth}
\newcommand{\panesize}{0.33\linewidth}
\newcommand{\halfpanesize}{0.175\linewidth}

\begin{figure*}[htpb]
    \centering
    \begin{subfigure}{0.1\linewidth}
        \caption{}
        \label{fig:single_membrane}
    \end{subfigure}
    \begin{subfigure}{\barsize}
        \hspace{1.0cm}
    \end{subfigure}
    \begin{subfigure}{\panesize}
        \includegraphics[width=\linewidth]{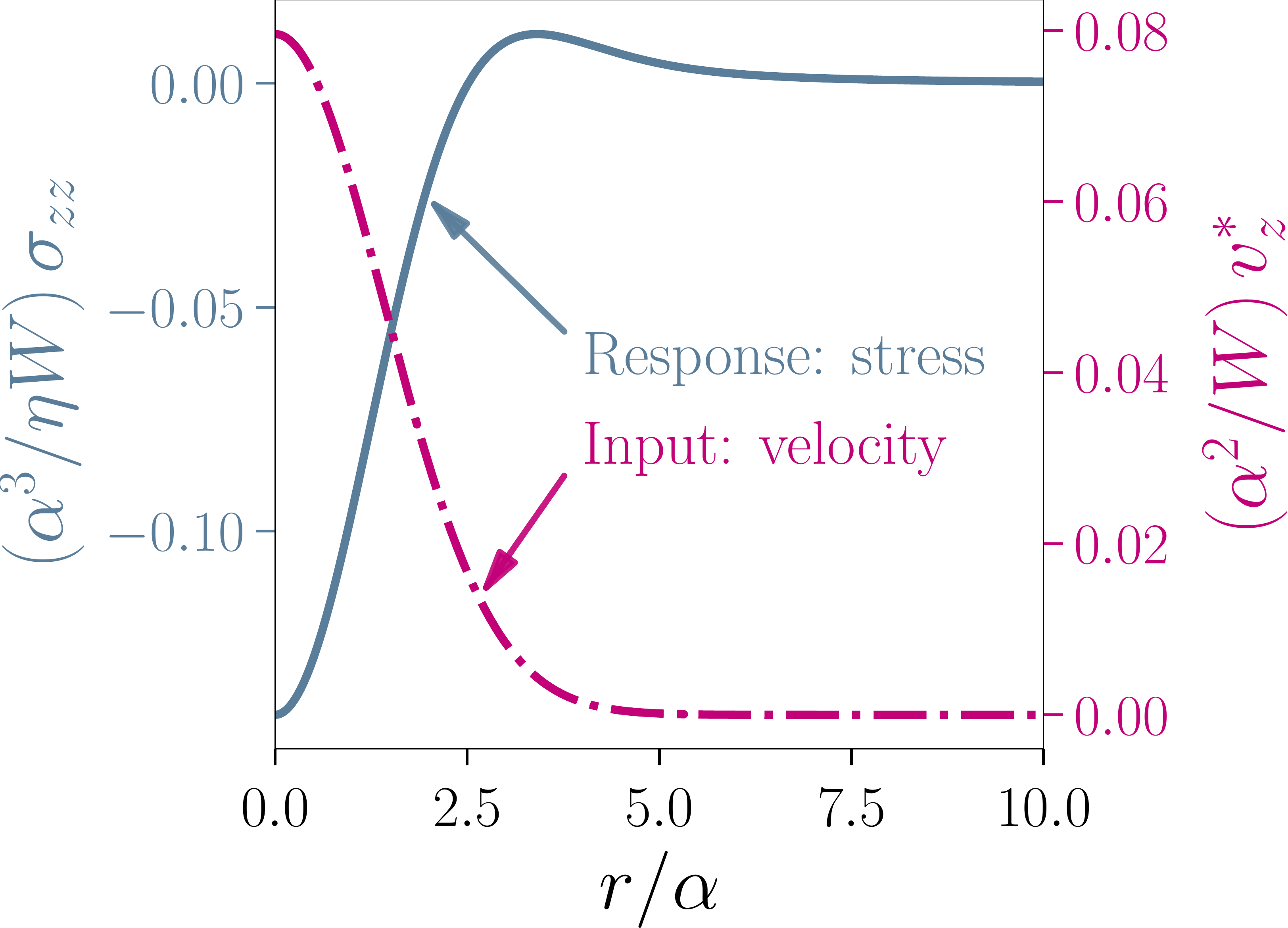}
    \end{subfigure}
    \begin{subfigure}{\halfpanesize}
        \includegraphics[width=\linewidth]{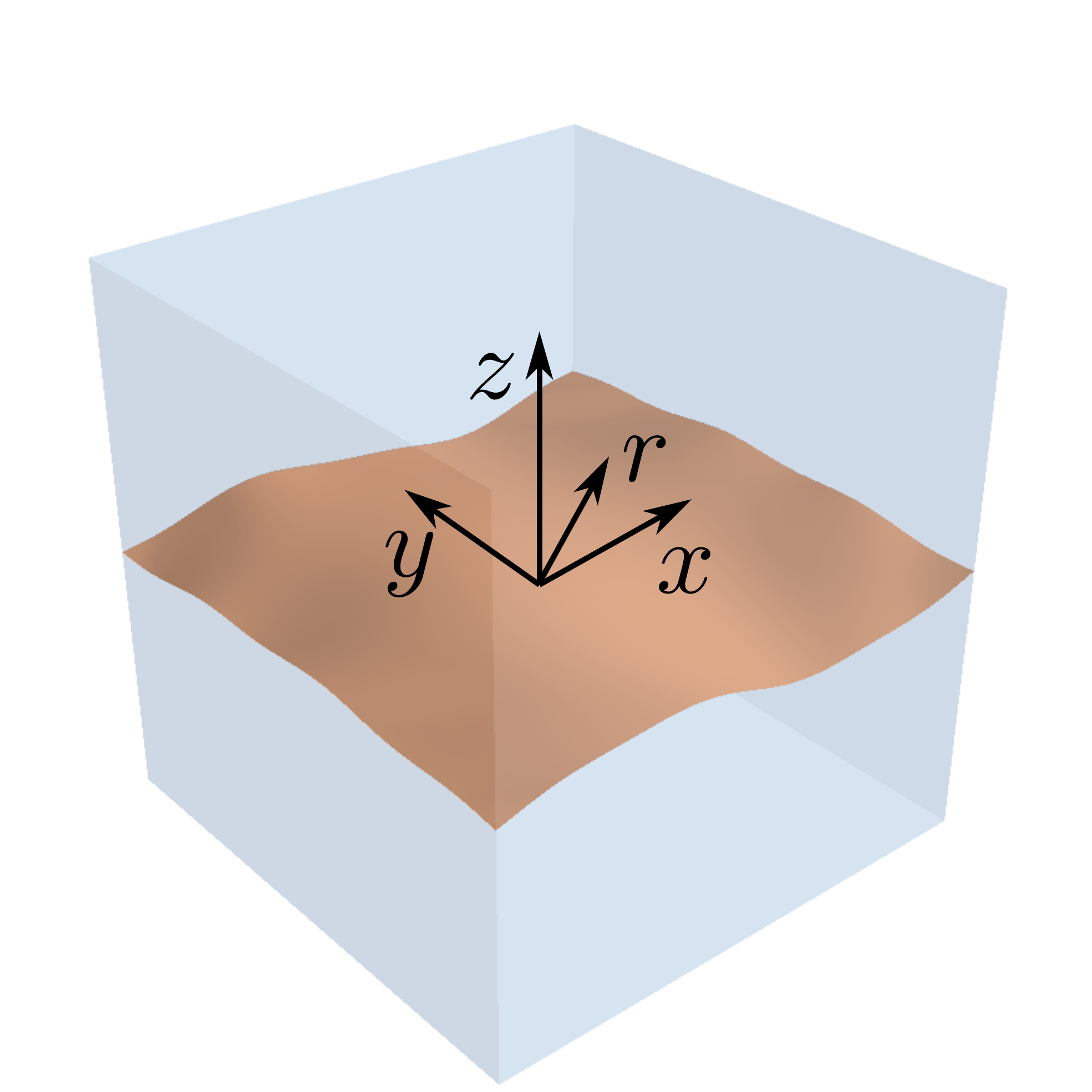}
        \includegraphics[width=\linewidth]{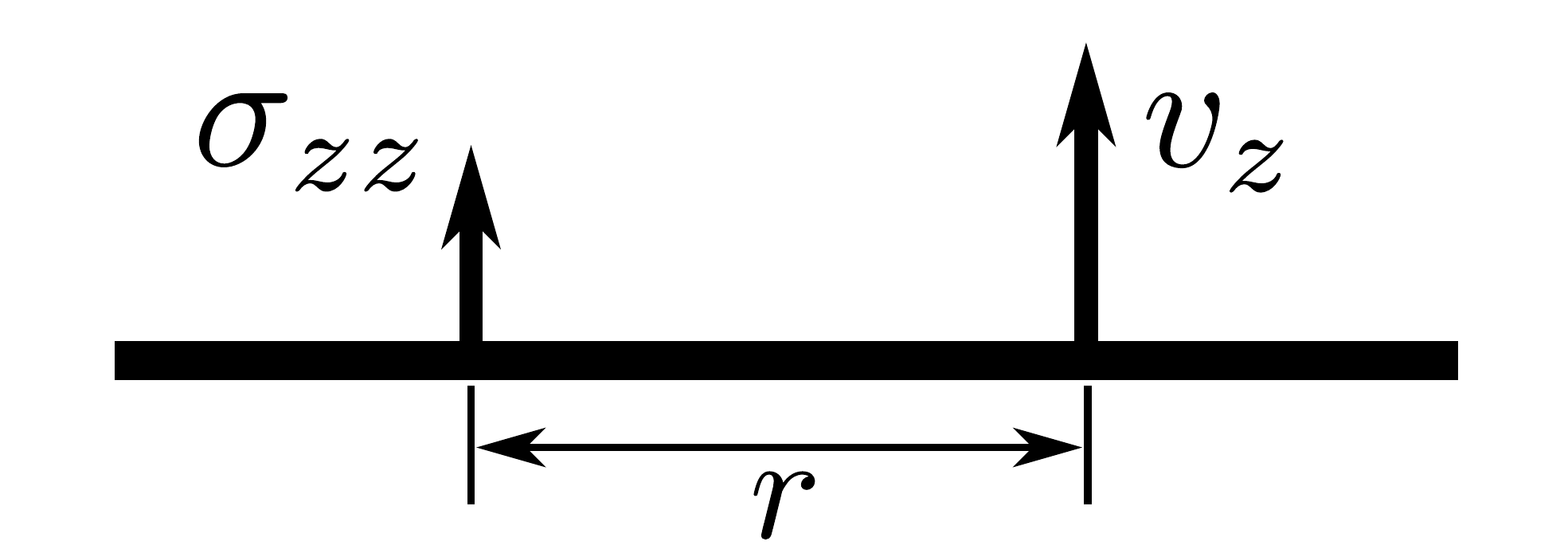}
    \end{subfigure}
    \begin{subfigure}{\panesize}
        \includegraphics[width=\linewidth]{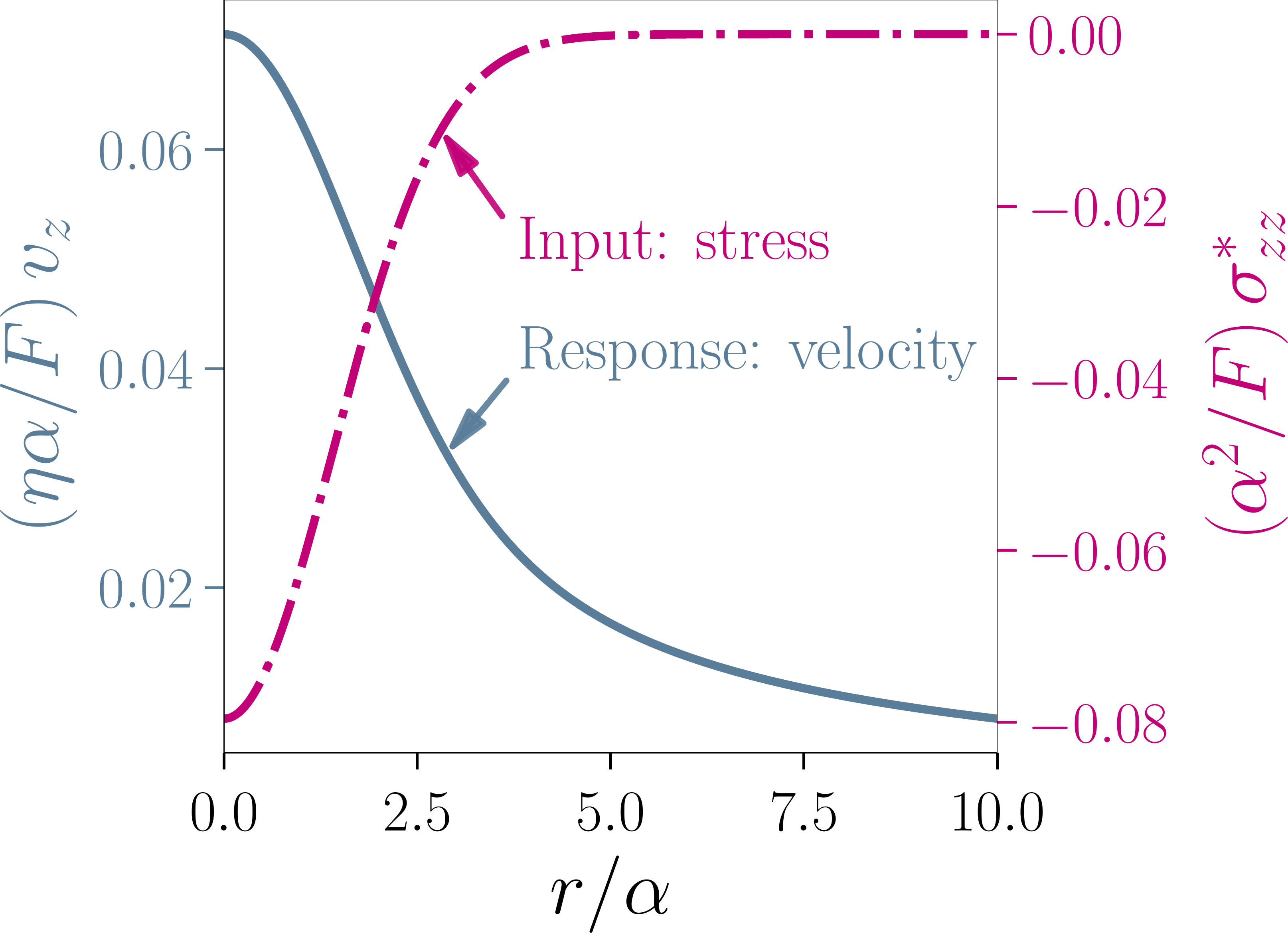}
    \end{subfigure}\\
    \begin{subfigure}{0.1\linewidth}
        \caption{}
        \label{fig:parallel_membranes}
    \end{subfigure}
    \begin{subfigure}{\barsize}
        \includegraphics[width=\linewidth]{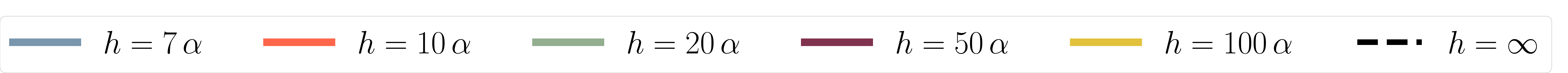}
    \end{subfigure}
    \begin{subfigure}{\panesize}
        \includegraphics[width=\linewidth]{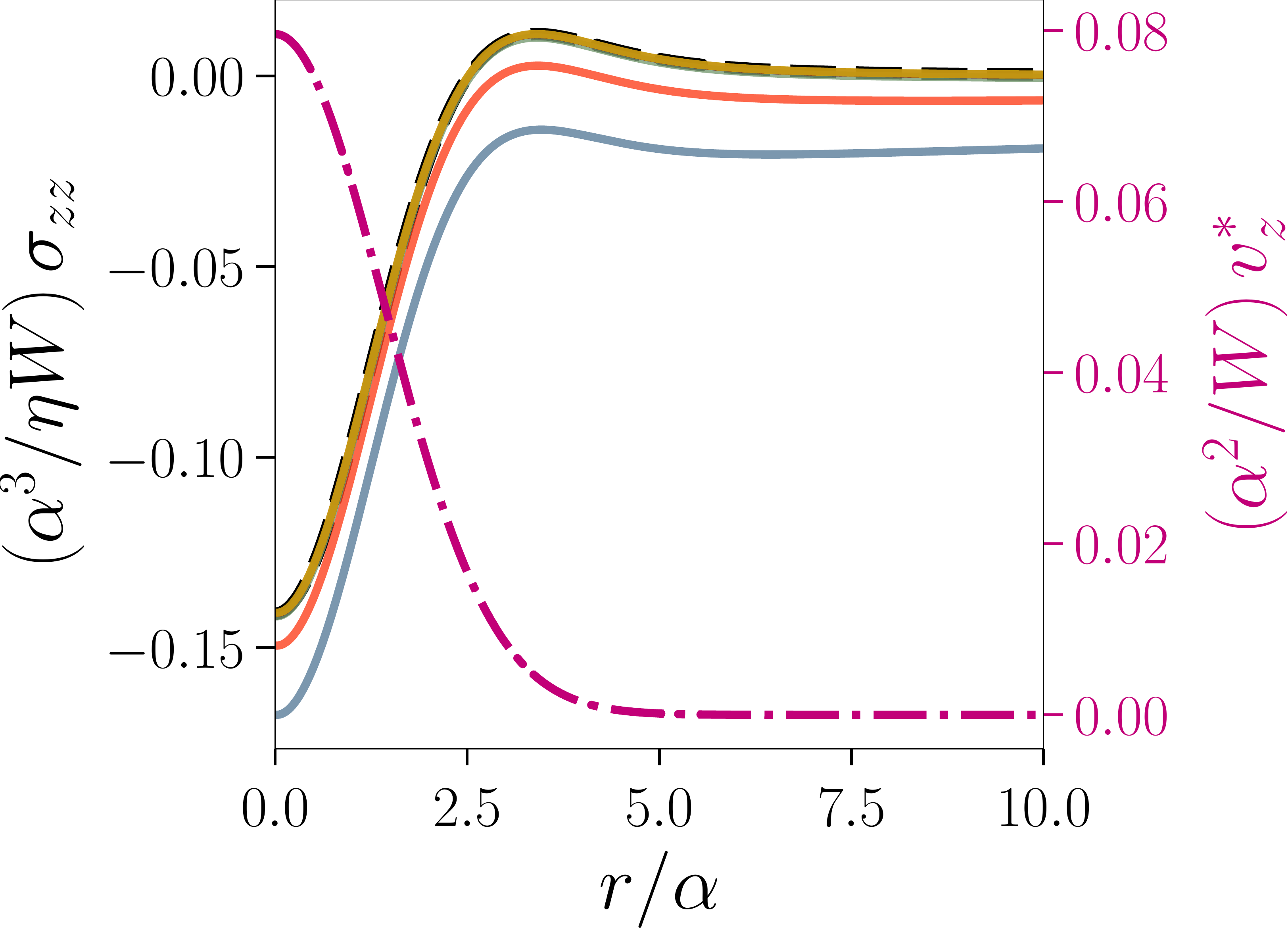}\\
        \hspace{1cm}\\
        \includegraphics[width=\linewidth]{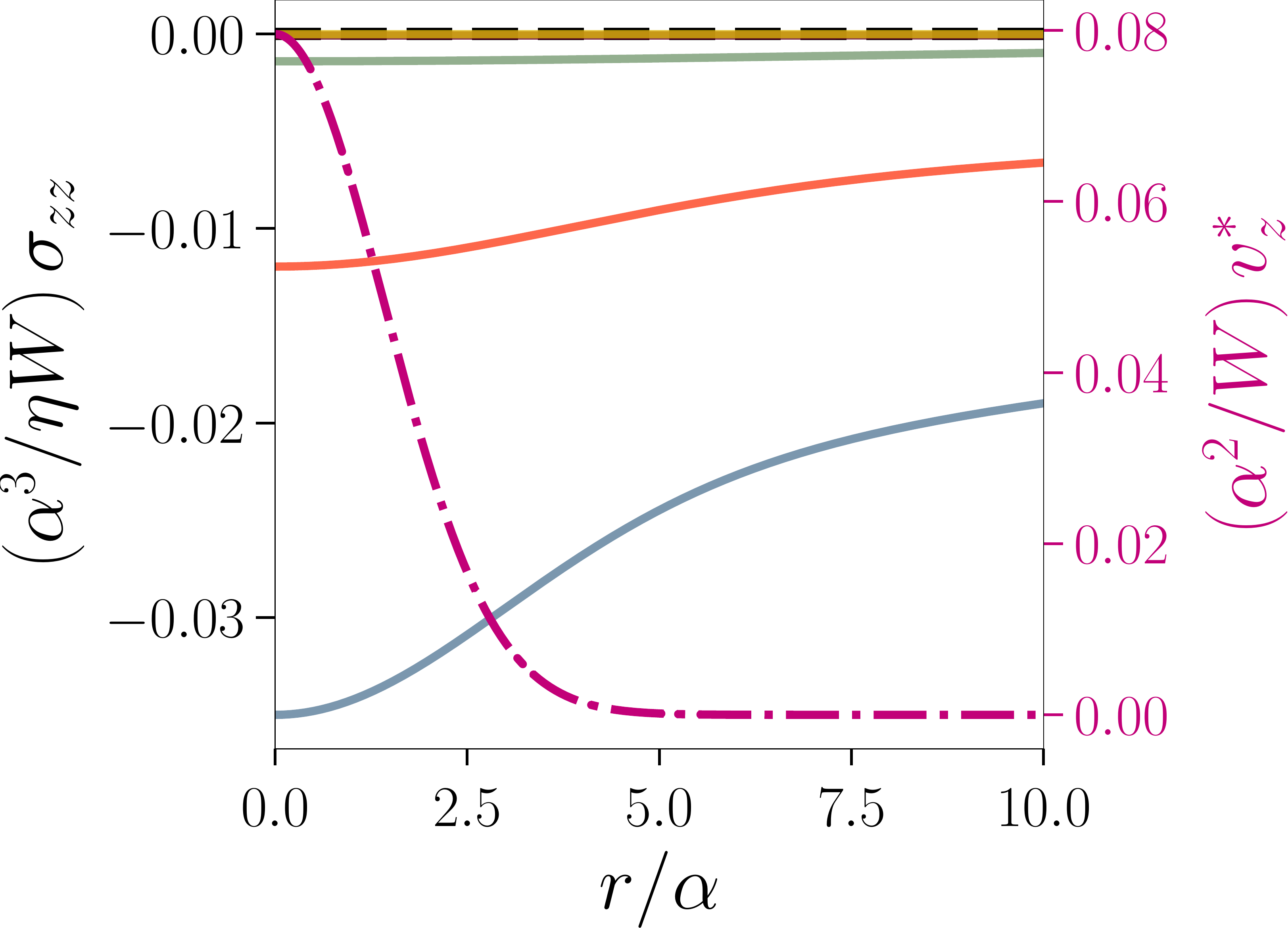}
    \end{subfigure}
    \begin{subfigure}{\halfpanesize}
        \includegraphics[width=\linewidth]{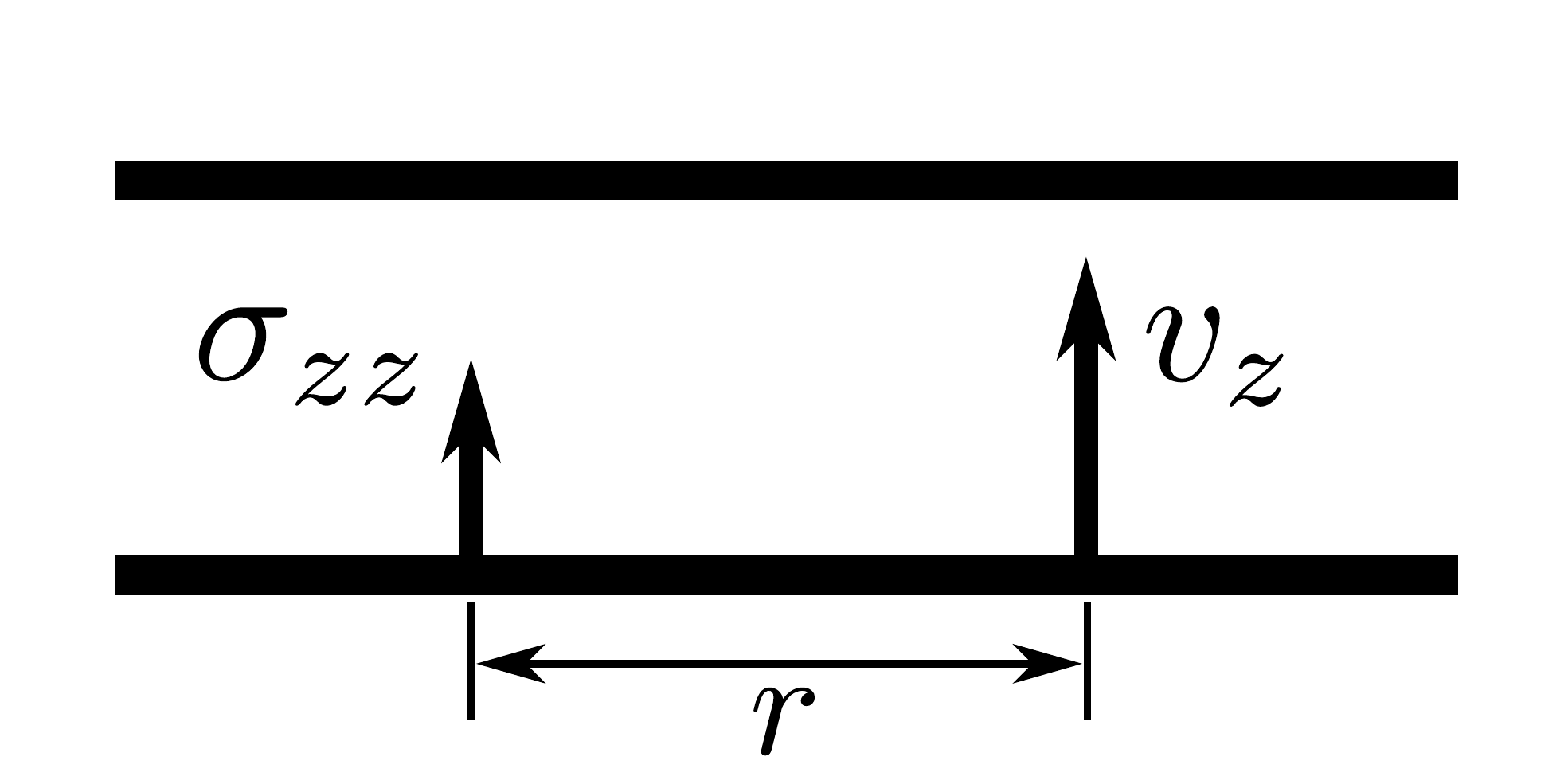}\\
        \hspace{1cm}\\
        \includegraphics[width=\linewidth]{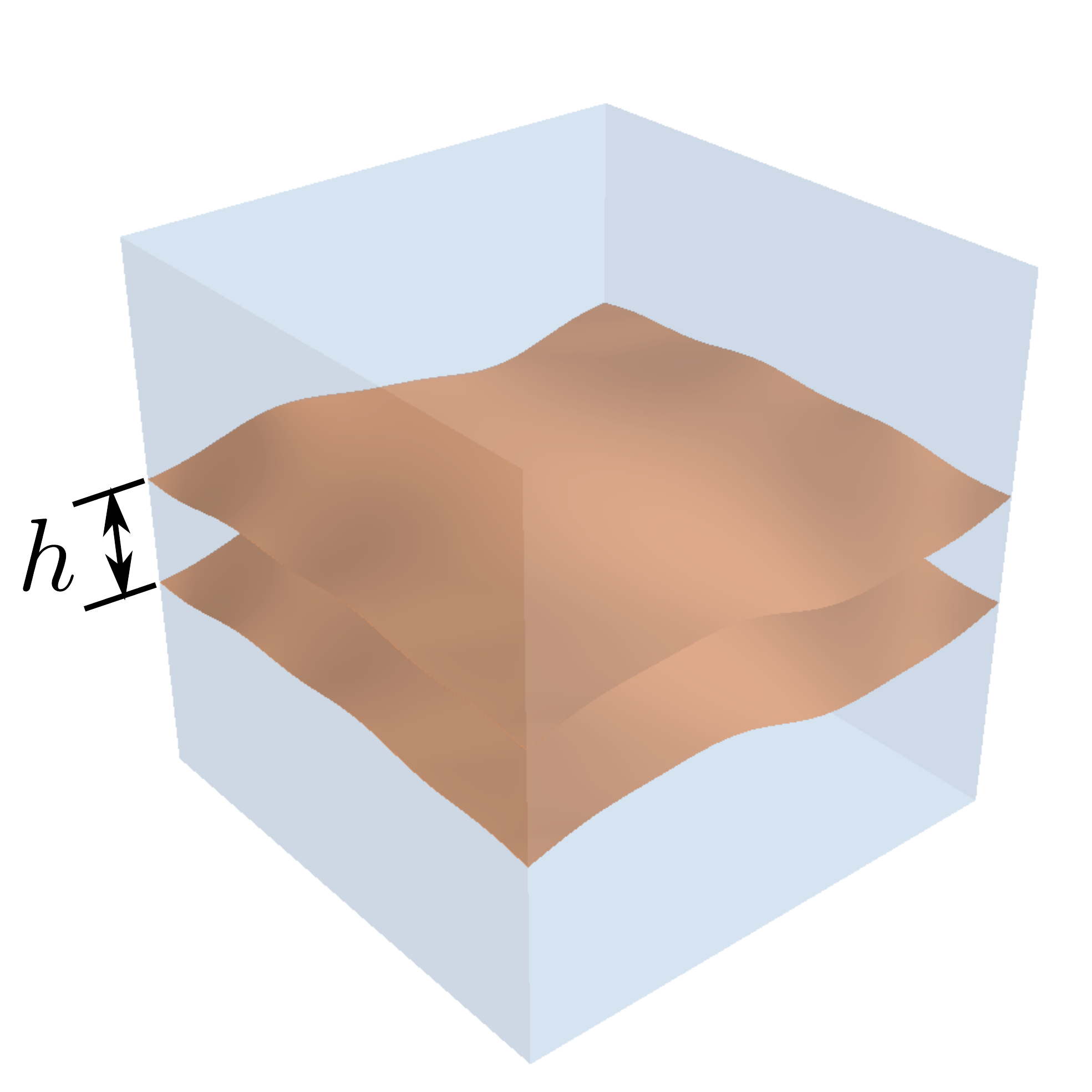}
        \includegraphics[width=\linewidth]{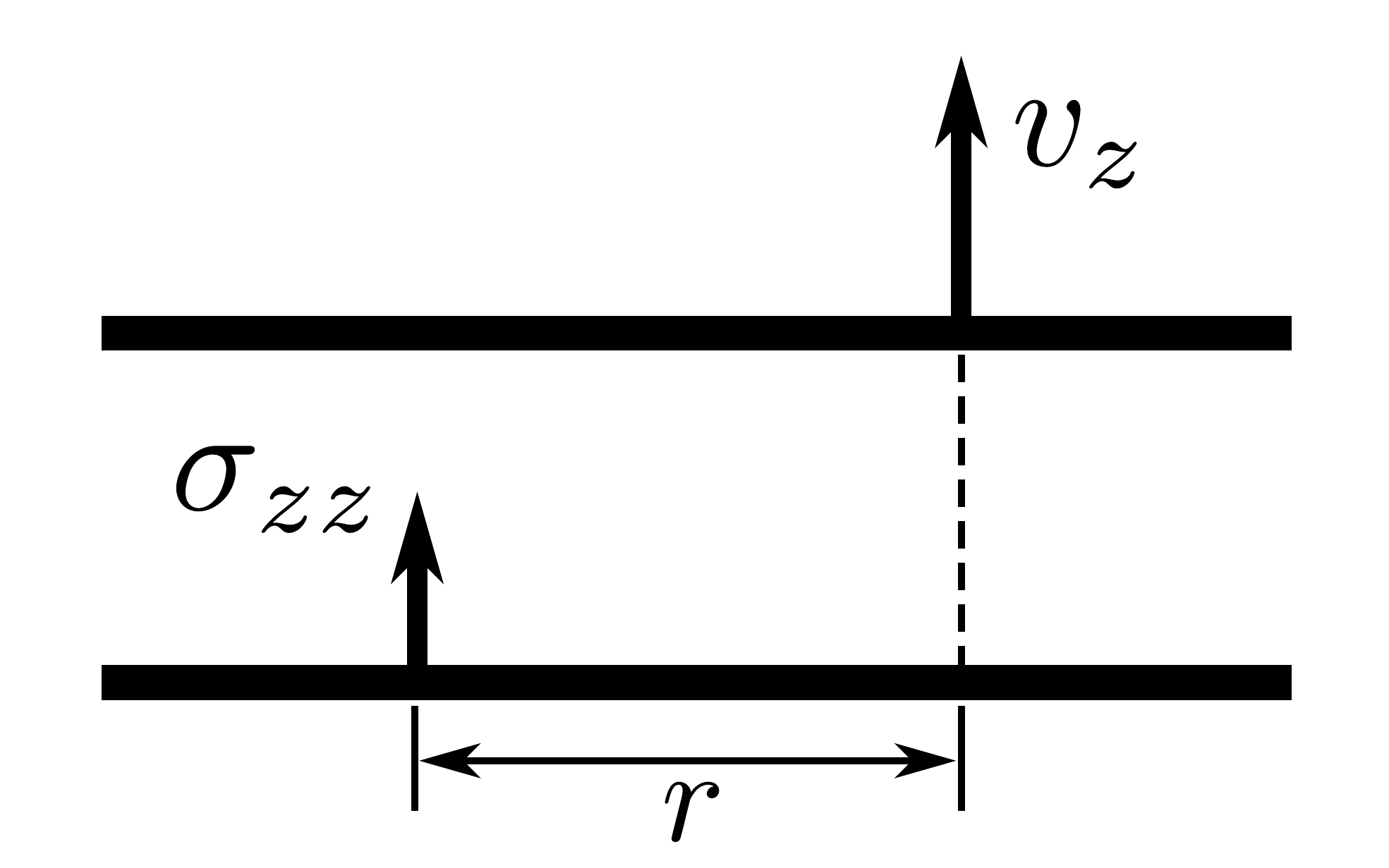}
    \end{subfigure}
    \begin{subfigure}{\panesize}
        \includegraphics[width=\linewidth]{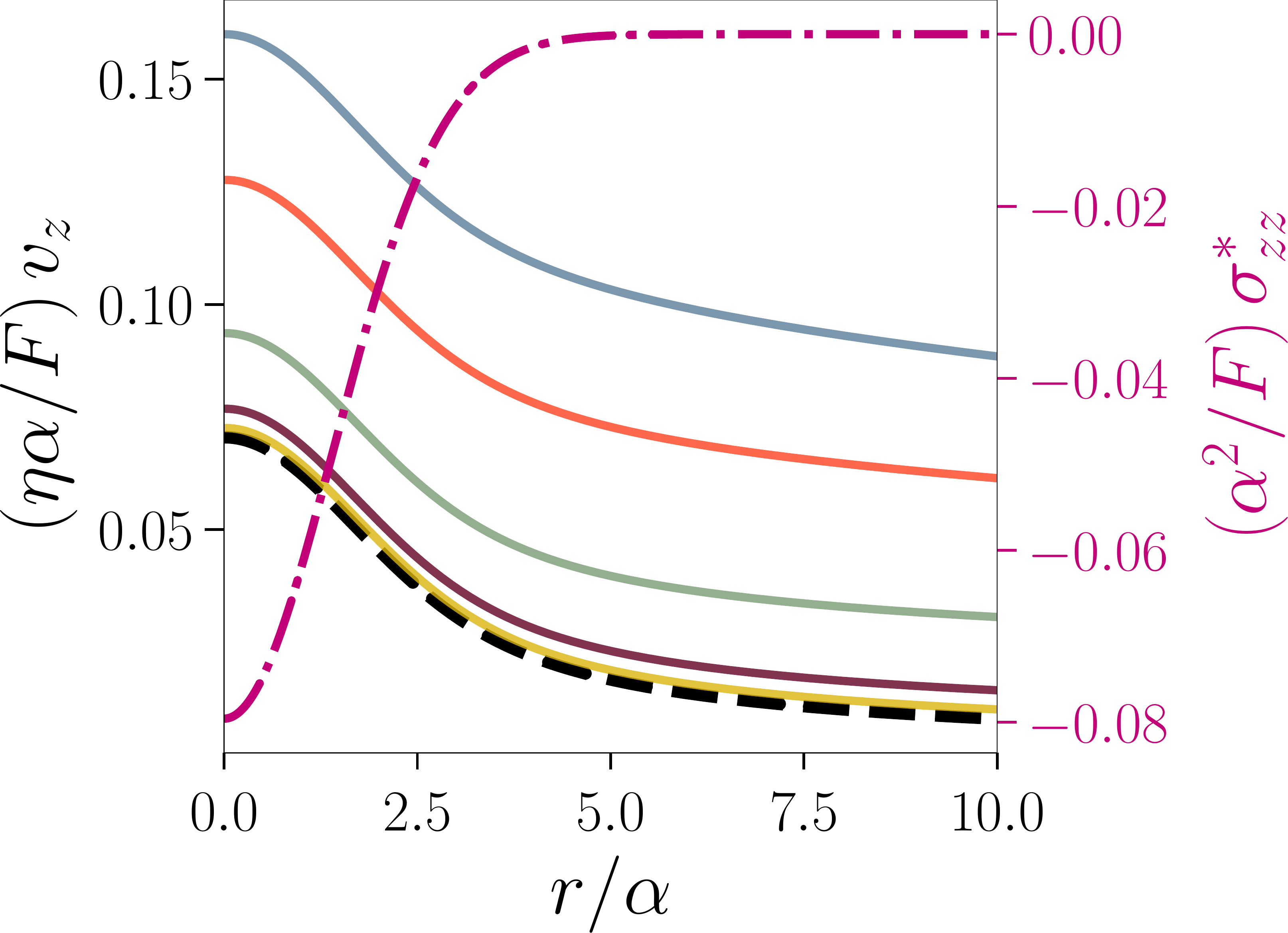}\\
        \hspace{1cm}\\
        \includegraphics[width=\linewidth]{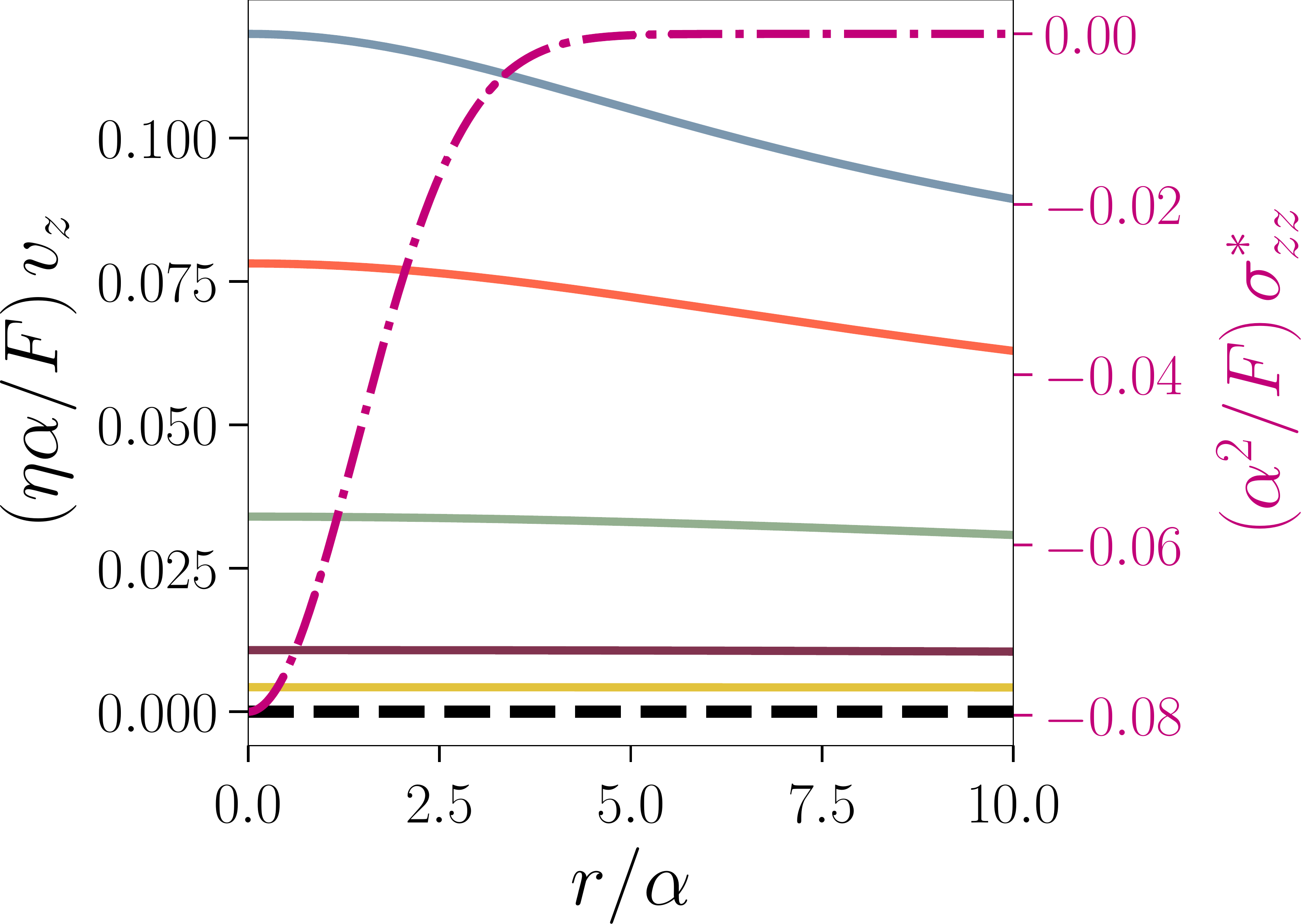}
    \end{subfigure}\\
    \begin{subfigure}{0.1\linewidth}
        \caption{}
        \label{fig:curved_membrane}
    \end{subfigure}
    \begin{subfigure}{\barsize}
        \includegraphics[width=0.66\linewidth]{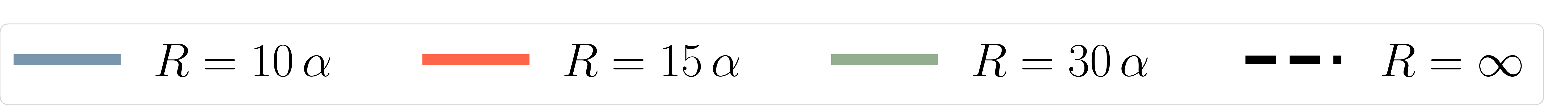}
    \end{subfigure}
    \begin{subfigure}{\panesize}
        \includegraphics[width=\linewidth]{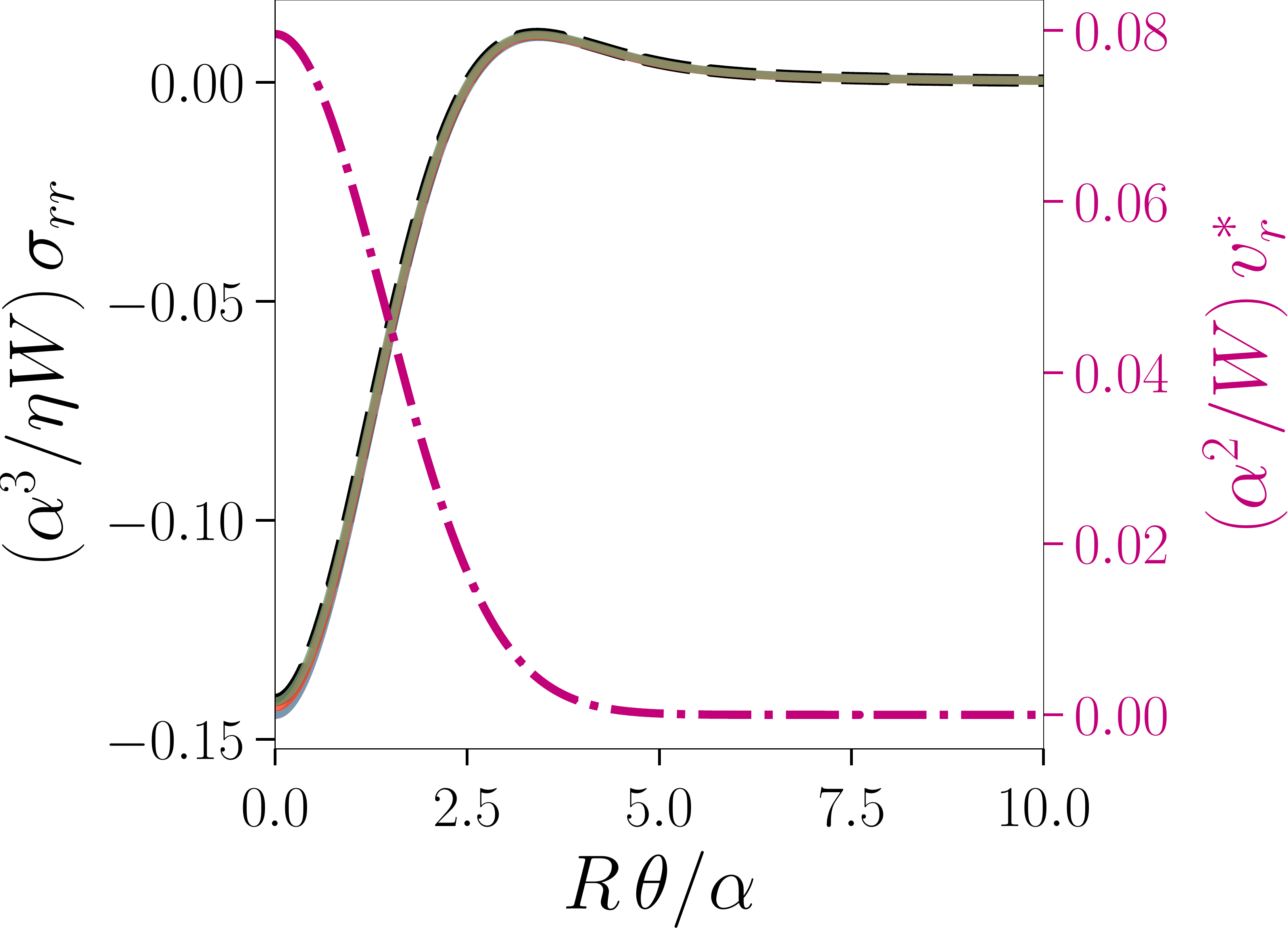}
    \end{subfigure}
    \begin{subfigure}{\halfpanesize}
        \includegraphics[width=\linewidth]{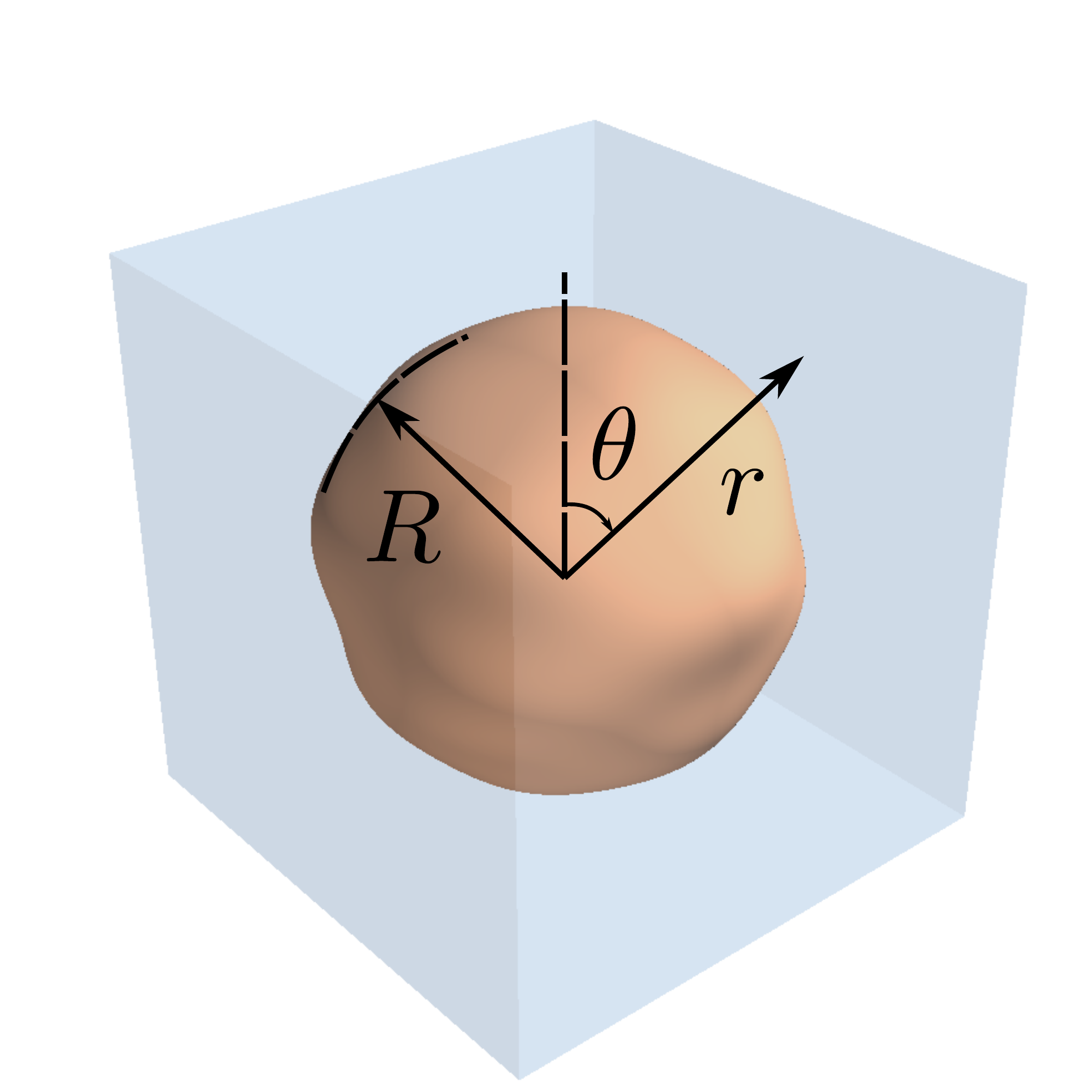}
        \includegraphics[width=\linewidth]{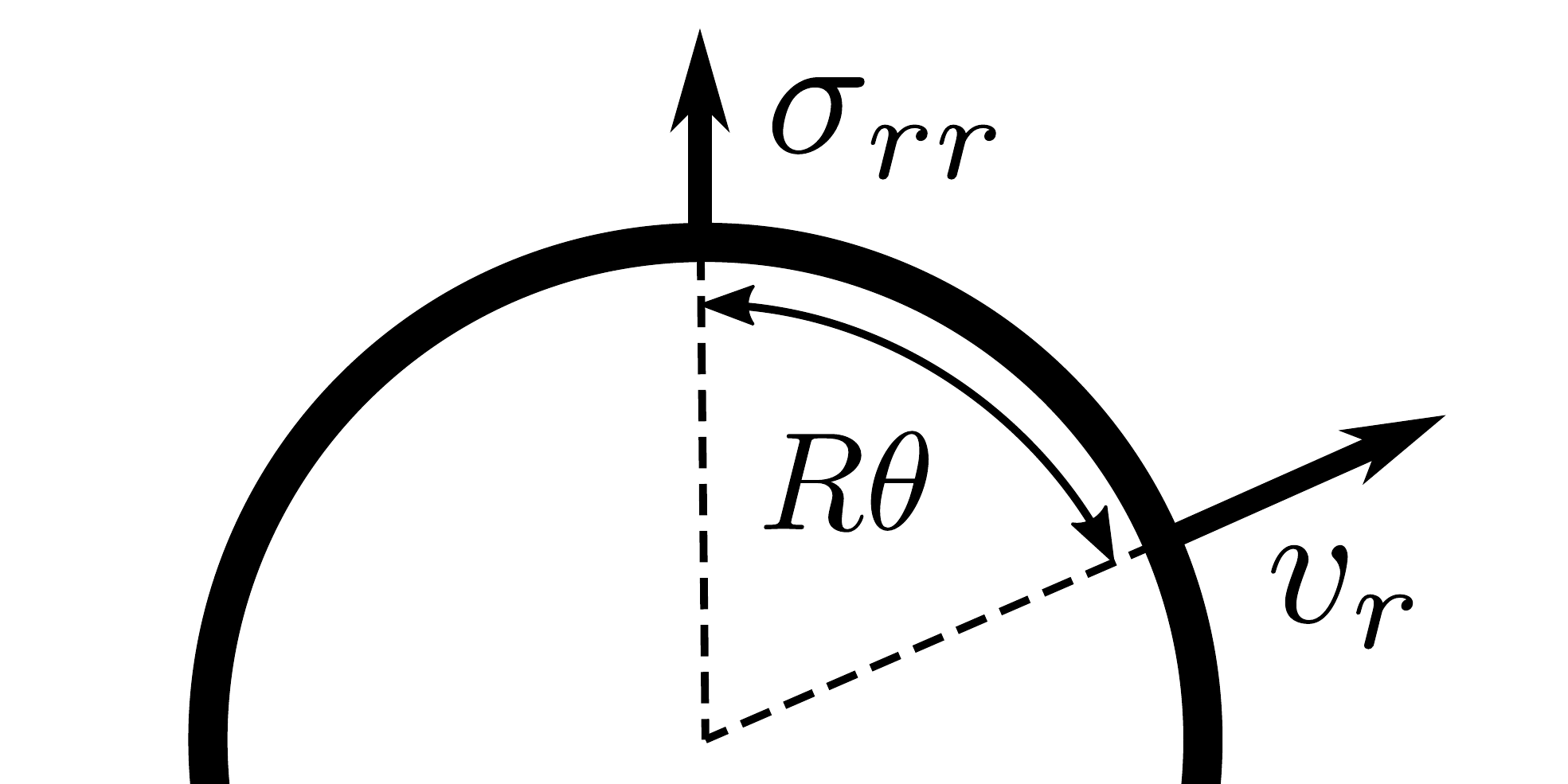}
    \end{subfigure}
    \begin{subfigure}{\panesize}
        \includegraphics[width=\linewidth]{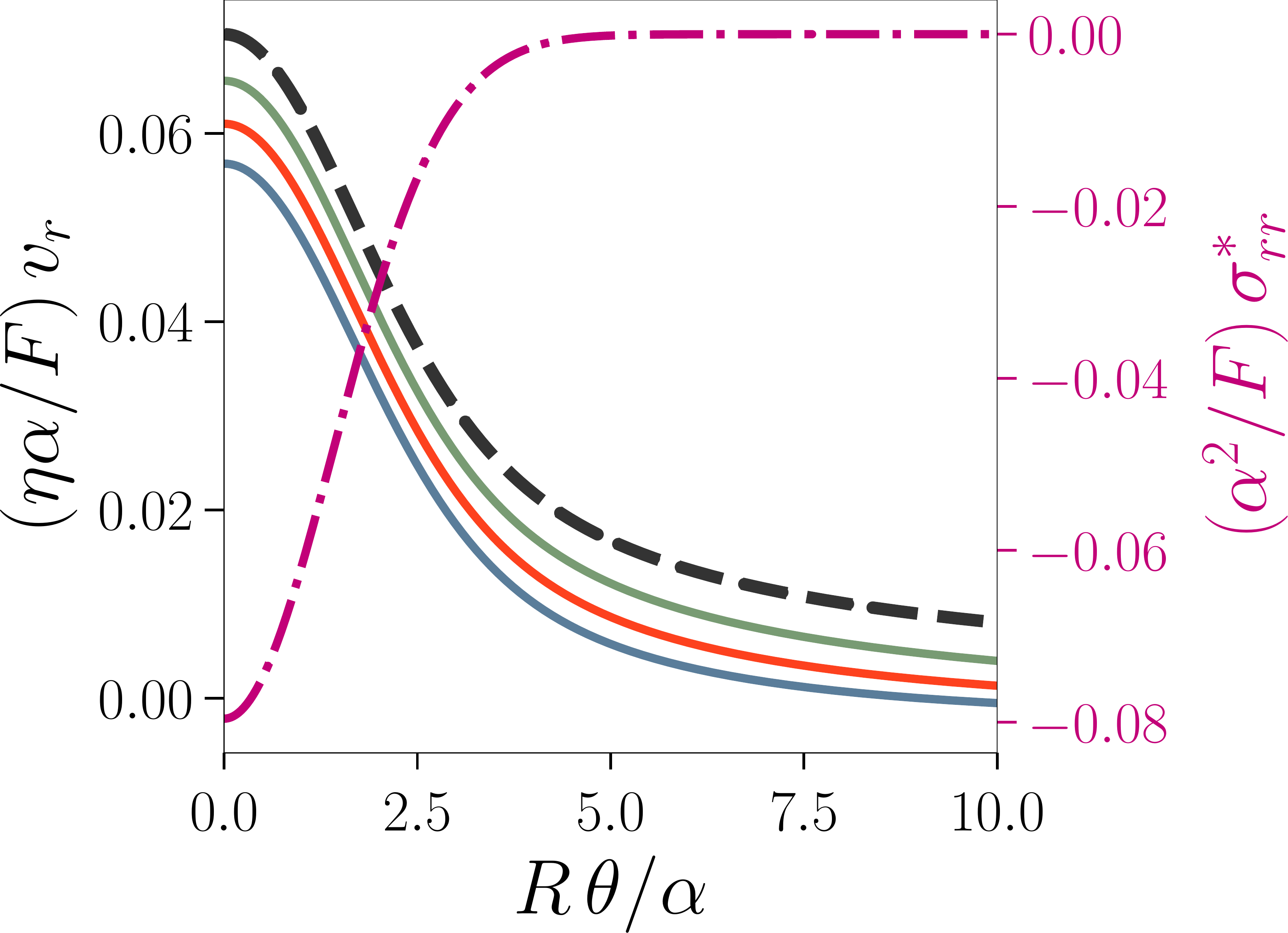}
    \end{subfigure}
%    \end{figure*}
%    \begin{figure*}
%    \ContinuedFloat
%    \centering
    \caption{\footnotesize Stress/velocity distribution on the surface of the membrane in response to a Gaussian velocity/stress boundary conditions (Eq. (\ref{eq:gaussian_bc})) for membranes suspended in a solvent with the viscosity $\eta$. Results are given for (\textbf{a}) a single planar membrane, (\textbf{b}) a pair of parallel planar membranes with the given separations, $h$, and (\textbf{c}) spherical vesicles of given radii. For parallel membranes, the boundary conditions are applied on the plane at $z=0$, and velocity and stress distribution are given either on the same plane, or on the opposing one (note the schematics in the middle column). For the spherical vesicles, rotationally-symmetric Gaussian boundary condition is applied at $\theta=0$. For all cases, the applied boundary conditions are also shown on the second axis (magenta dot-dashed lines). For comparison, results from ({a}), corresponding to $h\rightarrow \infty$ or $R\rightarrow \infty$, are reproduced in other plots (black dashed lines).}
    \label{fig:membrane_hydro}
\end{figure*}

%\clearpage

\begin{figure*}[htbp]
    \centering
    \begin{subfigure}{0.05\linewidth}
        \caption{}
        \label{fig:compiled_friction_0.1}
    \end{subfigure}
    \begin{subfigure}{0.44\linewidth}
        \includegraphics[width=\linewidth]{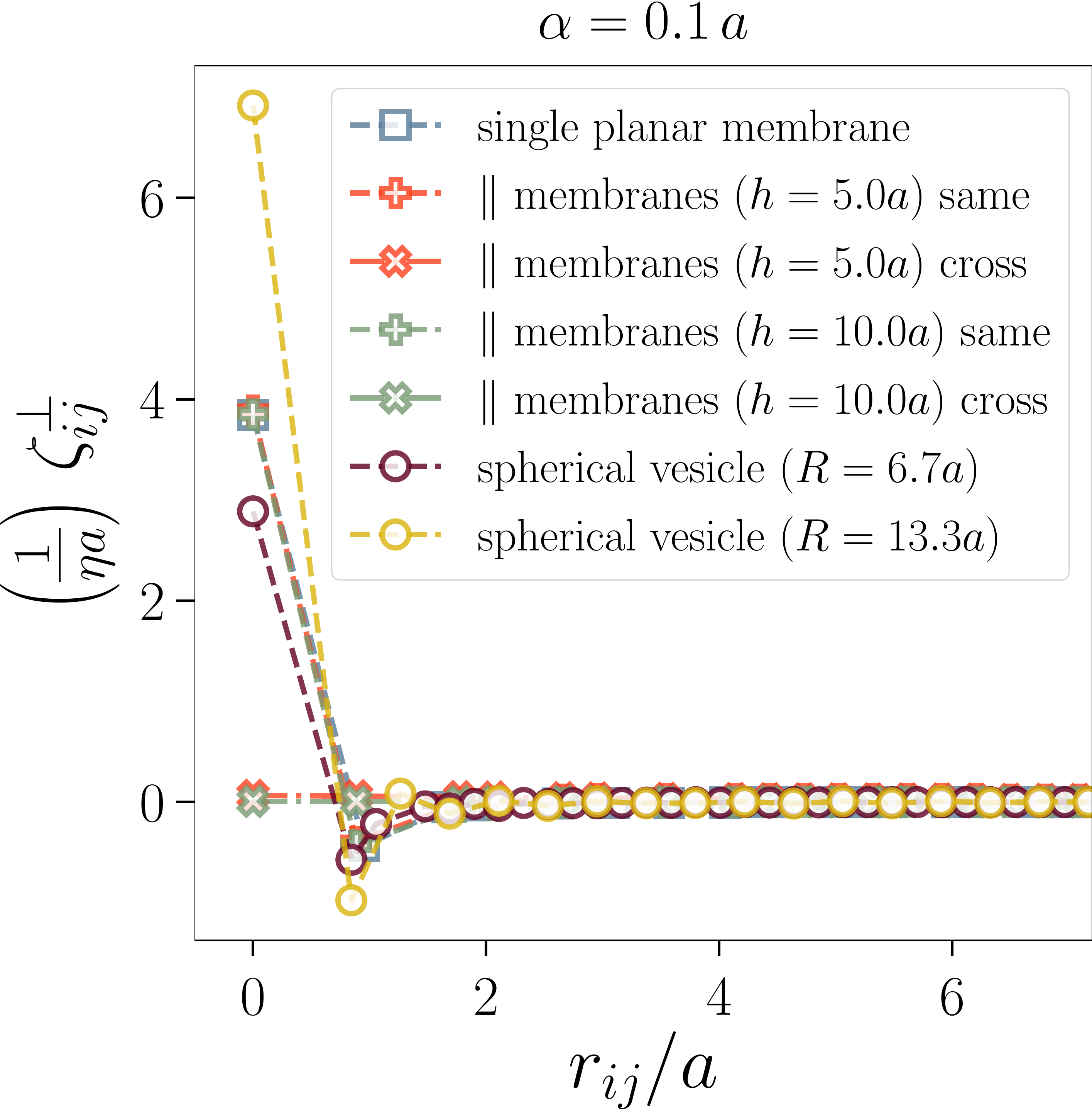}
    \end{subfigure}
    \begin{subfigure}{0.05\linewidth}
        \caption{}
        \label{fig:compiled_diffusion_0.1}
    \end{subfigure}
    \begin{subfigure}{0.44\linewidth}
        \includegraphics[width=\linewidth]{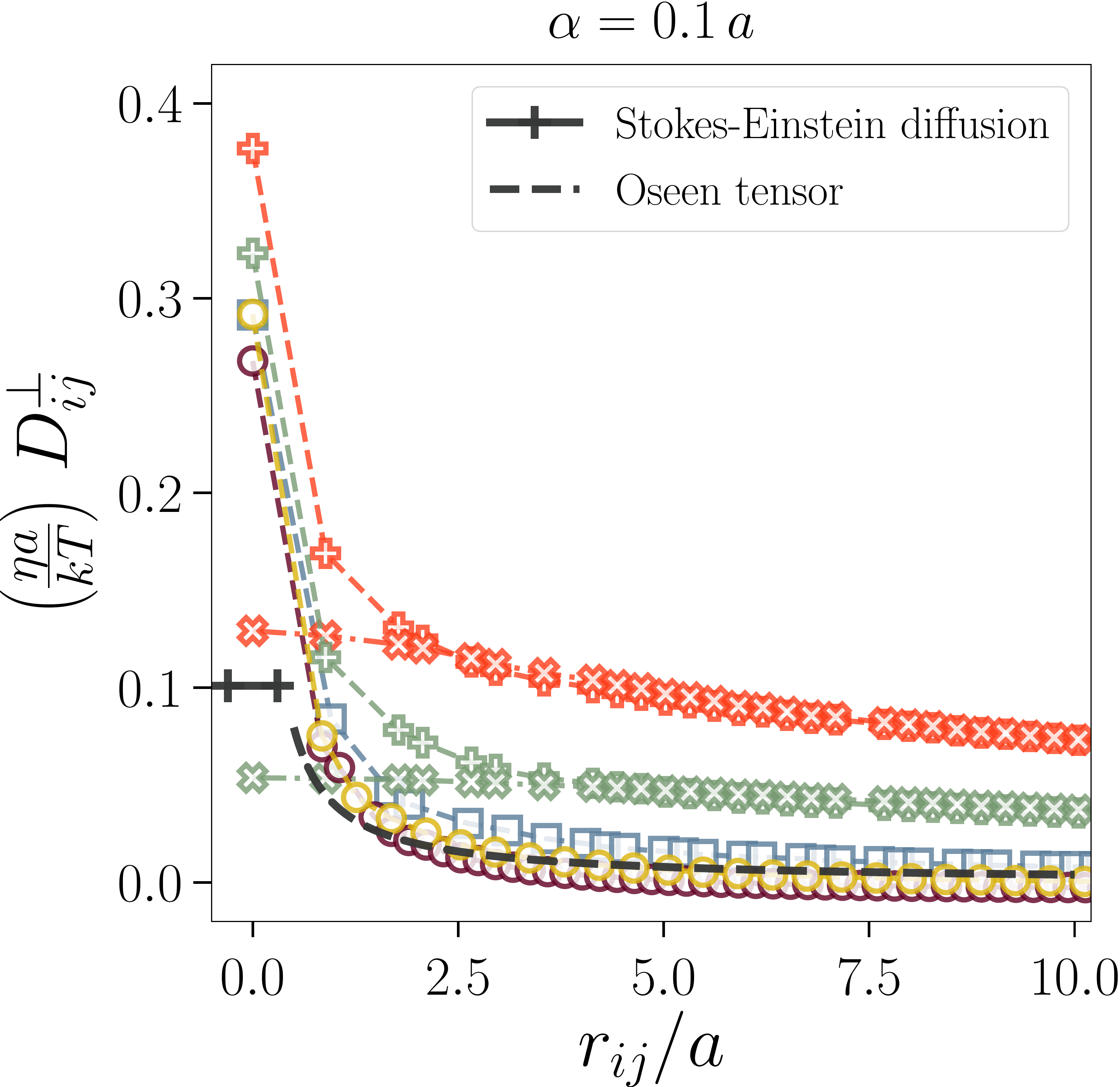}
    \end{subfigure}\\
    \hspace{1.0cm}\\
    \hspace{1.0cm}\\
    \begin{subfigure}{0.05\linewidth}
        \caption{}
        \label{fig:compiled_friction_0.5}
    \end{subfigure}
    \begin{subfigure}{0.44\linewidth}
        \includegraphics[width=\linewidth]{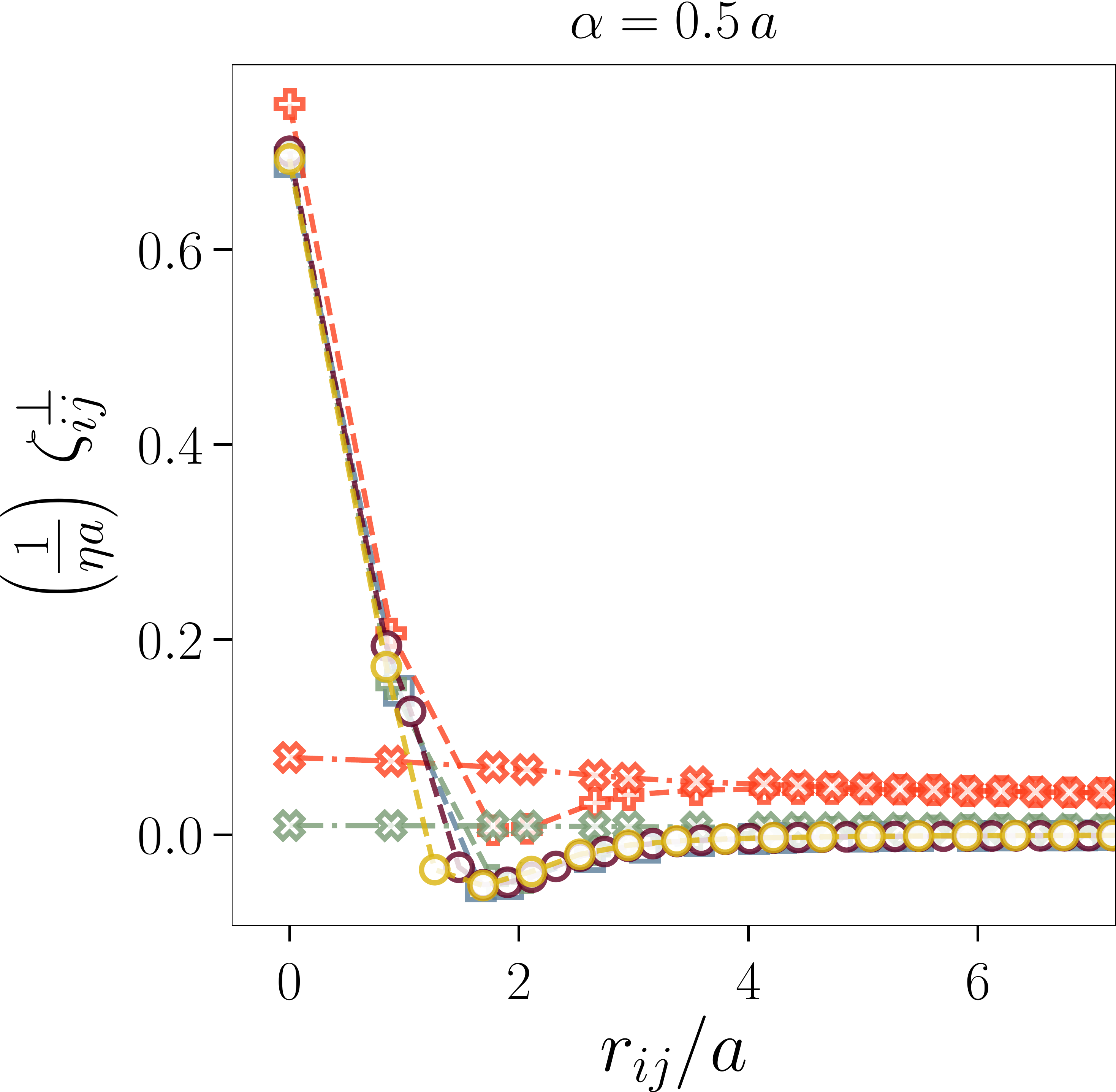}
    \end{subfigure}
    \begin{subfigure}{0.05\linewidth}
        \caption{}
        \label{fig:compiled_diffusion_0.5}
    \end{subfigure}
    \begin{subfigure}{0.44\linewidth}
        \includegraphics[width=\linewidth]{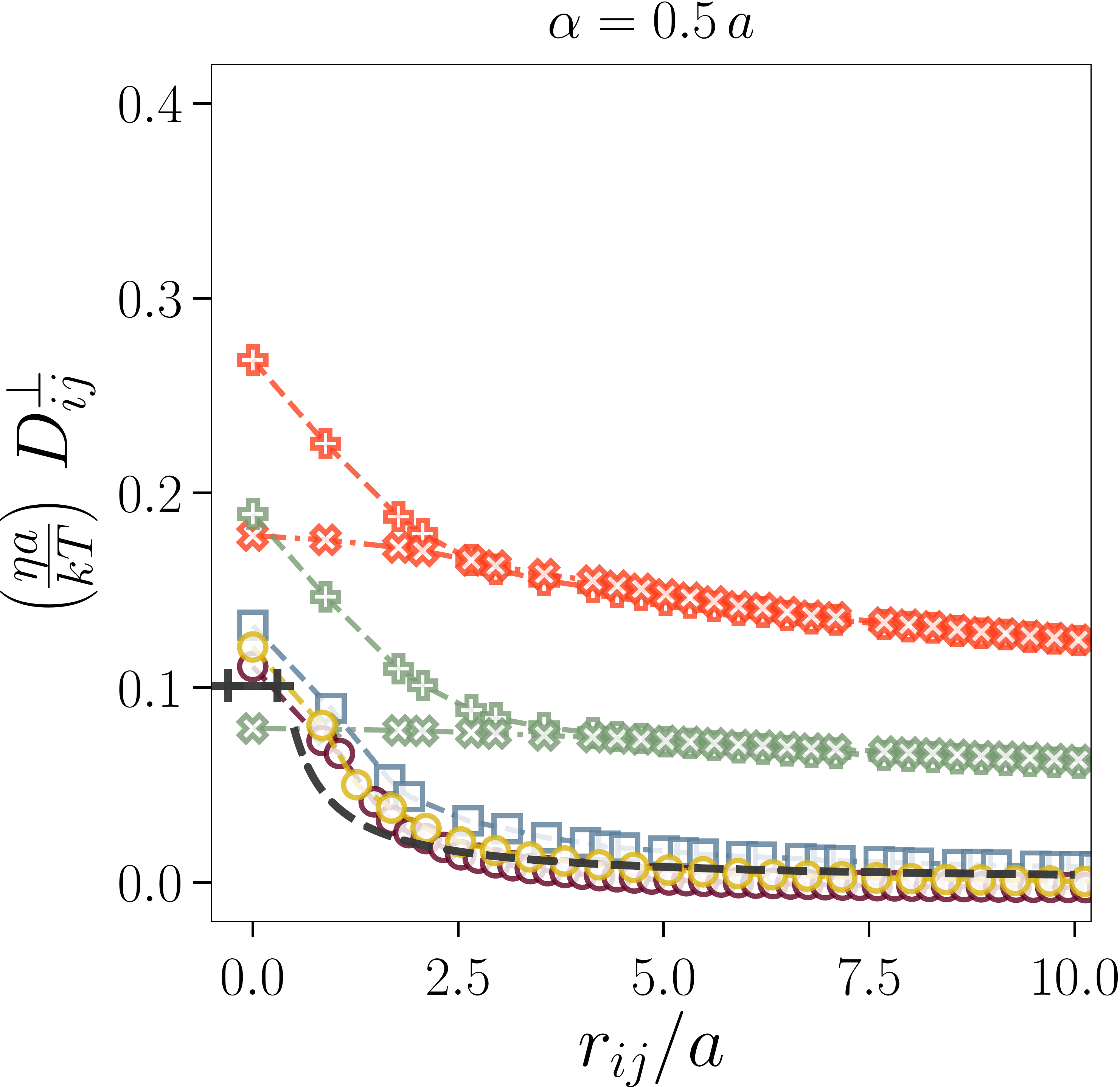}
    \end{subfigure}
    \caption{\footnotesize Compilation of numerical values of the out-of-plane components of (\textbf{a}, \textbf{c}) friction and (\textbf{b}, \textbf{d}) diffusion tensors, as a function of pairwise particle distances, $r_{ij}$, calculated on a hexagonal assembly of particles with the lattice parameter $a$. Results are given for a single planar membrane, two sets of parallel membranes with the given separation, and two spherical vesicles with the given radii. The pairwise distances are measured similar to the schematics in Fig. \ref{fig:membrane_hydro}. For these calculations, the parameter $\alpha$ is chosen equal to $0.1 a$ in (\textbf{a}) and (\textbf{b}) and $0.5 a$ in (\textbf{c}) and (\textbf{d}). Local diffusion coefficient from Stokes-Einstein relations (based on the effective radius of a particle on the surface), as well as hydrodynamic interactions predicted by the Oseen tensor are given for reference. The legend is shared between plots.}
    \label{fig:compiled_friction_diffusion}
\end{figure*}
%\clearpage

\clearpage

\end{document}